\documentclass[11pt,reqno]{amsart}

\usepackage{amsxtra}
\usepackage{amsfonts}
\usepackage{amsmath,amssymb,stmaryrd} 
\usepackage[english]{babel} 
\usepackage{graphicx} 
\usepackage{mathrsfs} 
\usepackage{layout}
\usepackage{color}

\newcommand{\N}{\mathbb{N}} 

\newcommand{\ba}{\begin{eqnarray}} 
\newcommand{\ea}{\end{eqnarray}}
\newcommand{\be}{\begin{equation}}
 \newcommand{\ee}{\end{equation}}
\newcommand{\bdm}{\begin{displaymath}}
\newcommand{\edm}{\end{displaymath}} 
\newcommand{\brr}{\begin{array}}
\newcommand{\err}{\end{array}}

\newcommand{\bea}{\begin{eqnarray}}
\newcommand{\eea}{\end{eqnarray}}
\newtheorem{definition}{Definition}[section]

\newcommand{\bml}{\begin{gather}} \newcommand{\eml}{\end{gather}}

\newtheorem{teo}{Theorem}[section]
\newtheorem{prop}{Proposition}[section]
\newtheorem{rem}{Remark}[section]

\newcommand{\spaz}{\vspace{.5cm} \noindent}

\numberwithin{equation}{section}

\newcommand{\R}{{\mathbb R}}

\newcommand{\pa}{\partial}

\begin{document}
\begin{center}
{\Large\textbf{A K}}{\Large\textbf{ac}} {\Large\textbf{M}}{\Large\textbf{odel for}} {\Large\textbf{F}}{\Large\textbf{ermions }}
\end{center}
\vspace{0.5cm}
\begin{center}
M. Colangeli\footnote{Dipartimento di Elettronica e Telecomunicazioni - Politecnico di Torino, Italy   \\ \emph{e-mail:} matteo.colangeli@polito.it},
F. Pezzotti\footnote{ Dipartimento di Matematica ``G. Castelnuovo'' -  ``Sapienza'', Universit$\grave{\text{a}}$ di Roma, Italy \\ \emph{e-mail:} pezzotti@mat.uniroma1.it},
M. Pulvirenti\footnote{Dipartimento di Matematica ``G. Castelnuovo'' -  ``Sapienza'', Universit$\grave{\text{a}}$ di Roma, Italy \\  \emph{e-mail:} pulvirenti@mat.uniroma1.it }
\end{center}
\vspace{0.5cm}
\begin{center}
\textbf{Abstract}
\end{center}
{\footnotesize We introduce a stochastic $N$-particle system and show that, as $N\to \infty$, an effective description ruled by the homogeneous fermionic Uehling-Uhlenbeck equation is recovered. The particle model we consider is the same as the Kac model for the homogeneous Boltzmann equation with an additional exclusion constraint taking into account the Pauli Exclusion Principle.

}

\tableofcontents

\section{Introduction}
\label{sec:sec1}
\setcounter{equation}{0}    
\def\theequation{1.\arabic{equation}}

One of the most important and challenging mathematical problems in Kinetic Theory is the rigorous derivation of the kinetic equations  from the basic mechanical laws. The first fundamental result in this direction was obtained in 1975 by O.  Lanford \cite{LANFORD}, who derived the Boltzmann equation for a system of hard spheres, for short times, in the Low-Density (or Boltzmann-Grad) limit  (see also \cite{S1}, \cite{U}, \cite{Pulv}, \cite{GSRT} and references therein). A similar result was also obtained for particle systems interacting via a two-body short-range, smooth potential \cite{King}, \cite{GSRT}, \cite{PSS}.

The only global in time result, refers to the special situation of an expanding cloud of a rare gas in the vacuum \cite{IP}.

On the other hand, a dense gas of weakly  interacting particles (weak-coupling limit) is expected to be described by the Landau equation. In this case there are no rigorous results. We only mention a very preliminary consistency result \cite{BPS}. 

Quantum systems are expected to be described by suitable Boltzmann equations in both Boltzmann-Grad and Weak-Coupling limits.  In the first case the Boltzmann equation is just the classical one, with the full quantum cross-section associated with the interaction potential. In the second, more interesting, case, the Boltzmann equation (U-U equation in the sequel) differs from the classical one because it takes into account the effects of the Bose-Einstein or Fermi-Dirac statistics. It  was heuristically introduced by Nordheim (1928) in \cite{nord} and Uehling and Uhlenbeck(1933) in \cite{UU}.

Concerning the rigorous derivation of the U-U equation starting  from an $N$-particle system evolving according to the Schr\"odinger equation, only formal or partial results are available up to now (see \cite{BCEP1},  \cite{ESY}, \cite{BCEP}, \cite{BCEP3}, \cite{Pulvi}, \cite{BCEP4}).

The U-U equation reads as

\begin{equation}
\left(
\partial_t + v \cdot \nabla_x
\right)
f=Q_{\theta}(f,f,f),
\end{equation}
where\begin{align}
&
\!\!\!\!Q_{\theta}(f,f,f)(x,v)
=\int dv_1 \, \int d\omega \; B_{\theta}(v-v_1;\omega)
\\
&
 \nonumber
\!\!\!\!\left[f(x,v')f(x,v_1')(1+\alpha \theta
f(x,v))  (1+\alpha\theta f(x,v_1))
\right.
\left.
- f(x,v)f(x,v_1)(1+ \theta \alpha 
f(x,v'))(1+\alpha\theta f(x,v_1'))
\right] ,
\end{align}
where $f(x,v,t)$ is the probability distribution of a test particle in the classical phase space ( $(x,v,t) $ denote posistion, momentum and time) describing the time evolution of the Wigner transform of a quantum state.
Here $\theta=+1$ or $\theta=-1$,
for the Bose-Einstein or the Fermi-Dirac statistics and $\alpha=(2 \pi \hbar)^3$, where $\hbar$ is the Planck constant. 

Finally $(v,v_1) \to (v',v_1')$ is the transition due to an elastic collision with scattering vector $\omega \in S^2$ and $B$ is proportional to the symmetrized cross-section (associated with the pair interaction potential) in the Born approximation. More precisely,  assuming the interaction $\phi$ to be real and spherically symmetric, $B_{\theta}(v-v_1;\omega)$ is given by:
\begin{eqnarray}\label{Bexpression}
\frac{1}{8\pi^2\hbar^4}\, |(v-v_1)\cdot \omega|\, \left|\widehat{\phi}\left(\frac{\left|(v-v_1)\cdot \omega\right|}{\hbar}\right)+\theta \widehat{\phi}\left(\frac{\left|\omega\cdot[(v-v_1)\cdot \omega]-(v-v_1)\right|}{\hbar}\right)\right|^2\, \chi_{\mathcal{S}^2_{-}}(\omega),
\end{eqnarray}
where $\chi_{\mathcal{S}^2_{-}}(\cdot)$ is the characteristic function of the set
$\mathcal{S}^2_{-}=\{\omega:\ |\omega|=1\ \text{and}\ (v-v_1)\cdot\omega\leq 0\}$ (see e.g. \cite{DarioMario}).

In 1956 M. Kac proposed a stochastic particle model yielding, in a suitable scaling limit (of Mean-Field type),  the classical Boltzmann equation (see \cite{KacA}). The purpose was to understand the delicate passage from an $N$-particle system to a one-particle kinetic description, in an easier context. 

The model consists of a set of $N$ particles with velocities $V_N=(v_1 \dots v_N)$. The positions are ignored. The evolution is the following. At an exponential time pick a pair of particles (say $i$ and $j$), select a scattering vector 
 $\omega \in S^2$  and perform the transition $(v_i,v_j) \to (v_i',v_j')$ with the usual elastic collision rules. More precisely, if  $W^N(V_N,t)$ is a probability distribution, its time evolution obeys the following Master equation
\be
\partial_t W^N=\frac{1}{N}L_N^{\mathbf{K}} W^N \label{MEkac}
\ee
where
\be
L_N^{\mathbf{K}} W^N(V_N)=\sum_{1\leq i<j\leq N}\int_{\mathcal{S}^{2}} d\omega B(v_i-v_j; \omega)  \left[  W^N(V_N^{i.j}) - W^N(V_N) \right]
 \label{LNK}
\ee
and
$$
V_N^{i.j}=\{v_1,...,v_i',...,v_j',...,v_N\}.
$$

It is possible to show that, in the limit $N \to \infty$, the $k$-particle marginals $f^N_k(V_k,t)$  of $W_N(V_N,t)$, converge to a sequence of marginals 
$f_k(V_k,t)$. Moreover, if  initially $W^N(\cdot, 0)=f_0^{\otimes N} $,  where $f_0$ is a one-particle distribution (namely the particles are initially independently distributed) then  $f_k (\cdot, t)=f ^{\otimes k} (t)$, where $f(t)$ solves the Boltzmann equation with cross-section $B$.

The Kac model has been widely investigated, see the recent paper  \cite{MM} and references quoted therein.

In the same spirit  we modify the Kac model including an exclusion constraint mimicing the Pauli exclusion principle with the scope of deriving the U-U equation for Fermions. The exclusion principle is implemented by introducing a grid of side $\delta$ in the one-particle phase space. Then we consider only admissible configurations, namely those exhibiting at most one particle per cell. The random transition $(v_i,v_j) \to (v_i',v_j')$ takes place only if the final configuration $V_N^{i.j}=\{v_1,...,v_i',...,v_j',...,v_N\}$ is still admissible.  Then we perform the limit $N \to \infty$, $ \delta \to 0$ with fixed $\alpha=N \delta^3$, $\alpha\in(0,1)$. In doing this we follow the Lanford strategy, namely we first derive a hierarchy of equations for the marginals $f^N_k(V_k,t)$ of the time evolution of an $N$-particle state. Such a derivation is straightforward, but tedious. The details are presented in the Appendix.  Then we bound, locally in time,  the series expansion expressing the solution of the hierarchy. We note that, due to the exclusion principle which gives us authomatically a bound on the density, we can treat arbitrary times by introducing a suitable family of norms. Finally we exploit the term by term convergence (see Section \ref{sec:sec7} below) by piling up a finite number of series expansions, each of them converging for a short time.

Our result is proven under suitable assumptions on the convergence of the initial data. In Section \ref{sec:sec8} we provide examples of initial states fulfilling the hypotheses of the main theorem. 

It may be worth to underline that our analysis, as well as the one suggested by the original Kac model, deals with the homogeneous U-U equation ($f(t,x,v)=f(t,v)$). Actually, the dynamics described by the Kac model is related to the interaction part of the popular numerical scheme called \emph{Direct Simulation Method} (see e.g. \cite{Pulv} for a mathematical description and \cite{PWZR} for the convergence). Therefore the results of the present paper could be of some interest for numerical problems associated with the simulation of the U-U equation. 

We notice that a non-homogeneous version of the fermionic Boltzmann equation with discrete velocities has been derived in \cite{DEP}  starting from a stochastic particle system on the lattice.

We finally remark that a model similar to the one considered in the present paper can be studied for Bosons as well. However, in this case, statistics induces particle concentration and the mathematical analysis is harder.

 \section{The model and the scaling limit}
\label{sec:sec2}
\setcounter{equation}{0}    
\def\theequation{2.\arabic{equation}}

Consider an $N$-particle system whose state space is $\R^{3N}$. 
A state of the system is a vector $V_N=\left(v_1,\dots, v_N\right)\in \R^{3N}$. We introduce a partition
of the one particle phase
-space $\R^3$, made by cubic cells $\Delta$ of side $\delta>0$ and denote by $\Delta(v_i)$ or $\Delta_i$ the cell associated with $v_i$, namely $v_i\in \Delta_i$.
Moreover, the occupation number $N_\Delta(V_N)$ of the cell $\Delta$ in the $N$-particle configuration $V_N=\left(v_1,\dots, v_N\right)$ can be written as:
\begin{equation}\label{eqN}
N_\Delta(V_N):=\sum_{\ell=1}^N \chi_\Delta(v_\ell),
\end{equation}
where $\chi_\Delta(\cdot)$ is the characteristic function of the cell $\Delta$.
Finally, for any $\ell,m\in \{1,\dots, N\}$, with $\ell\neq m$, we set:
\be
\left\{
  \begin{array}{ll}
\chi_\delta(v_\ell,v_m)=1, & \hbox{if $v_\ell$ and $v_m$ are in the same cell
}\\
\chi_\delta(v_\ell,v_m)=0, & \hbox{otherwise}
  \end{array}
  \right.
\ee
and $\overline{\chi}_\delta(v_\ell,v_m)=1-\chi_\delta(v_\ell,v_m)$. \\

A generic configuration $V_N=\{v_1\dots v_N\},\ v_i\in \R^3$ is said to be \textit{admissible} if \ \ $\overline{\chi}_\delta(V_N)=1$, where:
\begin{equation}\label{CHAR}
\overline{\chi}_\delta(V_N)=\prod_{1\leq \ell_1<\ell_2\leq N}\overline{\chi}_\delta(v_{\ell_1},v_{\ell_2}).
\end{equation}
We denote by $\mathcal{A}_\delta^N$ the set of all admissible configurations. On such a space we consider the jump process:
\be
\{v_1,...,v_i,...,v_j,...,v_N\}=V_N\rightarrow V_N^{i.j}=\{v_1,...,v_i',...,v_j',...,v_N\}
\ee
where
\be\label{collDEF}
\left\{
  \begin{array}{ll}
v'_i=v_i-\omega\left[ (v_i-v_j)\cdot \omega\right], 
\\
v'_j=v_j+\omega\left[ (v_i-v_j)\cdot \omega\right], 
  \end{array}
  \right.
\ee
are the outgoing velocities arising from an elastic collision with scattering vector $\omega$. Clearly, the transition $(v_i,v_j)\to (v'_i, v'_j)$ preserves total momentum and energy. 

The generator $\frac{1}{N}L_N^G$ of the process is given by:
\be
L_N^G \phi^N(V_N)=\sum_{1\leq i<j\leq N}\int_{\mathcal{S}^{2}} d\omega B(v_i-v_j; \omega)\overline{\chi}_\delta(v'_i,v'_j)(1-N_{\Delta'_i}(V_N))(1-N_{\Delta'_j}(V_N))\left[(\phi^N(V_N^{i.j})-\phi^N(V_N)\right], \label{LNGnumero}
\ee
where $\Delta'_i=\Delta(v'_i)$, $\mathcal{S}^{2}:=\{\omega\in \R^3:\ \ |\omega|=1 \}$
and the function $B$ is such that:
\be\label{Bsymm}
B(v_i-v_j;\omega)=B(|v_i-v_j|;\omega).
\ee
Due to the relationship between the function $B$ appearing in the U-U equation and the symmetrized cross-section of the interaction potential (see (\ref{Bexpression})), in all the paper, by a little abuse of language, we will always refer to the function $B$ as ''cross-section''. Precise assumptions on it will be stated later on.

Note that if $V_N\in \mathcal{A}_\delta^N$, then:
\be
 N_{\Delta'_i}(V_N)\in \{0,1\}\ \ \ \ \ \ \ \text{and}\ \ \ \ \ \ \ N_{\Delta'_j}(V_N)\in \{0,1\}.\label{numeri1}
\ee
Therefore, by (\ref{LNGnumero}) it follows that also $V_N^{i,j}\in \mathcal{A}_\delta^N$. 
\begin{rem}\label{NOtrivialTRANS}
It can be easily verified that the two trivial transitions $(v_i,v_j)\to (v'_i, v'_j)$ with $\Delta'_i=\Delta_i$, $\Delta'_j=\Delta_j$ and $\Delta'_i=\Delta_j$, $\Delta'_j=\Delta_i$  are \emph{not} taken into account by the above process (because $\overline{\chi}_\delta(v'_i,v'_j)(1-N_{\Delta'_i}(V_N))(1-N_{\Delta'_j}(V_N))=0$ in such situations).
\end{rem}

The time evolution of a symmetric probability distribution $W^N(V_N,t)$ describing the statistical state of the system 
is governed by the following Master equation:

\be
\partial_t W^N=\frac{1}{N}\left(L_N^G\right)^*W^N \label{MEn}
\ee
where $\left(L_N^G\right)^*$ is the adjoint of $L_N^G$, namely:
\begin{eqnarray}
&&\!\!\!\!\!\!\!\!\!\left(L_N^G\right)^* W^N(V_N)=\!\!\sum_{1\leq i<j\leq N}\int_{\mathcal{S}^{2}} d\omega B(v_i-v_j; \omega)\left[\overline{\chi}_\delta(v_i,v_j)(1-N_{\Delta_i}(V_N^{i.j}))(1-N_{\Delta_j}(V_N^{i.j}))W^N(V_N^{i.j})\right.+\nonumber\\
&&\qquad\qquad \qquad \ \ \ \ \ \ \ \ \ \ \ \ \ \ \ \ \ \ \ \ \left.-\overline{\chi}_\delta(v'_i,v'_j)(1-N_{\Delta'_i}(V_N))(1-N_{\Delta'_j}(V_N))W^N(V_N)\right].\label{LNnumero}
\end{eqnarray}
As we already noticed, due to the form of the generator $L_N^G$ in (\ref{LNGnumero}), the transition $V_N\to V_N^{i,j}$ is allowed if and only if both the departure and the arrival configurations are admissible. In fact, this is the crucial feature we are interested in and, by construction, it holds for $\left(L_N^G\right)^*$ as well.  Therefore, by choosing:
\begin{equation}\label{suppINDAT}
supp\,  W_0^N \subseteq 
\mathcal{A}_\delta^N,
\end{equation}
we are ensured that:
 \begin{equation}
\label{KacFERMIONS1}
 supp \ W^N(t) \subseteq \mathcal{A}_\delta^N,\ \ \  \forall\ \ t>0.
 \end{equation}
Instead of considering the Master equation (\ref{MEn}), we prefer to introduce the operator:
 \begin{eqnarray}
L_N W^N(V_N)=\sum_{1\leq i<j\leq N}\int_{\mathcal{S}^{2}} d\omega B(v_i-v_j; \omega)\left[\overline{\chi}_\delta(V_N)\, W^N(V_N^{i.j})-\overline{\chi}_\delta(V_N^{i.j})\, W^N(V_N)\right], \label{LN}
\end{eqnarray}
and consider the dynamics:
\be
\partial_t W^N=\frac{1}{N}L_N W^N, \label{ME}
\ee
where, with a little abuse of notation, we used the same symbol to denote the solution of equations (\ref{MEn}) and (\ref{ME}). In fact, that is just innocent since 
 (\ref{LN}) is almost equivalent to (\ref{LNnumero}) and, in particular,  property (\ref{KacFERMIONS1}) holds for (\ref{ME}) as well. The only difference between $\left(L_N^G\right)^*$ and $L_N$ concerns the two trivial transitions $(v_i,v_j)\to (v'_i, v'_j)$ with $\Delta'_i=\Delta_i$, $\Delta'_j=\Delta_j$ and $\Delta'_i=\Delta_j$, $\Delta'_j=\Delta_i$. In fact, 
in (\ref{LNnumero}) such transitions are \emph{not} taken into account 
while in (\ref{LN}) they are (because $\overline{\chi}_\delta(V_N^{i.j})=1$). Nevertheless, such transitions do not change the occupation numbers and, as a consequence, they are irrelevant for the dynamics.  Therefore, since we can choose freely one of the two expressions, from now on we will use (\ref{LNnumero}) which will be more convenient for our purposes.\\

The main goal of the present work is to investigate the limit as $N\to\infty$ of the stochastic dynamics presented above to recover, in a suitable sense, a one particle description ruled by the Uehling-Uhlenbeck equation. Simultaneously, we remove the grid by looking at the behavior as $\delta\to 0$ in such a way that $N\delta^3=\alpha\in(0,1)$.\\

\section{BBGKY hierarchy and formal asymptotics}
\label{sec:sec3}
\setcounter{equation}{0}    
\def\theequation{3.\arabic{equation}}

As usual in kinetic theory, we establish a hierarchy of equations (\emph{BBGKY hierarchy} in the sequel) for the marginals of the distribution $W^N(t)$ defined as:
$$
f_k^N(v_1,...,v_k,t)=\int W^N(v_1,...,v_N,t) dv_{k+1}...dv_{N},\ \ \ \ \ \ \ k=1,\dots, N,
$$
with $f_k^N\equiv 0$ for $k\geq N +1$.
Thus, by (\ref{ME}) and (\ref{LN}) we get:
\be
\partial_t f_k^N=\frac{1}{N}\sum_{1\leq i<j\leq N}\int dv_{k+1}...\int dv_{N} \int d\omega B(v_i-v_j; \omega)[\underbrace{\overline{\chi}_\delta(V_N) W^N(V_N^{i.j})}_{\rm G}-\underbrace{\overline{\chi}_\delta(V_N^{i.j}) W^N(V_N)}_{\rm L}] \label{LNrid}
\ee
where the contributions of the \textit{Gain} and \textit{Loss} terms, respectively $G$ and $L$, are made explicit. After some easy but quite tedious calculations, we get the desired BBGKY hierarchy:
\be
\partial_t f_k^N=\frac{1}{N}L_{k}^{N} f_{k}^N+\frac{1}{N}\sum_{s=1}^{2}L_{k,k+s}^{N} f_{k+s}^N+\sum_{s=1}^{3}C_{k,k+s}^{N} f_{k+s}^N \label{bbgky}
\ee
where
\begin{eqnarray}
&&L_{k}^{N} :=L_{k}^{N,+}-L_{k}^{N,-},\ \  \ \ \ \ L_{k,k+s}^{N} :=L_{k,k+s}^{N,+}-L_{k,k+s}^{N,-},\ \ \text{for}\ \ s=1,2\ \nonumber\\
&& C_{k,k+s}^{N} :=C_{k,k+s}^{N,+}-C_{k,k+s}^{N,-},\ \ \text{for}\ \ s=1,2,3
 \label{opDEF}
\end{eqnarray}
The superscript $^+$ stands for the $gain$ contribution and the superscript $^-$ stands for the $loss$ contribution. The explicit expression of operator $L_k^N$ is:
\begin{eqnarray}
\left(L_k^{N} f_k^N\right)(V_k)=\sum_{1\leq i<j\leq k} \int d\omega B(v_i-v_j; \omega)\ [\underbrace{\overline{\chi}_\delta(V_k)
f_k^N(V_k^{i.j})}_{L_k^{N,+} f_k^N}-\underbrace{\overline{\chi}_\delta(V_k^{i,j})
f_k^N(V_k)}_{L_k^{N,-} f_k^N}]
\label{charDECOMPOSITIONgain16}
\end{eqnarray}
and it is easy to check that (\ref{charDECOMPOSITIONgain16}) is exactly the $k$-particle version of operator $L_N$ defined in (\ref{LN}).
Concerning the other ''$L$-operators'', we have:
\begin{eqnarray}
&&\left(L_{k,k+1}^{N} f_{k+1}^N\right)(V_k)=-(N-k)\!\!\sum_{1\leq i<j\leq k}\int dv_{k+1}\int d\omega\,  B_{i.j}^{\omega}\ f_{k+1}^N(V_{k+1}^{i.j})
\times\nonumber\\
&&\ \ \    \times [\underbrace{
\ \overline{\chi}_\delta(V_k)\, 
\left(\chi_\delta(v_{i},v_{k+1})+\chi_\delta(v_{j},v_{k+1})\right)}_{L_{k,k+1}^{N,+} f_{k+1}^N}-\underbrace{\overline{\chi}_\delta(V_k^{i.j}) \, 
\left(\chi_\delta(v'_{i},v_{k+1})+\chi_\delta(v'_{j},v_{k+1})\right)}_{L_{k,k+1}^{N,-} f_{k+1}^N}]
\label{charDECOMPOSITIONgainLIN6}
\end{eqnarray}
and
\begin{eqnarray}
&&\left(L_{k,k+2}^{N} f_{k+2}^N\right)(V_k)=2(N-k)(N-k-1)\sum_{1\leq i<j\leq k}\int dv_{k+1}\int dv_{k+2} \int d\omega\,  B_{i.j}^{\omega}\ f_{k+2}^N(V_{k+2}^{i.j})\ \times\nonumber\\
&&\quad \qquad
\times \ 
[\underbrace{\overline{\chi}_\delta(V_k)\ 
\chi_\delta(v_{i},v_{k+1})\chi_\delta(v_{j},v_{k+2})}_{L_{k,k+2}^{N,+} f_{k+2}^N} - \underbrace{\overline{\chi}_\delta(V_k^{i.j})\ 
\chi_\delta(v'_{i},v_{k+1})\chi_\delta(v'_{j},v_{k+2})}_{L_{k,k+2}^{N,-} f_{k+2}^N}
],
\label{charDECOMPOSITIONgainNL5}
\end{eqnarray}
where we introduced the notation $B_{i.j}^{\omega}:=B(v_i-v_i;\omega)$.

On the other hand, the ''$C$-operators'' are given by:
\begin{eqnarray}
&&\!\!\!\!\!\!\left(C_{k,k+1}^{N} f_{k+1}^N\right)(V_k)=\nonumber\\
&&\!\!\!\!\!\!\ \ =\frac{N-k}{N}\sum_{1\leq i\leq k}\int dv_{k+1} \int d\omega \, B_{i.k+1}^{\omega}\ [\underbrace{\overline{\chi}_\delta(V_{k+1})\, 
f_{k+1}^N(V_{k+1}^{i.k+1})}_{C_{k,k+1}^{N,+} f_{k+1}^N}-\underbrace{\overline{\chi}_\delta(V_{k+1}^{i,k+1})\, f_{k+1}^N(V_{k+1})}_{C_{k,k+1}^{N,-} f_{k+1}^N} ],
\label{charDECOMPOSITIONgainNL8case3BIS}
\end{eqnarray}
\begin{eqnarray}
&&\left(C_{k,k+2}^{N} f_{k+2}^N\right)(V_k)=-\frac{(N-k)(N-k-1)}{N}\sum_{1\leq i\leq k}\int dv_{k+1}\int dv_{k+2} \int d\omega\,  B_{i.k+1}^{\omega}\ f_{k+2}^N(V_{k+2}^{i.k+1})\times\nonumber\\
&&  \times\ [\underbrace{\overline{\chi}_\delta(V_{k+1})\, 
\left(\chi_\delta(v_{i}, v_{k+2})+\chi_\delta(v_{k+2}, v_{k+1})\right)}_{C_{k,k+2}^{N,+} f_{k+2}^N}-\underbrace{\overline{\chi}_\delta(V_{k+1}^{i,k+1})\,
\left(\chi_\delta(v'_{i}, v_{k+2})+\chi_\delta(v_{k+2}, v'_{k+1})\right)}_{C_{k,k+2}^{N,-} f_{k+2}^N}]
\label{charDECOMPOSITIONgainNL17CASE3BIS}
\end{eqnarray}
and
 \begin{eqnarray}
&&\!\!\!\left(C_{k,k+3}^{N} f_{k+3}^N\right)(V_k)=\frac{(N-k)(N-k-1)(N-k-2)}{N}\sum_{1\leq i\leq k}\int dv_{k+1}\!\int dv_{k+2}\!\int dv_{k+3}\!\int d\omega\,  B_{i.k+1}^{\omega}\nonumber\\
&&\!\!\!  [\underbrace{\overline{\chi}_\delta(V_{k+1})\chi_\delta(v_{k+2}, v_{k+1})\chi_\delta(v_{i},v_{k+3})f_{k+3}^N(V_{k+3}^{i,k+1})}_{C_{k,k+3}^{N,+} f_{k+3}^N}
-\underbrace{\overline{\chi}_\delta(V_{k+1}^{i,k+1})\chi_\delta(v_{k+2}, v'_{k+1})\chi_\delta(v'_{i},v_{k+3})f_{k+3}^N(V_{k+3})}_{C_{k,k+3}^{N,-} f_{k+3}^N}].\nonumber\\
&&
\label{loss33}
\end{eqnarray}

The computations needed to recover (\ref{bbgky}) are deferred to the Appendix.\\

As already mentioned, we are interested in a kind of mean-field asymptotics as $N \rightarrow \infty$, $\delta \rightarrow 0$, with $N \delta^3= \alpha$.

Since, for any $v,w\in \R^3$, it holds

\be
N\int dw\, \chi_\delta(v,w)\, \varphi(w)\rightarrow \alpha\varphi(v),\ \ \ \ \ for\ any\ \varphi\in C_b^0(\R^3),
\ee
being $C_b^0(\R^3)$ the space of continuous and uniformly bounded functions on $\R^3$, 
the formal limit of eq. (\ref{bbgky}) yields:

\be
\partial_t f_k^\infty=C_{k,k+1}f_{k+1}^\infty+C_{k,k+2}f_{k+2}^\infty+C_{k,k+3}f_{k+3}^\infty \label{boltz}
\ee
where:
\bea
\!\!\!\left(C_{k,k+1} f_{k+1}^\infty\right)(V_k)&=& \sum_{i=1}^k\int dv_{k+1}\int d\omega \, B^\omega_{i,k+1}\  [f_{k+1}^\infty(V_{k+1}^{i,k+1})-f_{k+1}^\infty(V_{k+1})] \label{c1} \\
\!\!\!\left(C_{k,k+2} f_{k+2}^\infty\right)(V_k)&=& -\alpha\sum_{i=1}^k\int dv_{k+1}\int d\omega B_{i,k+1}^\omega\  [f_{k+2}^\infty(V_{k+1}^{i,k+1},v_i)+f_{k+2}^\infty(V_{k+1}^{i,k+1},v_{k+1}) +\nonumber\\
&&\!\!\!\qquad \ \ \ \ \ - f_{k+2}^\infty(V_{k+1},v_{i}')-f_{k+2}^\infty(V_{k+1},v_{k+1}')] \label{c2} \\
\!\!\!\left(C_{k,k+3} f_{k+3}^\infty\right)(V_k)&=& \alpha^2\sum_{i=1}^k\int dv_{k+1}\int d\omega B_{i,k+1}^\omega\ 
[f_{k+3}^\infty(V_{k+1}^{i,k+1},v_{k+1},v_{i}) - f_{k+3}^\infty(V_{k+1},v_{k+1}',v_{i}')]\nonumber\\
&& \label{c3}
\eea
and the sequence $\{f_k^\infty\}_{k\geq 1}$ is the limit, in a suitable sense (to be specified), of the sequence of marginal distributions $\{f_k^N\}_{k\geq 1}^N$.

Actually, it is immediate to observe that, due to the symmetry of the distributions $f_k^\infty$ with respect to any permutation of variables,
\bea
\left(C_{k,k+3} f_{k+3}^\infty\right)(V_k)&=& \alpha^2\sum_{ i=1}^k\int dv_{k+1}\int d\omega B_{i,k+1}^\omega\ 
[f_{k+3}^\infty(v_1,...,v_i',...,v_k,v_{k+1}',v_{k+1},v_{i}) +\nonumber\\
&-& f_{k+3}^\infty(v_1,...,v_i,...,v_k,v_{k+1},v_{k+1}',v_{i}')]=0. \label{c30}
\eea
Thus, the hierarchy we are going to recover in the limit $N\rightarrow \infty$, $\delta\rightarrow 0$, with $N \delta^3 =\alpha$, is indeed:
\be
\partial_t f_k^\infty=C_{k,k+1}f_{k+1}^\infty+C_{k,k+2}f_{k+2}^\infty.\label{boltzTRUE}
\ee
We refer to (\ref{boltzTRUE}) (and also (\ref{boltz})) as \emph{Boltzmann (or U-U) hierarchy} since it is easy to check that assuming 
\be
 f_k^\infty(V_k,t)=f_k^\infty(t)=\left(f(t)\right)^{\otimes k},\ \ \  \text{for some}\ \ f(v_1,t)\label{boltzTRUEsolfact}
\ee
to solve (\ref{boltzTRUE}) with a certain initial datum $f_0^{\otimes k}$, one finds that $f(t)$ solves the 
(homogeneous) U-U
equation:
\bea
\partial_t f&=&\int dv_{2}\int d\omega \, B^\omega_{1,2}\  [f(v'_1,t)f(v_{2}',t)-f(v_1,t)f(v_{2},t)]+\nonumber\\
&-& \alpha\int dv_{2}\int d\omega\,  B_{1,2}^\omega\  [f(v_1,t)f(v'_1,t)f(v'_2,t)+f(v'_1,t)f(v_2,t)f(v'_2,t)]+\nonumber\\
&+& \alpha\int dv_{2}\int d\omega \, B_{1,2}^\omega\  [f(v_1,t)f(v'_1,t)f(v_2,t)+ f(v_1,t)f(v_2,t)f(v'_2,t)], \label{UU}  
\eea
with initial datum $f_0$. Notice that (\ref{UU}) can be rewritten as:
\bea
\partial_t f&=&\int dv_{2}\int d\omega\,  B^\omega_{1,2}\  f(v'_1,t)f(v_{2}',t)\left(1-\alpha f(v_1,t)\right)\left(1-\alpha f(v_2,t)\right)+\nonumber\\
&-&\int dv_{2}\int d\omega\,  B^\omega_{1,2}\ f(v_1,t)f(v_{2},t)\left(1-\alpha f(v'_1,t)\right)\left(1-\alpha f(v'_2,t)\right)]
\label{UU1} 
\eea
which is the usual form in which the U-U equation is presented. 
On the other hand, if $f(t)$ solves (\ref{UU1}) with some initial datum $f_0$, it can be easily verified that the $k$-particle function
$\left(f(t)\right)^{\otimes k}$
solves (\ref{boltzTRUE}) with initial datum $f_0^{\otimes k}$. \\

\section{The main result}
\label{sec:sec4}
\setcounter{equation}{0}    
\def\theequation{4.\arabic{equation}}

\subsection{Assumptions and statement of the main result}
We make the following hypotheses on the cross-section $B$:
\begin{itemize}
\item[$i)$] there exists $M>0$, not depending on $N$, such that:
$$
 B(v;\omega)=0,\ \ \ \text{if}\ \ |v|>M, \ \ \ \ \ \ \text{for any}\ \ \omega\in \mathcal{S}^2
$$
\item[$ii)$] $B(v;\omega)$ is continuous 
\end{itemize}
Clearly the above assumptions imply that
there exists a constant $C_1>0$, not depending on $N$, such that:
\begin{equation}\label{LIMITATEZZAb}
\sup_{\omega\in \mathcal{S}^2}\sup_{v\in \R^3} B(v;\omega)<C_1<+\infty
\end{equation}
We notice that assumption $ii)$ is compatible with the form (\ref{Bexpression}) of real cross-sections while assumption $i)$ is not. 

As regard to the initial data, we assume the initial probability distribution $W_0^N$ to verify the following properties:
\begin{itemize}
\item[$\mathbf{1.}$] $supp\, W_0^N\subseteq \mathcal{A}_\delta^N$,
\item[$\mathbf{2.}$] there exists a 
family of symmetric probability distributions $\{f_k^\infty(0)\}_k$ 
such that, for all $k$,
\begin{equation}
\int dv_k\ f_k^\infty(v_1, \dots, v_k,0)=f_{k-1}^\infty(v_1, \dots, v_{k-1},0),
\end{equation}
and
\begin{equation}\label{ass2}
\sup_{V_k\in K}\left|f_k^N(V_k,0)- f_k^\infty(V_k,0)\right|\to 0, \ \text{as $N\to \infty,\ \delta\to 0,\ N\delta^3=\alpha$},
\end{equation}
for any compact set $K\subset \mathcal{A}^k$, where
\begin{eqnarray}\label{outOFdiagTEO}
\mathcal{A}^k:=\{V_k\in \R^{3k}:\ \ v_\ell\neq v_m,\ \ \ \text{for any}\ \ \ell\neq m,\ \ \ell, m=1,2,\dots, k\}.
\end{eqnarray}
\item[$\mathbf{3.}$]  
for all $k=1,2,\dots$, there exists a constant $z_1>0$,
not depending on $N$, such that:
\begin{eqnarray}\label{LinftyfN}
f_k^N(V_k,0)\leq (z_1)^k
\end{eqnarray}
\item[$\mathbf{4.}$]  
for all $k=1,2,\dots$, there exists a constant $z_2>0$  such that:
\begin{eqnarray}\label{Linftyflim}
f_k^\infty(V_k,0)\leq z_2\ \left(\frac{1}{\alpha}\right)^k
\end{eqnarray}
\item[$\mathbf{5.}$]
$f_k^\infty(0)\in C^0(\R^{3k})$, for all $k=1,2,\dots$
\end{itemize}
Observe that the convergence required in hypothesis $\mathbf{2.}$ is a natural notion of convergence compatible with the admissibility assumption on $W_N^0$ and the continuity of $f_k^\infty(0)$. In fact, such conditions prevent convergence on the diagonals $v_\ell = v_m$. In the sequel, we will refer to hypothesis $\mathbf{2.}$
saying that, for any $k$, $f_k^N(0)$ converges to $f_k^\infty(0)$ uniformly \emph{outside the diagonals}.
\vspace{0.2cm}

The main result we show is the following:
\begin{teo}\label{main}
Let the cross-section $B$ satisfy the above assumptions $i)$ and $ii)$ and 
the initial probability distribution $W_0^N$ verify properties $\mathbf{1.}$-$\mathbf{5.}$ 
Then, for any $t\geq 0$ it holds:
\begin{eqnarray}\label{convergence}
f_k^N(t)\rightarrow f_k^\infty(t) \ \ \text{in}\ \ L^1_{loc}(
\R^{3k}),
 \ \ \ \text{as} \ \ N\to \infty,\ \delta\to 0,\ N\delta^3=\alpha,
\end{eqnarray}
where
the sequence $\{f_k^\infty(t)\}_k$ is the unique $L^\infty$-solution of the Boltzmann hierarchy (\ref{boltzTRUE}) with initial datum $\{f_k^\infty(0)\}_k$.
\end{teo}

\vspace{0.5cm}

In Section \ref{sec:sec8} we will give two examples of initial data 
satisfying the above assumptions.  
Moreover, thanks to the Hewitt-Savage Theorem (see \cite{HS}), we can characterize the limiting distributions $\{f_k^\infty(0)\}_k$ in terms of solutions of the U-U equation (\ref{UU1}).  In fact, by the above theorem we know that there exists a unique probability measure $\nu$ on the space of one-particle probability densities 
$$
M^1_+(\R^3):=\left\{f:\R^3\to \R^+\ \ s. \ t.\ \ 
f\in L^1(\R^3),\ \ \ \int dv\,  f(v)=1\right\}
$$
 such that:
\begin{eqnarray}\label{hewittGEN2teo}
f_k^\infty(V_k,0)=\int_{M^1_+(\R^3)} d\nu(f_0) \ f_0^{\otimes k} (V_k).
\end{eqnarray}
In other words, each limiting distribution $f_k^\infty(0)$ can be tought as a statistical mixture of factorized densities and the statistical correlations are described by the measure $\nu$. Indeed,  the completely factorized distribution $f_k^\infty(V_k,0)=(f_{in})^{\otimes k}(V_k)$ 
corresponds to the case $\nu(f_0)=\delta(f_{in}-f_0)$ in which there are no correlations and statistical independence holds. This is what is usually known as \emph{hypothesis of molecular chaos} (see e.g. \cite{Pulv}).
In Section \ref{sec:sec8} we will present a situation in which $\nu$ cannot be a $\delta$ measure. Indeed, 
$\nu$ will be some  ''spread'' measure whose dispersion depends on $\alpha$ and goes to zero as $\alpha\to 0$. On the other hand, 
we will analyze even a situation in which asymptotic factorization holds and the measure $\nu$ is indeed a Dirac measure centered in the initial one particle datum $f_{in}$ we chose. 	\\
On the basis of the previous considerations, our Theorem \ref{main} can be reformulated as follows:
\begin{teo}\label{mainREF}
Let the cross-section $B$ satisfy the above assumptions $i), ii)$ and the initial probability distribution $W_0^N$ verify properties $\mathbf{1.}-\mathbf{5.}$ Let $\nu$ be the unique measure satisfying (\ref{hewittGEN2teo}). 
Then, for any $t\geq 0$ it holds:
\begin{eqnarray}\label{convergenceREF}
f_k^N(t)\rightarrow  \int_{M^1_+(\R^3)} d\nu(f_0) \ (f(t))^{\otimes k} \ \ \text{in}\ \ L^1_{loc}(
\R^{3k}),
 \ \ \ \text{as} \ \ N\to \infty,\ \delta\to 0,\ N\delta^3=\alpha,
\end{eqnarray}
where $f(t)$ is the unique $L^\infty$-solution of the U-U equation (\ref{UU1}) with initial datum $f_0$.\\
In particular, if $\nu=\delta(f_{in}-f_0)$ for some $f_{in}\in M^1_+(\R^3)$, namely, $f_k^\infty(0)$ factorizes,  convergence (\ref{convergenceREF}) yields:
\begin{eqnarray}\label{convergenceREFfact}
f_k^N(t)\rightarrow   (f(t))^{\otimes k} \ \ \text{in}\ \ L^1_{loc}(
\R^{3k}),
 \ \ \ \text{as} \ \ N\to \infty,\ \delta\to 0,\ N\delta^3=\alpha,
\end{eqnarray}
where $f(t)$ is the unique $L^\infty$-solution of the U-U equation (\ref{UU1}) with initial datum $f_{in}$. In other words, 
propagation of chaos holds for any time $t>0$ provided that the hypothesis of molecular chaos holds at time $t=0$.
\end{teo}

The strategy we are going to follow is based on the perturbative argument introduced by O. Lanford to derive the classical Boltzmann equation for the hard-sphere dynamics (see \cite{LANFORD}). For each $k=1,2,\dots$ we obtain a perturbative expansion for both the $k$-particle marginal $f_k^N(t)$, solving the BBGKY hierarchy (\ref{bbgky}), and the solution $f_k^\infty(t)$ of the Boltzmann hierarchy (\ref{boltzTRUE}). Those expressions are just derived by iterating the Duhamel formula. Then, the first step will be to show that, within a sufficiently short time interval,  both expansions are uniformly estimated in suitable norms by a converging series. 
The second step will be to show that term by term convergence holds for any $t\geq 0$. The convergence proof
will be achieved by piling up a finite number of series expansions which will be controlled by means of a priori estimates following by the exclusion principle.

\section{Short time estimates}
\label{sec:sec5}
\setcounter{equation}{0}    
\def\theequation{5.\arabic{equation}}

In order to proceed with the above program, the first tool we need are suitable estimates on both the operators appearing in the BBGKY hierarchy (\ref{bbgky}) and those appearing in the Boltzmann hierarchy (\ref{boltz}).

\subsection{Operator estimates}
\label{subOP}
We define the following norm:
\begin{eqnarray}
||g_k||_{ \delta,k}:=\sup_{\substack{\Delta_1\dots \Delta_k
}}\frac{1}{\delta^{3k}}\int_{\Delta_1} dv_1\dots \int_{\Delta_k} dv_k \left| g_k(V_k)\right|,\ \ \ \text{for}\ \ \ g_k:\R^{3k}\to \R,\ \ \ k=1,2\dots
\label{NORMdelta}
\end{eqnarray}
Note that for any function $g_k:\R^{3k}\to \R$ which is supported on the set $\mathcal{A}_\delta^k$ of $k$-particle admissible configurations, the supremum in (\ref{NORMdelta}), over all $k$-sequences of cells $\Delta_1,\dots \Delta_k$, can be replaced by the supremum over  all $k$-sequences of cells $\Delta_1,\dots \Delta_k$ such that $\Delta_\ell\neq \Delta_m$ for any pair of different indices $\ell, m\in \{1,\dots,k\}$.
Moreover, it is immediate to show that, for any $\delta_0>0$:
\begin{eqnarray}\label{normREM3}
|| g_k||_{ \delta,k}\leq || g_k||_{ \delta_0,k}\leq || g_k||_{\infty},\ \ \ \forall\ \delta>\delta_0.
\end{eqnarray}

We are introducing this kind of ($\delta$-dependent) norm because, thanks to the exclusion mechanism,
we are able to get a natural global in time control on $||f_k^N(t)||_{\delta,k}$, which is needed to prove the 
uniform boundedness 
estimates to hold for arbitrary times (see Section 6 below). 

Now, let us consider the hierarchy:
\begin{eqnarray}\label{BBGKY}
\pa_t f_k^N&=&\frac{1}{N}\left(L_k^N\, f_k^N\right)
+\sum_{s=1}^2 \frac{1}{N}\left(L_{k,k+s}^N\, f_{k+s}^N\right)+ 
  \sum_{s=1}^3\left(C_{k,k+s}^N\, f_{k+s}^N\right).
\end{eqnarray}
and show that all operators appearing there are suitably bounded with respect to the norm (\ref{NORMdelta}). Let us start from $L_k^N$, namely:
\begin{eqnarray}\label{Lk}
\left(L_k^N\, f_k^N\right)(v_1, \dots, v_k;t)=\sum_{1\leq i<j\leq k} \int d\omega\, B_{i,j}^{\omega} \left[\overline{\chi}_\delta(V_k)f_k^N(V_k^{i,j},t)-\overline{\chi}_\delta(V_k^{i,j})f_k^N(V_k,t)\right].
\end{eqnarray}
Integrating (\ref{Lk}) over a certain $k$-sequence of cells $\Delta_1\dots \Delta_k$, 
we obtain:
\begin{eqnarray}\label{Lkcell}
&&\frac{1}{\delta^{3k}}\int_{\Delta_1} dv_1\dots \int_{\Delta_k} dv_k \left|\left(L_k^N\, f_k^N\right)(V_k,t)\right|\leq\nonumber\\
&&\leq \frac{1}{\delta^{3k}}\sum_{1\leq i<j\leq k}  \int_{\Delta_1} dv_1\dots \int_{\Delta_k} dv_k\ \int d\omega\, B_{i,j}^{\omega} \left(f_k^N(V_k^{i,j},t)+f_k^N(V_k,t)\right).
\end{eqnarray}
For the loss part we immediately get:
\begin{eqnarray}\label{LkcellLOSS}
\frac{1}{\delta^{3k}}\sum_{1\leq i<j\leq k}  \int_{\Delta_1} dv_1\dots \int_{\Delta_k} dv_k\ \int d\omega\, B_{i,j}^{\omega} \ f_k^N(V_k,t)\leq 4\pi\, C_1 \, k^2|| f_{k}^N(t)||_{ \delta, k},
\end{eqnarray}
where $C_1$ is the constant appearing in (\ref{LIMITATEZZAb}). 
On the other hand, the gain part is:
\begin{eqnarray}\label{LkcellGAIN}
 \frac{1}{\delta^{3k}}\sum_{1\leq i<j\leq k}  \int d\omega\, \int_{\Delta_1} dv_1\dots \int_{\R^3} dv_i \ \chi_{\Delta_i}(v_i)\dots  \int_{\R^3} dv_j \ \chi_{\Delta_j}(v_j)\dots \int_{\Delta_k} dv_k\  B_{i,j}^{\omega} \ f_k^N(V_k^{i,j},t).
 \end{eqnarray}
Since, by (\ref{collDEF}), we know that  $|v_i-v_{j}|= |v'_i-v'_{j}|$, by (\ref{Bsymm}) it follows that $B(v_i-v_{j};\omega)=B(v'_i-v'_{j};\omega)$. Moreover, again by 
(\ref{collDEF}), we can easily verify that the Jacobian of the transformation $(v_i,v_j)\to (v'_i, v'_j)$ is unitary. Therefore, (\ref{LkcellGAIN}) yields:
\begin{eqnarray}\label{Lkcell1}
 \frac{1}{\delta^{3k}}\sum_{1\leq i<j\leq k} \int d\omega\, \int_{\Delta_1} dv_1\dots \int_{\R^3} dv_i \ \chi_{\Delta_i}(v'_i)\dots  \int_{\R^3} dv_j \ \chi_{\Delta_j}(v'_j)\dots \int_{\Delta_k} dv_k\  B_{i,j}^{\omega} \ f_k^N(V_k,t).
\end{eqnarray}
Notice that:
\begin{eqnarray}\label{Lkcell1setA}
\int_{\R^3} dv_i \ \chi_{\Delta_i}(v'_i) \int_{\R^3} dv_j \ \chi_{\Delta_j}(v'_j)=\int\int_{A_{ij}^\omega}dv_i\, dv_j
\end{eqnarray}
where
\begin{eqnarray}\label{Lkcell1setAdef}
 A_{ij}^\omega=
 \{v_i, v_j\in \R^3\times \R^3|\ \ v'_i\in \Delta_i,\ \ v'_j\in \Delta_j \}.
 \end{eqnarray}
 By the relation (\ref{collDEF}), it follows that, for any $\omega\in \mathcal{S}^2$, the set $ A^\omega_{ij}$ is surely covered by a finite number of pairs of 
 cells of volume $\delta^3$. In other words, there exists $n_0<+\infty$ such that:
 \begin{eqnarray}\label{Lkcell1setApair}
 A^\omega_{ij}\subset \bigcup_{\eta_i\in \mathcal{I}_i}\Delta_{\eta_i}^\omega\times\bigcup_{\eta_j\in \mathcal{I}_j}\Delta_{\eta_j}^\omega,
 \ \ \ \ \ \ \ \ 
 \text{where} \ \ \ \  |\mathcal{I}_i|=|\mathcal{I}_j|=n_0
 \end{eqnarray}
 Thus, the term in (\ref{Lkcell1}) is less or equal then:
 \begin{eqnarray}\label{Lkcell1setAest}
G_{n_0}:= \frac{1}{\delta^{3k}}\, \sum_{1\leq i<j\leq k} \ \sum_{\eta_i\in \mathcal{I}_i}\sum_{\eta_j\in \mathcal{I}_j}\int d\omega\, \int_{\Delta_1} dv_1\dots \int_{\Delta_{\eta_i}^\omega} dv_i \, \dots  \int_{\Delta_{\eta_j}^\omega} dv_j \, \dots \int_{\Delta_k} dv_k\  B_{i,j}^{\omega} \ f_k^N(V_k,t),
\end{eqnarray}
for which we get:
\begin{eqnarray}\label{Lkcell1setAestFIN}
G_{n_0}\leq 4\pi\, C_1 \, n_0^2\, k^2|| f_{k}^N(t)||_{ \delta, k}.
\end{eqnarray}
Therefore, by (\ref{LkcellLOSS}) and (\ref{Lkcell1setAestFIN}) we obtain: 
\begin{eqnarray}\label{Lkcell2}
&&|| L_k^N\, f_{k}^N(t)||_{ \delta, k}
\leq C_B \, k^2|| f_{k}^N(t)||_{ \delta, k},
\end{eqnarray}
where, from now on, we denote by $C_B$ any positive constant only depending on $B$.
Moreover, defining the $k$-particle evolution operator as:
\begin{eqnarray}\label{flowDEF}
S_k^N(t):=e^{\frac{L_k^N\, t}{N}},\ \ \ \ \ \ \ \ t>0 
\end{eqnarray}
by (\ref{Lkcell2}) we get:
\begin{eqnarray}\label{LkCONTinNEWnorm}
||S_k^N(t)\, g_k||_{ \delta,k}\leq 
e^{\frac{C_B\, k^2\ t}{N}}||g_k||_{\delta, k}\leq e^{C_B\, k\ t}||g_k||_{\delta, k},
\end{eqnarray}
for any $g_k:\R^{3k}\to \R$ such that $|| g_k||_{ \delta, k}<+\infty$. 
\\

Let us proceed with the estimates of the other operators appearing in (\ref{BBGKY}).
\\
We start by $L_{k,k+1}^{N}$ and, in particular, we consider the gain term:
\begin{eqnarray}
&&\left(L_{k,k+1}^{N,+} f_{k+1}^N\right)(V_k,t)=-(N-k)\sum_{1\leq i<j\leq k}\int dv_{k+1}\int d\omega\,  B_{i.j}^{\omega}\ f_{k+1}^N(V_{k+1}^{i.j},t)\times\qquad\qquad\nonumber\\
&&\qquad \qquad \qquad\qquad\qquad\times \ \overline{\chi}_\delta(V_k)
\left(\chi_\delta(v_{i},v_{k+1})+\chi_\delta(v_{j},v_{k+1})\right).
\label{charDECOMPOSITIONgainLIN6bis}
\end{eqnarray}
By integrating over a $k$-sequence $\Delta_1\dots \Delta_k$ of cells, 
we get:
\begin{eqnarray}
&&\frac{1}{\delta^{3k}}\int_{\Delta_1} dv_1\dots \int_{\Delta_k} dv_k\left|\left(L_{k,k+1}^{N,+} f_{k+1}^N\right)(V_k,t)\right|\leq\qquad\qquad\qquad\qquad\qquad\qquad\qquad\qquad\qquad\nonumber\\
&&\leq\frac{(N-k)}{ \delta^{3k}}\sum_{1\leq i<j\leq k} \sum_{s=i,j}\int d\omega\,\int_{\Delta_1} dv_1\dots\int_{\R^3} dv_i\, \chi_{\Delta_i}(v_i)\dots  \int_{\R^3} dv_j\, \chi_{\Delta_j}(v_j)\dots\nonumber\\
&&\qquad\qquad\qquad\qquad \qquad\qquad\qquad\qquad\ \ \ \dots\int_{\Delta_k} dv_k\int_{\Delta_s} dv_{k+1}\,  B_{i,j}^{\omega} \ f_{k+1}^N(V_{k+1}^{i.j},t).
\label{newESTnew}
\end{eqnarray}
Therefore, arguing as before, we find:
\begin{eqnarray}
&&\frac{1}{\delta^{3k}}\int_{\Delta_1}  dv_1\dots \int_{\Delta_k} dv_k\left|\left(L_{k,k+1}^{N,+} f_{k+1}^N\right)(V_k,t)\right|\leq\qquad\qquad\qquad\qquad\qquad\qquad\qquad\nonumber\\
&&\leq 
 \frac{(N-k)}{\delta^{3k}}\, \sum_{1\leq i<j\leq k}  \sum_{s=i,j}\, \sum_{\eta_i\in \mathcal{I}_i}\sum_{\eta_j\in \mathcal{I}_j}\int d\omega\, \int_{\Delta_1} dv_1\dots \int_{\Delta_{\eta_i}^\omega} dv_i \, \dots  \int_{\Delta_{\eta_j}^\omega} dv_j  \dots\nonumber\\
 &&\qquad\qquad\qquad\qquad \qquad\qquad\qquad\quad\dots \int_{\Delta_k} dv_k\int_{\Delta_s} dv_{k+1}\, \  B_{i,j}^{\omega} \ f_{k+1}^N(V_{k+1},t).
\label{newEST1}
\end{eqnarray}
Reminding that $N\delta^3=\alpha$ 
and using the boundedness assumption we made on $B$, we obtain:
\begin{eqnarray}
\frac{1}{\delta^{3k}}\int_{\Delta_1}  dv_1\dots \int_{\Delta_k} dv_k\left|\left(L_{k,k+1}^{N,+} f_{k+1}^N\right)(V_k,t)\right|\leq 8\pi\, C_1 \, n_0^2\, k^2\, \alpha\ || f_{k+1}^N(t)||_{ \delta, k+1}.
\label{newEST1new}
\end{eqnarray}
Next, let us consider the loss term:
\begin{eqnarray}
&&\left(L_{k,k+1}^{N,-} f_{k+1}^N\right)(V_k,t)=-(N-k)\sum_{1\leq i<j\leq k}\int dv_{k+1}\int d\omega \, B_{i.j}^{\omega}\ f_{k+1}^N(V_{k+1},t)\times\qquad\nonumber\\
&& \qquad \qquad\qquad \qquad\qquad\times \ \overline{\chi}_\delta(V_k^{i.j}) \, 
\left(\chi_\delta(v'_{i},v_{k+1})+\chi_\delta(v'_{j},v_{k+1})\right)
\label{loss22EST}
\end{eqnarray}
By integrating over a $k$-sequence $\Delta_1\dots \Delta_k$ of cells, 
we get:
\begin{eqnarray}
&&\frac{1}{\delta^{3k}}\int_{\Delta_1} dv_1\dots \int_{\Delta_k} dv_k\left|\left(L_{k,k+1}^{N,-} f_{k+1}^N\right)(V_k,t)\right|\leq
\nonumber\\
&&\leq
\frac{(N-k)}{ \delta^{3k}}\sum_{1\leq i<j\leq k} 
\int d\omega\,\int_{\Delta_1} dv_1\dots\int_{\Delta_i} dv_i
\dots  
\int_{\Delta_j} dv_j
\dots
\nonumber\\
&&\qquad \qquad\qquad \qquad \qquad \ \ \ \dots
\int_{\Delta_k} dv_k\int_{\R^3} dv_{k+1}\ \chi_\delta(v'_{i},v_{k+1})\   B_{i,j}^{\omega} \ f_{k+1}^N(V_{k+1},t) +\nonumber\\
&& + \, \frac{(N-k)}{ \delta^{3k}}\sum_{1\leq i<j\leq k} 
\int d\omega\,\int_{\Delta_1} dv_1\dots\int_{\Delta_i} dv_i
\dots  \int_{\Delta_j} dv_j
\dots\nonumber\\
&&\qquad\qquad \qquad\qquad \qquad\ \ \ \dots\int_{\Delta_k} dv_k\int_{\R^3} dv_{k+1}\ \chi_\delta(v'_{j},v_{k+1})\   B_{i,j}^{\omega} \ f_{k+1}^N(V_{k+1},t).
\label{newESTnewLOSS}
\end{eqnarray}
Now, following the same argument discussed in (\ref{Lkcell1setA})-(\ref{Lkcell1setApair}), we infer that, for any $\omega\in \mathcal{S}^2$, the sets:
\begin{eqnarray}\label{Lkcell1setAdefL}
 A_{i}^{\omega,j}=
 \{v'_i\in \R^3 |\ \ v_i \in \Delta_i,\ \ v_j\in \Delta_j \} \ \ \ \ \ \ \ \text{and}\ \ \ \ \ \ \  A_{j}^{\omega,i}=
 \{v'_j\in \R^3 |\ \ v_i \in \Delta_i,\ \ v_j\in \Delta_j \}
 \end{eqnarray}
are surely covered by a finite number of 
 cells of volume $\delta^3$. In other words:
 \begin{eqnarray}\label{Lkcell1setApairL}
 A^{\omega,j}_{i}\subset \bigcup_{\eta_i\in \mathcal{I}_i}\Delta^{\omega'}_{\eta_i}
 \ \ \ \ \ \ \ \ \text{and}\ \ \ \ \ \ \ \ A^{\omega,i}_{j}\subset \bigcup_{\eta_j\in \mathcal{I}_j}\Delta^{\omega'}_{\eta_j},\ \ \ 
 \text{where} \ \ \ \  |\mathcal{I}_i|=|\mathcal{I}_j|=n_0
 \end{eqnarray}
 Therefore, for any $v'_i\in A^{\omega,j}_{i}$, denoting by $\Delta'_i$ the cell of the grid containing $v'_i$, we have that: 
$$
\Delta'_i\subset \bigcup_{\eta_i\in \mathcal{I}_i}\Delta^{\omega'}_{\eta_i}
$$
and the same holds for $v'_j$, i.e.
$$
\Delta'_j\subset \bigcup_{\eta_j\in \mathcal{I}_j}\Delta^{\omega'}_{\eta_j}
$$
Thus, (\ref{newESTnewLOSS}) yields:
\begin{eqnarray}
&&\frac{1}{\delta^{3k}}\int_{\Delta_1} dv_1\dots \int_{\Delta_k} dv_k\left|\left(L_{k,k+1}^{N,-} f_{k+1}^N\right)(V_k,t)\right|\leq\qquad\qquad\qquad\qquad\qquad\qquad\qquad\qquad\qquad\nonumber\\
&&\leq\frac{(N-k)}{ \delta^{3k}}\sum_{1\leq i<j\leq k} \sum_{\eta_i\in \mathcal{I}_i}
\int d\omega\,\int_{\Delta_1} dv_1\dots\int_{\Delta_i} dv_i\dots  \int_{\Delta_j} dv_j\dots\nonumber\\
&&\qquad\qquad \qquad\qquad\qquad\qquad\qquad \ \ \ \dots\int_{\Delta_k} dv_k\int_{\Delta_{\eta_i}^{\omega'}} dv_{k+1}\   B_{i,j}^{\omega} \ f_{k+1}^N(V_{k+1},t) +\nonumber\\
&&+\, \frac{(N-k)}{ \delta^{3k}}\sum_{1\leq i<j\leq k} 
\sum_{\eta_j\in \mathcal{I}_j}
\int d\omega\,\int_{\Delta_1} dv_1\dots\int_{\Delta_i} dv_i\, \dots  \int_{\Delta_j} dv_j\dots\nonumber\\
&&\qquad\qquad \qquad\qquad\qquad\qquad\qquad \ \ \ \dots\int_{\Delta_k} dv_k\int_{\Delta_{\eta_j}^{\omega'}} dv_{k+1}\   B_{i,j}^{\omega} \ f_{k+1}^N(V_{k+1},t)
\label{newESTnewLOSS11}
\end{eqnarray}
implying:
\begin{eqnarray}
\frac{1}{\delta^{3k}}\int_{\Delta_1} dv_1\dots \int_{\Delta_k} dv_k\left|\left(L_{k,k+1}^{N,-} f_{k+1}^N\right)(V_k,t)\right|\leq 8\pi\, C_1 \, n_0\, k^2\, \alpha\ 
 \left\| f_{k+1}^N(t)\right\|_{ \delta,k+1}.
\label{Lk+11normVIIante}
\end{eqnarray}
Therefore, by (\ref{newEST1new}) and (\ref{Lk+11normVIIante}) it follows that:
\begin{eqnarray}
\left\|L_{k,k+1}^{N} f_{k+1}^N(t)\right\|_{ \delta,k}\leq   C_B \ k^2\ \alpha\ \left\| f_{k+1}^N\right\|_{ \delta,k+1}.
\label{Lk+11normVII}
\end{eqnarray}

The $k+2$-particle operator $L_{k,k+2}^{N}$ defined by (\ref{charDECOMPOSITIONgainNL5}) can be estimated using the same arguments discussed above and
we get:
\begin{eqnarray}\label{Lk+2normIV}
||L_{k,k+2}^{N}\, f_{k+2}^N(t)||_{\delta, k}\leq  C_B\, k^2\, \alpha^2\,  ||f_{k+2}^N(t)||_{ \delta, k+2}.
\end{eqnarray}

Let us look now at the $C$'s operators appearing in the BBGKY hierarchy (\ref{BBGKY}). Concerning operator $C_{k,k+1}^N$ and, in particular, its gain part, by (\ref{charDECOMPOSITIONgainNL8case3BIS}) it follows that:
\begin{eqnarray}\label{Ck+1norm}
&&\frac{1}{\delta^{3k}}\int_{\Delta_1} dv_1\dots \int_{\Delta_k} dv_k\left|\left(C_{k,k+1}^{N,+}\, f_{k+1}^N\right)(V_{k},t)\right|\leq \nonumber\\
&&\leq \sum_{i=1}^k \frac{1}{\delta^{3k}}\int_{\Delta_1} dv_1\dots \int_{\Delta_k} dv_k\int dv_{k+1}\, \int d\omega\,  B^\omega_{i,k+1}
\ f_{k+1}^N(V_{k+1}^{i,k+1},t)=\nonumber\\
&&=\sum_{i=1}^k \sum_{\Delta_{k+1}\subset \R^3} \frac{1}{\delta^{3k}}\int d\omega\,  \int_{\Delta_1} dv_1\dots\int_{\R^3} dv_i\, \chi_{\Delta_i}(v_i)\dots \qquad\qquad\qquad \nonumber\\
&&\qquad\qquad \qquad\qquad \dots\int_{\Delta_k} dv_k\int_{\R^3} dv_{k+1}\  \chi_{\Delta_{k+1}}(v_{k+1})\ B^\omega_{i,k+1}
\ f_{k+1}^N(V_{k+1}^{i,k+1},t),
\end{eqnarray}
where, as before, $\Delta_1\dots\Delta_k$ is a fixed $k$-sequence of cells. 
However, due to our (compact support) assumption $i)$ on the cross-section $B$ (see Theorem \ref{main}), we know that:
$$
B_{i,k+1}^\omega=B(v_i-v_{k+1};\omega)=0,\ \ \ \ \text{if}\ \ |v_i-v_{k+1}|>M.
$$
Therefore, (\ref{Ck+1norm}) yields:
\begin{eqnarray}\label{Ck+1normTRISa}
&&\frac{1}{\delta^{3k}}\int_{\Delta_1} dv_1\dots \int_{\Delta_k} dv_k\left|\left(C_{k,k+1}^{N,+}\, f_{k+1}^N\right)(V_{k},t)\right|\leq \nonumber\\
&&\leq  \sum_{i=1}^k \sum_{\Delta_{k+1}\in \mathcal{C}^M_{\Delta_i}} \frac{1}{\delta^{3k}}\int d\omega\,  \int_{\Delta_1} dv_1\dots\int_{\R^3} dv_i\, \chi_{\Delta_i}(v_i)\dots \qquad\qquad\qquad \nonumber\\
&&\qquad\qquad \qquad\qquad \dots\int_{\Delta_k} dv_k\int_{\R^3} dv_{k+1}\  \chi_{\Delta_{k+1}}(v_{k+1})\ B^\omega_{i,k+1}
\ f_{k+1}^N(V_{k+1}^{i,k+1},t),
\end{eqnarray}
where
$$
 \mathcal{C}^M_{\Delta_i}:=\{\Delta\subset \R^3 |\  |v_i-v_{k+1}|\leq M,\ \ \forall\ \, (v_i,v_{k+1})\in \Delta_i\times\Delta\}
$$
and, as it can be easily verified, there exists a constant $0<C_2<+\infty$ such that:
\begin{eqnarray}\label{orderOFmag}
| \mathcal{C}^M_{\Delta_i}|= \frac{C_2\, M^3}{\delta^3},
\ \ \ \ \ \ \text{for any cell $\Delta_i$}.
\end{eqnarray}
Now,
by the change of variables $v_i\to v'_i,\ v_{k+1}\to v'_{k+1}$, the r. h. s. of (\ref{Ck+1normTRISa}) gives rise to:
\begin{eqnarray}\label{Ck+1normDELTA}
 && \sum_{i=1}^k \sum_{\Delta_{k+1}\subset \mathcal{C}^M_{\Delta_i}} \frac{1}{\delta^{3k}}\int d\omega\,  \int_{\Delta_1} dv_1\dots\int_{\R^3} dv_i\, \chi_{\Delta_i}(v'_i)\dots \qquad\qquad\qquad \nonumber\\
&&\qquad\qquad \qquad\qquad \dots\int_{\Delta_k} dv_k\int_{\R^3} dv_{k+1}\  \chi_{\Delta_{k+1}}(v'_{k+1})\ B^\omega_{i,k+1}
\ f_{k+1}^N(V_{k+1},t).
\end{eqnarray}
Following the same argument discussed in (\ref{Lkcell1setA})-(\ref{Lkcell1setApair}), we get:
\begin{eqnarray}\label{CestII}
\frac{1}{\delta^{3k}}\int_{\Delta_1} dv_1\dots \int_{\Delta_k} dv_k\left|\left(C_{k,k+1}^{N,+}\, f_{k+1}^N\right)(V_{k},t)\right|\leq
4\pi\, C_1\, \left(C_2\, M^3\right)\, n_0^2\, k\, || f_{k+1}^N(t)||_{ \delta, k+1}.
\end{eqnarray}
On the other hand, by (\ref{charDECOMPOSITIONgainNL8case3BIS}) we easily infer that the loss term $C_{k,k+1}^{N,-}$ 
can be estimated as:
\begin{eqnarray}\label{CestIIL}
\frac{1}{\delta^{3k}}\int_{\Delta_1} dv_1\dots \int_{\Delta_k} dv_k\left|\left(C_{k,k+1}^{N,-}\, f_{k+1}^N\right)(V_{k},t)\right|\leq
4\pi\, C_1\, \left(C_2\, M^3\right)\, k\, || f_{k+1}^N(t)||_{ \delta, k+1},
\end{eqnarray}
thus, by (\ref{CestII}) and (\ref{CestIIL}) it follows that:
\begin{eqnarray}
\left\|C_{k,k+1}^{N} f_{k+1}^N(t)\right\|_{ \delta,k}\leq   C_B \ k\  \left\| f_{k+1}^N\right\|_{ \delta,k+1}.
\label{Ck+1normIV}
\end{eqnarray}

The estimates for operators $C_{k,k+2}^{N}$ and $C_{k,k+3}^{N}$ are achieved essentially by interpolating the arguments used to estimate operators $L$'s and those used for $C_{k,k+1}^N$. Indeed, by (\ref{charDECOMPOSITIONgainNL17CASE3BIS}), (\ref{loss33}) and (\ref{opDEF}) one gets:
\begin{eqnarray}\label{Ck+2norm}
||C_{k,k+2}^{N}\, f_{k+2}^N(t)||_{\delta,k}\leq C_{B}\, k\, \alpha || f_{k+2}^N(t)||_{\delta,k+2},
\end{eqnarray}
and
\begin{eqnarray}\label{Ck+3norm}
||C_{k,k+3}^{N}\, f_{k+3}^N(t)||_{ \delta, k}\leq  C_{B}\, k\,  \alpha^2\, || f_{k+3}^N(t)||_{\delta,k+3}.
\end{eqnarray}
\vspace{0.2cm}

On the basis of the computations we did to prove estimates (\ref{Lkcell2}),
(\ref{Lk+11normVII}), (\ref{Lk+2normIV}) on operators $L_{k}^N$, $L_{k,k+1}^N$, $L_{k,k+2}^N$ and estimates (\ref{Ck+1normIV}), (\ref{Ck+2norm}) 
(\ref{Ck+3norm}) on operators $C_{k,k+1}^N$, $C_{k,k+2}^N$, $C_{k,k+3}^N$, it can be easily verified that, under assumptions $i)$ and $ii)$ on the cross section $B$, such estimates holds with respect to the $L^\infty$-norm as well. Indeed, we have:
\begin{eqnarray}
&&|| L_k^N\, f_{k}^N(t)||_{ \infty}
\leq C_B \, k^2|| f_{k}^N(t)||_{ \infty}, \label{LkestINFTY}
\end{eqnarray}
\begin{eqnarray}
&&\left\|L_{k,k+1}^{N} f_{k+1}^N\right\|_{\infty}\leq  C_B\, k^2\,  \alpha\, \left\| f_{k+1}^N\right\|_{\infty},
\label{Lk+1estLinfty}
\end{eqnarray}
\begin{eqnarray}
&&\left\|L_{k,k+2}^{N} f_{k+2}^N\right\|_{\infty}\leq C_B \, k^2\,  \alpha^2\, \left\| f_{k+2}^N\right\|_{\infty},
\label{Lk+2estLinfty}
\end{eqnarray}
\begin{eqnarray}\label{Ck+1normINFTY}
&&||\left(C_{k,k+1}^{N}\, f_{k+1}^N\right)(t)||_{\infty}\leq C_B\, k\,   ||f_{k+1}^N(t)||_{\infty},
\end{eqnarray}
\begin{eqnarray}\label{Ck+2normINFTY}
&&||\left(C_{k,k+2}^{N}\, f_{k+2}^N\right)(t)||_{\infty}\leq C_B\,  k\, \alpha\,    ||f_{k+2}^N(t)||_{\infty},
\end{eqnarray}
\begin{eqnarray}\label{Ck+3normINFTY}
&&||\left(C_{k,k+3}^{N}\, f_{k+3}^N\right)(t)||_{\infty}\leq C_B\,  k\, \alpha^2\,    ||f_{k+3}^N(t)||_{\infty}.
\end{eqnarray}

By (\ref{LkestINFTY}), for the flow $S_k^N(t):= e^{\frac{L_k^N\, t}{N}}$ we have:
\begin{eqnarray}\label{LkCONT}
||S_k^N(t)\, g_k||_{ \infty}\leq 
e^{\frac{C_B\, k^2\ t}{N}}||g_k||_{\infty}\leq e^{C_B\, k\ t}||g_k||_{\infty},
\end{eqnarray}
for any $g_k:\R^{3k}\to \R$ such that $|| g_k||_{ \infty}<+\infty$. 
\vspace{0.2cm}\\

Let us look now at the Boltzmann hierarchy (\ref{boltz}), namely:
\begin{eqnarray}\label{BOLTZtrue}
\pa_t f_k^\infty=  C_{k,k+1}\, f_{k+1}^\infty+C_{k,k+2}\, f_{k+2}^\infty+C_{k,k+3}\, f_{k+3}^\infty
\end{eqnarray}
where $C_{k,k+1}$, $C_{k,k+2}$ and $C_{k,k+3}$ are defined by (\ref{c1}), (\ref{c2}) and (\ref{c3}) respectively. 
We already observed that operator $C_{k,k+3}$ gives indeed no contribution, leading to the hierarchy (\ref{boltzTRUE}). Nevertheless, here and henceforth we consider the form (\ref{boltz}) of the Boltzmann hierarchy (instead of (\ref{boltzTRUE})) since this arises
from the BBGKY hierarchy (\ref{bbgky}) in the limit $N\to+\infty,\delta\to 0, N\delta^3=\alpha>0$.

It is easy to check that estimates (\ref{Ck+1normINFTY}), (\ref{Ck+2normINFTY}) and (\ref{Ck+3normINFTY})
for operators $C_{k,k+1}^N$, $C_{k,k+2}^N$ and $C_{k,k+3}^N$
hold for operators $C_{k,k+1}$, $C_{k,k+2}$ and $C_{k,k+3}$ as well. 
In fact, by (\ref{c1}), (\ref{c2}) and (\ref{c3}) we get:
\begin{eqnarray}
&&\!\!\!\!\!\left|\left(C_{k,k+1} f_{k+1}^\infty\right)(V_k)\right|\leq \sum_{1\leq i\leq k}\int dv_{k+1}\int d\omega \ B^\omega_{i,k+1} \ \, \left[f_{k+1}^\infty(V_{k+1}^{i,k+1})+f_{k+1}^\infty(V_{k+1})\right]\nonumber \label{c1ST} \\
&&\!\!\!\!\!\left|\left(C_{k,k+2} f_{k+2}^\infty\right)(V_k)\right|\leq \alpha\sum_{1\leq i\leq k}\int dv_{k+1}\int d\omega \ B_{i,k+1}^\omega\, \left[f_{k+2}^\infty(V_{k+1}^{i,k+1},v_i)+f_{k+2}^\infty(V_{k+1}^{i,k+1},v_{k+1}) \right.+\nonumber\\
&& \qquad\qquad\qquad\ \ \ \  \ \ \ \ \ \ \ \ \ \ \ \ \ \ \ \ \ \left. + f_{k+2}^\infty(V_{k+1},v_{i}')+f_{k+2}^\infty(V_{k+1},v_{k+1}')\right]\nonumber\label{c2ST} \\
&&\!\!\!\!\!\left|\left(C_{k,k+3} f_{k+3}^\infty\right)(V_k)\right|\leq \alpha^2\!\!\sum_{1\leq i\leq k}\int dv_{k+1}\int d\omega \ B_{i,k+1}^\omega \left[f_{k+3}^\infty(V_{k+1}^{i,k+1},v_{k+1},v_i)+f_{k+3}^\infty(V_{k+1},v_{k+1}',v'_i)\right],\nonumber
\end{eqnarray}
that, under the assumptions $i)$ and $ii)$ on $B$, imply:
\begin{eqnarray}\label{estimatesREC1bol}
&&||C_{k,k+1} \, f_{k+1}^\infty(t)||_{L^\infty(\R^{3k})}\leq C_{B}\, k \,  || f_{k+1}^\infty(t)||_{L^{\infty}(\R^{3(k+1)})},
\end{eqnarray}
\begin{eqnarray}\label{estimatesRERandC2bol}
&&||C_{k,k+2}\, f_{k+2}^\infty(t)||_{L^\infty(\R^{3k})}\leq C_{B}\, k \, \alpha\,  
 || f_{k+2}^\infty(t)||_{L^\infty(\R^{3(k+2)})}
\end{eqnarray}
and
\begin{eqnarray}\label{Ck+3normINFTYlim}
&&||\left(C_{k,k+3}\, f_{k+3}^\infty\right)(t)||_{\infty}\leq C_B\,  k\, \alpha^2\,    ||f_{k+3}^\infty(t)||_{\infty}.
\end{eqnarray}

\subsection{Uniform boundedness for $f_k^N(t)$ (short time)}
\label{UNIFboundN}

By (\ref{BBGKY}), (\ref{flowDEF}) and the Duhamel formula we find that, for any $k=1,\dots, N$:
\begin{eqnarray}\label{BBGKYsol}
f_k^N(t)&=&S_k^N(t)f_k^N(0)+\int_0^t dt_1 \, S_k^N(t-t_1)\, \sum_{s=1}^2 \frac{1}{N}\left(L_{k,k+s}^N\, f_{k+s}^N\right)(t_1)+ \nonumber\\
&+& \int_0^t dt_1 \, S_k^N(t-t_1)\, \sum_{s=1}^3\left(C_{k,k+s}^N\, f_{k+s}^N\right)(t_1).
\end{eqnarray}
\vspace{0.1cm}

Iterating the Duhamel formula in (\ref{BBGKYsol}) 
we get the following expansion:
\begin{eqnarray}\label{exSUM}
&&f_k^N(t)=\sum_{n=0}^{+\infty}\ \, \sum_{\substack{j(1)\dots j(n)
}} \sum_{O^N}
\int d\mathbf{t}_n	\, 
S_k^N(t-t_1)\, O_{k,k+i(1)}^N
\, S_{k+i(1)}^N(t_1-t_2)O_{k+i(1),k+i(2)}^N
\dots \nonumber\\
&&\qquad\qquad\qquad\qquad\qquad\qquad \qquad\qquad \qquad \dots 
O_{k+i(n-1),k+i(n)}^N\, 
S_{k+i(n)}^N(t_n)f_{k+i(n)}^N(0), 
\end{eqnarray}
where
\begin{eqnarray}
&& \imath)\ \ \int d\mathbf{t}_n:= \int_0^t dt_1\dots\int_0^{t_{n-1}}dt_{n} \nonumber\\
&& \imath\imath)\ \ 
\sum_{\substack{j(1)\dots j(n)}}:=\sum_{\substack{(j(1),\dots, j(n))\in \{1,2,3\}^{ n}}}
\ \ \ \ \ \text{and}\ \ \ \ \ i(m):=\sum_{\ell=1}^m j(\ell),\ \ \forall\ \, m=1,\dots, n\label{exSUMnotations3}\\
&&\imath \imath\imath )\ \ \sum_{O^N}:=\sum_{O_{k,k+i(1)}^N\dots O_{k+i(n-1),k+i(n)}^N}\ \ \ \text{where, setting $i(0)=0$, for any $m=1,\dots,n$ we have}:\nonumber\\
&& \ \cdot\ \ O_{k+i(m-1),k+i(m)}^N
\in \left\{\frac{1}{N}L_{k+i(m-1),k+i(m-1)+1}^N, C_{k+i(m-1),k+i(m-1)+1}^N\right\}\ \ \nonumber\\
&&\qquad  \qquad \text{if}\ \  i(m)-i(m-1)=j(m)=1\nonumber\\
&&\   \cdot\ \  O_{k+i(m-1),k+i(m)}^N\in \left\{\frac{1}{N}L_{k+i(m-1),k+i(m-1)+2}^N, C_{k+i(m-1),k+i(m-1)+2}^N\right\}\ \ \nonumber\\
&&\qquad \qquad \text{if}\  \ i(m)-i(m-1)=j(m)=2\nonumber\\
&&\ \cdot\ \  O_{k+i(m-1),k+i(m)}^N=C_{k+i(m-1),k+i(m-1)+3}^N\ \ \ \text{if}\ \  i(m)-i(m-1)=j(m)=3,\label{exSUMnotations2}
\end{eqnarray}
Notice that, since $f_k^N\equiv 0$ for $k\geq N+1$, the series in (\ref{exSUM}) is indeed a finite sum. \\

We are going to show that, within a suitable time interval, expansion (\ref{exSUM}) is uniformly bounded with respect to the norm $||\cdot||_{\delta,k}$ since it is dominated by a (geometric) converging series.  To this end, we introduce the following:

\begin{definition}\label{remOPNORM}
For any $k\in \N^*$ and $\delta>0$, we define:
\begin{eqnarray}\label{OPNORMspace}
\mathcal{H}_k^\delta:=\{f_k:\R^{3k}\to \R,\ \ s.t.\ ||f_k||_{\delta,k}<+\infty\}
\end{eqnarray}
Thus, for any operator $\mathcal{O}_k^N\in \{
\frac{1}{N}L_{k,k+1}^N,\frac{1}{N}L_{k,k+2}^N, C_{k,k+1}^N, C_{k,k+2}^N,  C_{k,k+3}^N \}$, we denote by:
\begin{eqnarray}\label{OPNORM}
||\mathcal{O}_k^N||:=\sup_{f_{k+\ell}\in \mathcal{H}_{k+\ell}^\delta}\frac{|| \mathcal{O}_k^N f_{k+\ell}^N||_{\delta,k}}{|| f_{k+\ell}^N||_{\delta,k+\ell}},\ \ \ell=1,2,3
\end{eqnarray}
namely, $||\cdot||$ is the operator norm of $\mathcal{O}_k^N: \mathcal{H}_{k+\ell}^\delta\to \mathcal{H}_{k}^\delta$, with $\ell=1,2,3$.
\end{definition}
By (\ref{exSUM}) we know that the series we need to control is:
\begin{eqnarray}\label{exSUMest}
\sum_{n=0}^{
+\infty} \ \sum_{\substack{j(1)\dots j(n)
}} \sum_{O^N}
\left\|\int d\mathbf{t}_n	\, 
S_k^N(t-t_1)\, O_{k,k+i(1)}^{N}
\dots 
S_{k+i(n)}^N(t_n)f_{k+i(n)}^N(0) \right\|_{\delta,k}
\end{eqnarray}
Using estimate
(\ref{LkCONTinNEWnorm}) for the flow $S_k^N(t)$, we find that:
\begin{eqnarray}\label{exSTRINGest}
&&\left\|\int d\mathbf{t}_n	\, 
S_k^N(t-t_1)\, O_{k,k+i(1)}^{N}
\dots 
S_{k+i(n)}^N(t_n)f_{k+i(n)}^N(0) \right\|_{\delta,k}\leq\nonumber\\
&& \leq \int d\mathbf{t}_n	 \, e^{C_B\, k\, (t-t_1)}\left\|O_{k,k+i(1)}^N\right\|
\dots 
\left\|O_{k+i(n-1),k+i(n)}^N\right\|
e^{C_B\, (k+i(n))\, t_n}\left\|f_{k+i(n)}^N(0)\right\|_{\delta,k+i(n)}\leq\nonumber\\
&&\leq\frac{t^n}{n!}\, e^{C_B\, (k+i(n))\, t}\left\|O_{k,k+i(1)}^N\right\|
\dots
\left\|O_{k+i(n-1),k+i(n)}^N\right\|
\left\|f_{k+i(n)}^N(0)\right\|_{\delta,k+i(n)}.\nonumber
\end{eqnarray}
Now, thanks to estimates (\ref{Lk+11normVII}), (\ref{Lk+2normIV}), (\ref{Ck+1normIV}), (\ref{Ck+2norm}) and (\ref{Ck+3norm}) for operators $O^N$'s, we know that there exists a positive constant $C_{\alpha,B}$ such that:
\begin{eqnarray}\label{exSTRINGestI}
&&\left\|\int d\mathbf{t}_n	\, 
S_k^N(t-t_1)\, O_{k,k+i(1)}^N
\, S_{k+i(1)}^N(t_1-t_2)
\dots 
O_{k+i(n-1),k+i(n)}^N
S_{k+i(n)}^N(t_n)f_{k+i(n)}^N(0)\right\|_{\delta,k}\leq \nonumber\\
&&\leq \frac{t^n}{n!}\, e^{C_B\, (k+i(n))\, t}\, C_{\alpha,B}^n\, k(k+i(1))\dots\left(k+i(n-1)\right)\left\|f_{k+i(n)}^N(0)\right\|_{\delta,k+i(n)}.
\end{eqnarray}
and since, by construction, $i(m)\leq 3m$ for any $m=1,\dots, n-1$, we obtain:
\begin{eqnarray}\label{exSTRINGestII}
&&\left\|\int d\mathbf{t}_n	\, 
S_k^N(t-t_1)\, O_{k,k+i(1)}^N
\, S_{k+i(1)}^N(t_1-t_2)
\dots 
O_{k+i(n-1),k+i(n)}^N
S_{k+i(n)}^N(t_n)f_{k+i(n)}^N(0)\right\|_{\delta,k}\leq \nonumber\\
&&\leq \frac{t^n}{n!}\, e^{C_B\, (k+3n)\, t}\, C_{\alpha,B}^n\, k(k+3)\dots\left(k+3n-3)\right)\left\|f_{k+i(n)}^N(0)\right\|_{\delta,k+i(n)}.
\end{eqnarray}
By inequality (\ref{normREM3}) and assumption (\ref{LinftyfN}) of the initial data it follows that, for any $k=1,2,\dots$, 
\begin{eqnarray}\label{LinftyfNDELTA}
||f_k^N(0)||_{\delta,k}\leq (z_1)^k,\ \ \ \ \forall\ \  \delta>0.
\end{eqnarray}
Then, (\ref{exSTRINGestI}) yields:
\begin{eqnarray}\label{exSTRINGestIInew}
&&\left\|\int d\mathbf{t}_n	\, 
S_k^N(t-t_1)\, O_{k,k+i(1)}^N
\, S_{k+i(1)}^N(t_1-t_2)
\dots 
O_{k+i(n-1),k+i(n)}^N
S_{k+i(n)}^N(t_n)f_{k+i(n)}^N(0)\right\|_{\delta,k}\leq \nonumber\\
&&\leq \frac{t^n}{n!}\, e^{C_B\, (k+3n)\, t}\, C_{\alpha,B}^n\, k(k+3)\dots\left(k+3n-3)\right)\, z_1^{k+3n}.
\end{eqnarray}
Let us consider the product $k(k+3)\dots\left(k+3n-3\right)$. We have:
\begin{eqnarray}\label{fattori}
 &&k(k+3)\dots\left(k+3n-3\right)\leq (k+3n)^n=\sum_{j=0}^n \frac{n!}{j!(n-j)!}\, k^j\, (3n)^{n-j}\leq\nonumber\\
 &&\qquad
 \leq \sum_{j=0}^n \frac{n!}{j!(n-j)!}\, \frac{k^j}{j!}\, (3n)^{n-j}\, j^j\leq
 e^k\, \sum_{j=0}^n \frac{n!}{j!(n-j)!}\, (3n)^{n-j}\, n^j=\nonumber\\
 &&\qquad = 
 e^k\, (3n)^{n}\, \left(\frac{4}{3}\right)^n=e^k\, (4n)^{n}.
 \end{eqnarray}
 Therefore, by (\ref{exSTRINGestIInew}) we get:
 \begin{eqnarray}\label{TERMINIk+3est13}
&&\left\|\int d\mathbf{t}_n	\, 
S_k^N(t-t_1)\, O_{k,k+i(1)}^N
\, S_{k+i(1)}^N(t_1-t_2)
\dots 
O_{k+i(n-1),k+i(n)}^N
S_{k+i(n)}^N(t_n)f_{k+i(n)}^N(0)\right\|_{\delta,k}\leq \nonumber\\
&&\leq\frac{t^n}{n!}\, e^{C_B\, (k+3n)\, t}\, (C_{\alpha,B})^n\, e^k\, (4n)^{n}\, z_1^{k+3n}=e^k\,   z_1^k\,  e^{C_B\, k\, t}\  \frac{n^n}{n!}\, \left(4\, C_{\alpha,B}\, z_1^3 \, t\,  e^{3C_B\, t}\right)^n,
\end{eqnarray}
that, by using the Stirling formula, yields:
 \begin{eqnarray}\label{TERMINIk+3est14}
&&\left\|\int d\mathbf{t}_n	\, 
S_k^N(t-t_1)\, O_{k,k+i(1)}^N
\, S_{k+i(1)}^N(t_1-t_2)
\dots 
O_{k+i(n-1),k+i(n)}^N
S_{k+i(n)}^N(t_n)f_{k+i(n)}^N(0)\right\|_{\delta,k}\leq \nonumber\\
&&\leq e^k\,    z_1^k\,  e^{C_B\, k\, t}\, \left(4\, e\, C_{\alpha,B}\, z_1^3\, t\,  e^{3C_B\, t}\right)^n.
\end{eqnarray}
As a consequence, the series in (\ref{exSUMest}) is estimated in the space $\mathcal{H}_k^\delta$ by:
\begin{eqnarray}\label{exSTRINGestIV}
e^k\,    z_1^k\,  e^{C_B\, k\, t}\, \sum_{n=0}^{+\infty} \ \sum_{\substack{j(1)\dots j(n)
}} \sum_{O^N}
\left(4\, e\, C_{\alpha,B}\,  z_1^3\, t\, e^{3C_B\, t}\right)^n,
\end{eqnarray}
and, since by definitions (\ref{exSUMnotations3}), (\ref{exSUMnotations2}) it follows that:
$$
\sum_{\substack{j(1)\dots j(n)
}} \sum_{O^N}\  (1)<
(3^n) \, 5^n=15^n,
$$
we find that (\ref{exSTRINGestIV}) is less or equal than:
\begin{eqnarray}\label{exSTRINGestIVbis}
e^k\,    z_1^k\,  e^{C_B\, k\, t}\, \sum_{n=0}^{+\infty} \left(60\, e\, C_{\alpha,B}\,  z_1^3\,  t\, e^{3C_B\, t}\right)^n.
\end{eqnarray}
Clearly, the above series is converging if:
 \begin{eqnarray}\label{exSTRINGestV}
t\, e^{3C_B\, t}<\frac{1}{\left(60\, e\, C_{\alpha,B}\, z_1^3\right)}:=\lambda_0.
\end{eqnarray}
Therefore, looking at the evolution up to some time $T>0$, we are guaranteed that the series (\ref{exSTRINGestIVbis}) is converging for $t<t_0$, where:
 \begin{eqnarray}\label{exSTRINGestVbis}
t_0:=e^{-3C_B\, T}\, \lambda_0.
\end{eqnarray}

\subsection{Uniform boundedness for the Boltzmann hierarchy (short time)}

Let us consider the Boltzmann hierarchy (\ref{BOLTZtrue}) 
with initial datum $\{f_k^\infty(0)\}_k$. For any $k=1,2,\dots$ we have:
\begin{eqnarray}\label{BOLTZsol}
f_k^\infty(t)=f_k^\infty(0)+\int_0^t dt_1 \, \sum_{s=1}^3\left(C_{k,k+s}\, f_{k+s}^\infty\right)(t_1),
\end{eqnarray}
where $C_{k,k+1}$, $C_{k,k+2}$ and $C_{k,k+3}$ are defined as in (\ref{c1}), (\ref{c2}), (\ref{c3}).

As before, iterating the Duhamel formula in (\ref{BOLTZsol}) we obtain a perturbative expansion of the form:
\begin{eqnarray}\label{exSUMboltz}
f_k^\infty(t)=\sum_{n=0}^{+\infty}\ \, \sum_{\substack{j(1)\dots j(n)
}} 
\, 
\frac{t^n}{n!}\, C_{k,k+i(1)}
\, C_{k+i(1),k+i(2)}
\dots 
C_{k+i(n-1),k+i(n)}
\, f_{k+i(n)}^\infty(0), 
\end{eqnarray}
where we used the notations already introduced in (\ref{exSUMnotations3}).

Thanks to estimates (\ref{estimatesREC1bol}), (\ref{estimatesRERandC2bol}) and (\ref{Ck+3normINFTYlim}) on operators $C_{k,k+1}$, $C_{k,k+2}$ and $C_{k,k+3}$ respectively, we can make calculations that are completely analogous to those we did to prove the uniform boundedness of the series associated with the BBGKY hierarchy (\ref{BBGKY}). Then, using assumption (\ref{Linftyflim}) for the initial datum,
it turns out that the series (\ref{exSUMboltz}) is converging in $L^\infty(\R^{3k})$ within a certain time interval.  In fact, it is uniformly bounded by:
\begin{eqnarray}\label{stimaLinf}
e^k\,   \left(\frac{1}{\alpha}\right)^{k}\, z_2\sum_{n=0}^{+\infty}\left(12\, e\, C_{\alpha,B}\,  \alpha^{-3}\, t\right)^n,
\end{eqnarray}
which is converging for $t\in [0,t_1)$, with:
\begin{eqnarray}\label{exSTRINGestVbisBOL}
t_1:=\frac{1}{\left(12\, e\, C_{\alpha,B}\, \alpha^{-3}\right)}.
\end{eqnarray}
By
just taking $t^*:=\min\{t_0,t_1\}$, with $t_0$ determined in (\ref{exSTRINGestVbis}),
we are guaranteed that for $t<t^*$ both expansions (that associated with $f_k^N(t)$ and that associated with $f_k^\infty(t)$) are uniformly bounded in $N$ and $\delta$.

\begin{rem}\label{unicitaBOLTZMANN}
The above argument shows that, for $t\in [0,t^*)$, the Boltzmann hierarchy  (\ref{boltz}) 
has a unique solution in the class of sequences $\{f_k\}_k$ such that, for any $k$,  $||f_k||_\infty\leq C^k$, for some $C>0$.
\end{rem}

\begin{rem}\label{ForPILING}
It is useful for future purposes to introduce other series expansions, different from (\ref{exSUM}) and (\ref{exSUMboltz}), for $f_k^N(t)$ and $f_k^\infty(t)$ respectively. \\
We set:
\begin{eqnarray}\label{eq1:REM5.3}
f_k^N(t)=\sum_{n=0}^{+\infty} \mathscr{T}_n^N(t)\, f_{k+n}^N(0),
\end{eqnarray}
where:
\begin{eqnarray}\label{eq2:REM5.3}
&&\!\!\!\!\!\!\mathscr{T}_n^N(t)\, f_{k+n}^N(0)=
\sum_{m=m_0(n)}^{n} \ \,  \sum_{\substack{j(1)\dots j(m)\in \{1,2,3\}^m\\ \\ \sum_r j(r)=n}}\, \sum_{O^N_{j(1)}\dots O^N_{j(m)}}\, 
\int d\mathbf{t}_m	
\, S_k^N(t-t_1)\, O_{j(1)}^N \, S_{k+j(1)}^N(t_1-t_2)\dots \nonumber\\
&&\qquad\qquad\qquad\qquad\qquad\qquad\qquad\qquad\qquad\qquad\qquad\dots O_{j(m)}^N\, S_{k+n}^N(t_m)f_{k+n}^N(0),
\end{eqnarray}
and $O_{j(1)}^N$ is any operator $L_{k,k+j(1)}^N, C_{k,k+j(1)}^N$ increasing the number of particles by $j(1)$. The other operators $O_{j(r)}^N$'s, with $r\geq 2$, are defined recursively and $\sum_{O^N_{j(1)}\dots O^N_{j(m)}}$ is the sum over all such possible choices (see (\ref{exSUMnotations3}) and (\ref{exSUMnotations2})). Finally, $m_0(n)$ is the smaller integer larger or equal than $\frac{n}{3}$.\\
In other words, here we expand by fixing the number ''$n$'' of particles created in the process, rather than the number of interactions as in (\ref{exSUM}).
\\
Analogously, we set:
\begin{eqnarray}\label{eq3:REM5.3}
f_k^\infty(t)=\sum_{n=0}^{+\infty} \mathscr{T}_n(t)\, f_{k+n}^\infty(0),
\end{eqnarray}
where:
\begin{eqnarray}\label{eq4:REM5.3}
&&\!\!\!\!\!\!\mathscr{T}_n(t)\, f_{k+n}^\infty(0)=
\sum_{m=m_0(n)}^{n} \ \,  \sum_{\substack{j(1)\dots j(m)\in \{1,2,3\}^m\\ \\ \sum_r j(r)=n}}\, 
\frac{t^m}{m!}	
\ \, O_{j(1)} \dots O_{j(m)}\, f_{k+n}^\infty(0),
\end{eqnarray}
and
$O_{j(1)}=C_{k,k+j(1)}$. 
The other operators $O_{j(r)}$'s, with $r\geq 2$, are defined recursively, namely, $O_{j(r)}=C_{k+\sum_{p=1}^{r-1} j(p),k+\sum_{p=1}^{r-1} j(p) \, +j(r)}$.

The previous analysis shows that the series (\ref{eq1:REM5.3}) and (\ref{eq3:REM5.3}) are 
bounded,  uniformly in $N$, by 
a series of the form:
\begin{eqnarray}\label{eq5:REM5.3}
\sum_{n=0}^{+\infty} \lambda^n,
\end{eqnarray}
where $\lambda<1$ provided that $||f_k^N(0)||_{\delta,k}\leq z^k$ and $||f_k^\infty(0)||_{\infty}\leq z^k$, for some $z>0$ and $t<\tau$ with $\tau$ sufficiently small, chosen accordingly to $z$.

\end{rem}

\section{Uniform global in time estimates}
\label{sec:sec6}
\setcounter{equation}{0}    
\def\theequation{6.\arabic{equation}}

We now show how to control $||f_k^N(t)||_{\delta,k}$ for any $t\geq 0$. The estimate we are going to prove is a direct consequence of the exclusion principle and this argument motivates the use of the norm $||\cdot||_{\delta,k}$.\\
We also show an a priori bound for $||f_k^\infty(t)||_\infty$ by using the Hewitt-Savage Theorem (see \cite{HS}).

\subsection{A priori estimates for $||f_k^N(t)||_{\delta,k}$} 
For a fixed a sequence of (different) cells $\Delta_1,\dots,\Delta_k$, the mean value of the product of the occupation numbers of $\Delta_1,\dots,\Delta_k$ at time $t$ is given by:
\begin{eqnarray}\label{number4}
\langle n^t(\Delta_1)\dots n^t(\Delta_k)\rangle = \int_{\R^{3N}}dV_N\ \sum_{i_1=1}^N \chi_{\Delta_1} (v_{i_1})\sum_{\substack{i_2=1\\i_2\neq i_1}}^N \chi_{\Delta_2} (v_{i_2})\dots \sum_{\substack{i_k=1,\\i_k\neq i_m,\\ m=1,\dots,k-1}}^N \chi_{\Delta_k} (v_{i_k})\ W_N(V_N,t),
\end{eqnarray}
and, since $n^t(\Delta_i)\in \{0,1\}$ for $i=1,\dots, k$, we have:
\begin{eqnarray}\label{number4a}
\langle n^t(\Delta_1)\dots n^t(\Delta_k)\rangle \leq 1.
\end{eqnarray}
By the symmetry the distribution $W_N$ under permutations of the indeces:
\begin{eqnarray}\label{number4b}
\langle n^t(\Delta_1)\dots n^t(\Delta_k)\rangle = N(N-1)\dots (N-k+1)\int_{\R^{3N}}dV_N\ \chi_{\Delta_1} (v_{1}) \chi_{\Delta_2} (v_{2})\dots  \chi_{\Delta_k} (v_{k})\ W_N(V_N,t),\nonumber
\end{eqnarray}
and, by the definition of $k$-particle marginal:
\begin{eqnarray}\label{number4d}
\langle n^t(\Delta_1)\dots n^t(\Delta_k)\rangle &= &N(N-1)\dots (N-k+1)\int_{\R^{3k}}dV_k\ \chi_{\Delta_1} (v_{1}) \chi_{\Delta_2} (v_{2})\dots  \chi_{\Delta_k} (v_{k})\ f_k^N(V_k,t)=\nonumber\\
&=&N(N-1)\dots (N-k+1)\int_{\Delta_1}dv_1\dots \int_{\Delta_k}dv_k\ f_k^N(V_k,t)\leq 1.
\end{eqnarray}
Therefore:
\begin{eqnarray}\label{margONE1deltaVIk}
\frac{1}{\delta^{3k}}\int_{\Delta_{1}}dv_1\dots \int_{\Delta_{k}}dv_k\  f_k^N(V_k,t)\leq \frac{1}{N(N-1)\dots (N-k+1)\ \delta^{3k}}.
\end{eqnarray}
Since $\delta^3=\alpha\, N^{-1}$, it follows that:
\begin{eqnarray}\label{eqT}
\frac{1}{N(N-1)\dots (N-k+1)\ \delta^{3k}}= \left(\frac{1}{\alpha}\right)^k\, (T_{k,N}),
\end{eqnarray}
where
\begin{eqnarray}\label{eqT1}
T_{k,N}:=\frac{N^k}{N(N-1)\dots (N-k+1)}.
\end{eqnarray}
Notice that, for $k\leq \frac{N}{2}$, we have:
\begin{eqnarray}\label{eqT2}
T_{k,N}=\frac{1}{(1)(1-\frac{1}{N})(1-\frac{2}{N})\dots (1-\frac{k-1}{N})}\leq\left( \frac{1}{1-\frac{k}{N}}\right)^k\leq 2^k.
\end{eqnarray}
On the other hand, if $k>\frac{N}{2}$, we have:
\begin{eqnarray}\label{eqT3}
T_{k,N}=\frac{N^k}{N(N-1)\dots (N-k+1)}=\frac{N^k\, (N-k)!}{N!}\leq \frac{N^k\, N^{N-k}}{N!}=\frac{N^N}{N!}\leq e^N< e^{2k}.
\end{eqnarray}
Therefore, for any $k=1,\dots, N$, we have:
\begin{eqnarray}\label{eqT4}
T_{k,N}<e^{2k},
\end{eqnarray}
and, by (\ref{margONE1deltaVIk}):
\begin{eqnarray}\label{margONE1deltaVIIk}
\sup_{\substack{\Delta_{1}\dots\Delta_{k},\\ \Delta_{m}\neq \Delta_{\ell}}}\frac{1}{\delta^{3k}}\int_{\Delta_{1}}dv_1\dots \int_{\Delta_{k}}dv_k\  f_k^N(V_k,t)=||f_k^N(t)||_{\delta,k}<\frac{e^{2k}}{\alpha^k},\ \ \ \forall\ \ t\geq 0.
\end{eqnarray}

 \subsection{A priori estimates for $||f_k^\infty(t)||_{L^\infty(\R^{3k})}$} 
 
By the assumptions we made on the initial limiting sequence $\{f_k^\infty(0)\}_k$ (see Theorem \ref{main}) it follows that, thanks to the Hewitt-Savage Theorem (see \cite{HS}), there exists a unique probability measure $\nu$ on the space $M^1_+(\R^3)$ of one-particle probability densities such that, for all $k$,
\begin{eqnarray}\label{hewittGEN2teoSEC7}
f_k^\infty(V_k,0)=\int_{M^1_+(\R^3)} d\nu(f_0) \ f_0^{\otimes k} (V_k).
\end{eqnarray}
In other words, the limiting distribution $f_k^\infty(0)$ can be tought as a statistical mixture of factorized distributions. As we already observed, the measure $\nu$ describes the correlations of the $k$-particle distribution $f_k^\infty(0)$.

Now, by property  (\ref{Linftyflim}) we know that $||f_k^\infty(0)||_\infty\leq \frac{z_2}{\alpha^k}$, for some positive constant $z_2>0$. We are going to show that such a boundedness condition implies a natural bound for any $f_0$ in (\ref{hewittGEN2teoSEC7}). More precisely, we have the following:
\begin{prop}\label{hewittPROPOSITION2}
If $f_k^\infty(0)$ is such that $||f_k^\infty(0)||_\infty\leq C/\alpha^k$,\ for some finite constants $C,\alpha>0$, and (\ref{hewittGEN2teoSEC7}) holds, then:
\begin{eqnarray}\label{boundLinfty}
||f_0||_\infty\leq \frac{1}{\alpha},\ \ \ \ \ \nu\ \, \text{a.e.}
\end{eqnarray}
\end{prop}
\textbf{Proof:}
\\
For any $v\in \R^3$, consider the function $\Phi_v:M^1_+(\R^3)\to \R^+$ defined as:
\begin{equation}\label{phi}
\phi_v(f_0)=f_0(v),\ \ \ \ \text{for\, all}\ \ f_0\in M^1_+(\R^3).
\end{equation}
Then, Proposition \ref{hewittPROPOSITION2} is proven once we show that:
\begin{equation}\label{phiBOUND}
||\phi_v||_{L^\infty(d\nu)}\leq \frac{1}{\alpha}.
\end{equation}
But inequality (\ref{phiBOUND}) is simply achieved since:
\begin{eqnarray}\label{phiBOUND2}
||\phi_v||_{L^\infty(d\nu)}&\!\!\!\!=\!\!&\lim_{p\to +\infty}||\phi_v||_{L^p(d\nu)}= \lim_{p\to +\infty}\left(\int d\nu(f_0)\, |\phi_v(f_0)|^p\right)^{\frac{1}{p}}=\lim_{p\to +\infty}\left(\int d\nu(f_0)\, f_0(v)^p\right)^{\frac{1}{p}}=\nonumber\\
&\!\!\!\!=\!\!&\lim_{p\to +\infty}\left(\int d\nu(f_0)\, \underbrace{f_0(v)\dots f_0(v)}_{p\ \text{times}}\right)^{\frac{1}{p}}=\lim_{p\to +\infty}\left(f_p^\infty( \underbrace{v\dots v}_{p\ \text{variables}})\right)^{\frac{1}{p}}\leq \nonumber\\
&\!\!\!\!\leq\!\!& \frac{1}{\alpha} \ \lim_{p\to +\infty} C^{\frac{1}{p}}=\frac{1}{\alpha}.
\end{eqnarray}
\hfill \qed\\

We notice that, as a byproduct of the above proposition, we can claim that:
\begin{eqnarray}\label{byPROD}
|| f_k^\infty(0)||_\infty\leq \left(\frac{1}{\alpha}\right)^k.
\end{eqnarray}
Now, for any time $t>0$, we define the $k$-particle distribution $f_k(V_k,t)$ as:
\begin{eqnarray}\label{HS1}
f_k(V_k,t)=\int_{M^1_+(\R^3)} d\nu(f_0) \ f_t^{\otimes k} (V_k).
\end{eqnarray}
where $\nu$ is the same measure appearing in (\ref{hewittGEN2teoSEC7}) and $f_t$ is the unique $L^\infty$-solution of the U-U equation (\ref{UU1}) with initial datum $f_0$. Of course, for $t=0$ we have:
$$
f_k(0)=f_k^\infty(0)
$$
It can be easily verified that, for all $k=1,2,\dots$, the distribution $f_k(t)$ is uniformly bounded. In fact, we have:
\begin{eqnarray}\label{HS2}
||f_k(t)||_\infty=|| f_t||_\infty^{k},
\end{eqnarray}
and, as proven in \cite{DOLB}, the bound (\ref{boundLinfty}) ensures that:
\begin{eqnarray}\label{HS3}
|| f_t||_\infty\leq \frac{1}{\alpha},
\end{eqnarray}
namely, an $L^\infty$ maximum principle holds for the U-U equation. 
Therefore, by (\ref{HS2}) and (\ref{HS3}) we get:
\begin{eqnarray}\label{HS4}
||f_k(t)||_\infty\leq \left(\frac{1}{\alpha}\right)^k,\ \ \ \ \text{for all}\ \ t\geq 0.
\end{eqnarray}

As observed by H. Spohn in \cite{S2},  the bounded family of distributions $\{f_k(t)\}_k$ is a solution of the Boltzmann hierarchy (\ref{boltzTRUE}) with initial datum $\{f_k^\infty(0)\}_k$.  Thus, since estimate (\ref{HS4}) holds, the uniqueness result we proved in Section \ref{sec:sec5} (see Remark \ref{unicitaBOLTZMANN}) ensures that:
\begin{eqnarray}\label{HS5}
f_k(t)=f_k^\infty(t),\ \ \ \ \text{for all}\ \ t\in [0,t^*),
\end{eqnarray}
where $t^*$ has been defined in Section \ref{sec:sec5}.
 As a consequence, the $L^\infty$-bound (\ref{HS4}) holds, up to time $t^*$, for all distributions $f_k^\infty(t)$, namely:
\begin{eqnarray}\label{HS6}
||f_k^\infty(t)||_\infty\leq \left(\frac{1}{\alpha}\right)^k,\ \ \ \ \text{for all}\ \  t\in [0,t^*).
\end{eqnarray}

By iterating the above argument (i.e., interpolation between Hewitt-Savage Theorem and the uniqueness result in a suitable class of distributions), we prove that estimate (\ref{HS6}) holds in fact for all $t\geq 0$, i.e.
\begin{eqnarray}\label{HS6iterGLOB}
||f_k^\infty(t)||_\infty\leq \left(\frac{1}{\alpha}\right)^k,\ \ \ \ \text{for all}\ \  t\geq 0.
\end{eqnarray}

\section{Convergence}
\label{sec:sec7}
\setcounter{equation}{0}    
\def\theequation{7.\arabic{equation}}

In this Section we exploit the term by term convergence by piling up a finite number of series expansions converging for a short time.

\subsection{''Piling up'' series expansions
}

In this paragraph we show that, thanks to the a priori estimates for $f_k^N(t)$ and $f_k^\infty(t)$, we can express those quantities in terms of a finite sum plus an arbitrarily small remainder for any $t\in [0,T]$, $T>0$ arbitrary and fixed.

We partition the interval $[0,t]$ into $s$ intervals of amplitude $\tau$, where $\tau$, according to Remark \ref{ForPILING}, is chosen in such a way that the series (\ref{eq1:REM5.3}) and (\ref{eq3:REM5.3}) are bounded by a converging geometric series $\sum_{n=0}^{+\infty}\lambda^n$ for $z=\frac{e^2}{\alpha}$ (see estimates (\ref{margONE1deltaVIIk}) and (\ref{HS6iterGLOB})).\\
Then, we can write:
\begin{eqnarray}\label{eq1:REM5.3PILING}
f_k^N(t)=\sum_{n_1=0}^{+\infty} \dots \sum_{n_s=0}^{+\infty}\mathscr{T}_{n_1}^N(\tau)\dots \mathscr{T}_{n_s}^N(\tau)\, f_{k+\sum_{p=1}^s n_p}^N(0).
\end{eqnarray}
Here and in the sequel we make use of the semigroup property for which $ \sum_{n=0}^{+\infty}\mathscr{T}_{n}^N(2\tau)f_{k+n}^N(0)=\sum_{n_1=0}^{+\infty}\sum_{n_2=0}^{+\infty}\mathscr{T}_{n_1}^N(\tau)\mathscr{T}_{n_2}^N(\tau)f_{k+n_1+n_2}^N(0)$. \\
Next, we introduce a sequence of cutoff $M_1<M_2<\dots M_s$, to be fixed later on, for which:
\begin{eqnarray}\label{eq2:REM5.3PILING}
&&\!\!\!\!
\!\!\!\!\!\!\!
f_k^N(t)
=f_k^N(s\tau)
=\sum_{n_1=0}^{M_1} \mathscr{T}_{n_1}^N(\tau) f_{k+n_1}^N((s-1)\tau)+ R_{k,M_1}^N=\nonumber\\
&&\!\!\!\!
\!\!\!\!\!\!\!
= R_{k,M_1}^N(t) + \sum_{n_1=0}^{M_1} \mathscr{T}_{n_1}^N(\tau) R_{k+n_1,M_2}^N((s-1)\tau)+\sum_{n_1=0}^{M_1}\sum_{n_2=0}^{M_2}\mathscr{T}_{n_1}^N(\tau)\mathscr{T}_{n_2}^N(\tau)f_{k+n_1+n_2}^N((s-2)\tau) =\nonumber\\
&&\!\!\!\!
\!\!\!\!\!\!\!
\dots\nonumber\\
&&\!\!\!\!
\!\!\!\!\!\!\!
\dots\nonumber\\
&&\!\!\!\!
\!\!\!\!\!\!\!
= R_{k,M_1}^N(t)+ \sum_{n_1=0}^{M_1} \!\!\mathscr{T}_{n_1}^N(\tau) R_{k+n_1,M_2}^N((s-1)\tau)+\dots+\sum_{n_1=0}^{M_1}\dots\!\!\sum_{n_{s-1}=0}^{M_{s-1}}\!\!\mathscr{T}_{n_1}^N(\tau)\dots \mathscr{T}_{n_{s-1}}^N(\tau)R_{k+\sum_{p=1}^{s-1} n_p,M_{s}}^N(\tau) \nonumber\\
&&\!\!\!\!
\!\!\!\!\!\!\!
+\sum_{n_1=0}^{M_1}\!\!\dots\!\!\sum_{n_{s}=0}^{M_{s}}\!\!\mathscr{T}_{n_1}^N(\tau)\dots \mathscr{T}_{n_{s}}^N(\tau)f_{k+\sum_{p=1}^{s} n_p}^N(0),
\end{eqnarray}
where, for any $j\geq k$,
\begin{eqnarray}\label{eq3:REM5.3PILING}
&&R_{j,
M_{\ell}}^N(s\tau-(\ell-1)\tau) =\sum_{n=M_\ell +1}^{+\infty} \mathscr{T}_{n}^N(\tau) f_{j+n}^N(s\tau-\ell\tau).
\end{eqnarray}
Moreover,
\begin{eqnarray}\label{eq4:REM5.3PILING}
&&||R_{j,
M_{\ell}}^N(s\tau-(\ell-1)\tau) ||_{\delta,j}\leq \sum_{n=M_\ell +1}^{+\infty} \lambda^n\leq c\ \,  \lambda^{M_\ell},\ \ \ \text{for some}\,\, c>0.
\end{eqnarray}
Let us look at the last remainder term in expression (\ref{eq2:REM5.3PILING}), i.e.
\begin{eqnarray}\label{eq5:REM5.3PILING}
\sum_{n_1=0}^{M_1}\dots\!\!\sum_{n_{s-1}=0}^{M_{s-1}}\!\!\mathscr{T}_{n_1}^N(\tau)\dots \mathscr{T}_{n_{s-1}}^N(\tau)R_{k+\sum_{p=1}^{s-1} n_p,M_{s}}^N(\tau).
\end{eqnarray}
It is estimated by:
\begin{eqnarray}\label{eq6:REM5.3PILING}
\sum_{n=0}^{\sum_{\ell=1}^{s-1} M_\ell} C^n\, (t-\tau)^n\, \lambda^{M_s}\leq \sum_{n=0}^{+\infty}(C\, t)^n\, \lambda^{\frac{M_s}{\sum_{\ell=1}^{s-1} M_\ell}} \lambda^n,\ \ \ \ \text{for some}\ \, C>0.
\end{eqnarray}
Now, choosing:
\begin{eqnarray}\label{eq7:REM5.3PILING}
M_s=2M_1\sum_{\ell=1}^{s-1}M_\ell,
\end{eqnarray}
by (\ref{eq6:REM5.3PILING}) we get:
\begin{eqnarray}\label{eq8:REM5.3PILING}
 \sum_{n=0}^{+\infty}(C\, t)^n\, \lambda^{\frac{M_s}{\sum_{\ell=1}^{s-1} M_\ell}} \lambda^n\leq \lambda^{M_1}\sum_{n=0}^{+\infty}(C\, t\, \lambda^{M_1})^n.
\end{eqnarray}
Therefore, choosing $M_1$ so large that:
\begin{eqnarray}\label{eq9:REM5.3PILING}
C \, t\, \lambda^{M_1}<1,
\end{eqnarray}
the remainder term (\ref{eq5:REM5.3PILING})  turns out to be bounded by:
\begin{eqnarray}\label{eq10:REM5.3PILING}
 \frac{ \lambda^{M_1}}{1-C\, t\, \lambda^{M_1}}.
 \end{eqnarray}
 All the other remainder terms in (\ref{eq2:REM5.3PILING}) can be estimated in the same way, so that:
 \begin{eqnarray}\label{eq11:REM5.3PILING}
f_k^N(t)=
\sum_{n_1=0}^{M_1}\dots\!\!\sum_{n_{s}=0}^{M_{s}}\!\!\mathscr{T}_{n_1}^N(\tau)\dots \mathscr{T}_{n_{s}}^N(\tau)f_{k+\sum_{p=1}^{s} n_p}^N(0)+\mathscr{E}_k^N(t),
\end{eqnarray}
where
 \begin{eqnarray}\label{eq12:REM5.3PILING}
\left\|\mathscr{E}_k^N(t)\right\|_{\delta,k}\leq c(t)\, \lambda^{M_1},
\end{eqnarray}
for some (time depending) constant $c(t)>0$. Notice that both constraint (\ref{eq9:REM5.3PILING}) and bound (\ref{eq12:REM5.3PILING}) are uniform in $N$.

Prooceding as above, we can show that the same expansion holds for the limiting distributions $f_k^\infty(t)$ as well, namely:
 \begin{eqnarray}\label{eq13:REM5.3PILING}
f_k^\infty(t)=
\sum_{n_1=0}^{M_1}\dots\!\!\sum_{n_{s}=0}^{M_{s}}\!\!\mathscr{T}_{n_1}(\tau)\dots \mathscr{T}_{n_{s}}(\tau)f_{k+\sum_{p=1}^{s} n_p}^\infty(0)+\mathscr{E}_k(t),
\end{eqnarray}
where
 \begin{eqnarray}\label{eq14:REM5.3PILING}
\left\|\mathscr{E}_k(t)\right\|_{\infty}\leq c(t)\, \lambda^{M_1}.
\end{eqnarray}

\subsection{Term by term convergence}

In this paragraph we are concerned with the analysis of the asymptotic behavior of the generic term $\mathscr{T}_{n_1}^N(\tau)\dots \mathscr{T}_{n_{s}}^N(\tau)f_{k+\sum_{p=1}^{s} n_p}^N(0)$ of the sum (\ref{eq11:REM5.3PILING}) in the limit $N\to \infty, \delta\to 0, N\delta^3=\alpha>0$. Thanks to the semigroup property, we can reduce the analysis to the quantity $\mathscr{T}_n^N(t)\, f_{k+n}^N(0)$ defined in (\ref{eq2:REM5.3}) and, by that definition, we are lead to look at the asymptotics of a string of the form:
\begin{eqnarray}\label{eq1TERMbyterm}
S_k^N(t-t_1)\, O_{j(1)}^N \, S_{k+j(1)}^N(t_1-t_2)\dots  O_{j(m)}^N\, S_{k+n}^N(t_m)f_{k+n}^N(0),
\end{eqnarray}
where the index $0\leq m\leq n$, the string $(j(1),\dots, j(m))$ and operators $O_{j(r)}^N$'s are characterized as in (\ref{eq2:REM5.3}).
As before, we define $ i(r)=\sum_{q=1}^r j(q)$ so that $i(m)=n$ and (\ref{eq1TERMbyterm}) yields:
\begin{eqnarray}\label{eq2TERMbyterm}
S_k^N(t-t_1)\, O_{k,k+i(1)}^N \, S_{k+i(1)}^N(t_1-t_2)\dots  O_{k+i(m-1), k+i(m)}^N\, S_{k+n}^N(t_m)f_{k+n}^N(0),
\end{eqnarray}
where we set $i(0)=i(-1)=0$ and $O_{k,k}^N=1$.

\subsubsection{Vanishing terms}
We focus on the vanishing terms first, namely, all the strings of the form (\ref{eq2TERMbyterm}) in which for some $q\in \{1,\dots, m\}$ we have:
\begin{eqnarray}\label{eq3TERMbyterm}
 O_{k+i(q-1), k+i(q)}^N=\frac{1}{N}L_{k+i(q-1), k+i(q)}^N.
\end{eqnarray}
By estimates (\ref{Lk+1estLinfty}) and (\ref{Lk+2estLinfty}), since all operators in the game are bounded in $L^\infty$ (see (\ref{Ck+1normINFTY})-(\ref{LkCONT})), we conclude that all such terms converge uniformly to zero as $N\to \infty, \delta\to 0, N\delta^3=\alpha>0$.

\subsubsection{Removing the $S^N$'s}
By the considerations done in the previous paragraph we are reduced to study strings of the form:
\begin{eqnarray}\label{eq4TERMbyterm}
S_k^N(t-t_1)\, C_{k,k+i(1)}^N \, S_{k+i(1)}^N(t_1-t_2)\dots  C_{k+i(m-1), k+i(m)}^N\, S_{k+n}^N(t_m)f_{k+n}^N(0).
\end{eqnarray}
Notice that the string of operators in (\ref{eq4TERMbyterm}) can be written as:
\begin{eqnarray}\label{eq5TERMbyterm}
&&\!\!\!\!\!\!\!\!
\sum_{r=0}^m C_{k,k+i(1)}^NC_{k+i(1),k+i(2)}^N\dots
\left(S_{k+i(r)}^N(t_r-t_{r+1})-I\right) \, C_{k+i(r),k+i(r+1)}^N \, S_{k+i(r+1)}^N(t_{r+1}-t_r)\dots S_{k+n}^N(t_m)
\nonumber\\
&&\!\!\!\!\! \!\!\!+\ 
C_{k,k+i(1)}^N \dots  C_{k+i(m-1), k+i(m)}^N.
\end{eqnarray}
By the $L^\infty$-boundedness of operators $C^N$ and $S^N$ (see (\ref{Ck+1normINFTY})-(\ref{LkCONT})), we conclude that all terms in the above sum are uniformly converging to zero thanks to the trivial estimate:
\begin{eqnarray}\label{eq6TERMbyterm}
\left\| (S_j^N(t)-I\right\|\leq C\,  \frac{j^2}{N}\, e^{C\, \frac{j^2}{N}\, t},
\end{eqnarray}
where we denoted by $\left\|\cdot\right\|$ the operator norm on $L^\infty(\R^{3j})$.
Therefore, the asymptotic behavior of (\ref{eq4TERMbyterm}) is the same as $C_{k,k+i(1)}^N \dots  C_{k+i(m-1), k+i(m)}^N\, f_{k+n}^N(0)$.

\subsubsection{Convergence for strings involving only operators $C^N$'s}

In this paragraph we shall prove that the string:
\begin{eqnarray}\label{eq7TERMbyterm}
 C_{k,k+i(1)}^N \dots  C_{k+i(m-1), k+i(m)}^N\, f_{k+n}^N(0),
\end{eqnarray}
converges pointwise to the corresponding one:
\begin{eqnarray}\label{eq8TERMbyterm}
 C_{k,k+i(1)}\dots  C_{k+i(m-1), k+i(m)}\, f_{k+n}^\infty(0),
\end{eqnarray}
where the operators $C$'s are those involved in the Boltzmann hierarchy (\ref{boltz}).  We are going to write (\ref{eq8TERMbyterm}) in a different way, which will be more convenient for the forthcoming considerations. First, we write  the action of operators $C_{k,k+2}$ and $C_{k,k+3}$ in the following equivalent form:
\bea
\!\!\!\left(C_{k,k+2} f_{k+2}^\infty\right)(V_k)\!\!\!\!&=&\!\!\!\! -\alpha\sum_{i=1}^k\int\! dv_{k+1}\!\!\int \!dv_{k+2}\!\!\int\! d\omega B_{i,k+1}^\omega\!\!\left\{  f_{k+2}^\infty(V_{k+2}^{i,k+1})\, [\delta(v_{k+2}-v_i)+\delta(v_{k+2}-v_{k+1})] +\right.\nonumber\\
&&\!\!\!\quad\qquad\qquad \qquad\qquad  \ \ \ \ \ -\left. f_{k+2}^\infty(V_{k+2})\, [\delta(v_{k+2}-v'_i)+\delta(v_{k+2}-v'_{k+1})]\right\}
 \label{c2N} 
 \eea
 \bea
\!\!\!\left(C_{k,k+3} f_{k+3}^\infty\right)\!(V_k)\!\!\!\!&=&\!\! \!\!\!\alpha^2\sum_{i=1}^k\!\int\! dv_{k+1}\!\!\int \!dv_{k+2}\!\!\int \!dv_{k+3}\!\!\int \!d\omega B_{i,k+1}^\omega\!\!
\left\{f_{k+3}^\infty(V_{k+3}^{i,k+1}) \delta(v_{k+2}-v_{k+1}) \delta(v_{k+3}-v_{i}) +\right.\nonumber\\
&&\!\!\!\quad \qquad \qquad \qquad\qquad \ \ \ \ \ \left.- f_{k+3}^\infty(V_{k+3})\delta(v_{k+2}-v'_{k+1}) \delta(v_{k+3}-v'_{i})\right\}.\label{c3N}
\eea
Note that in considering (\ref{eq8TERMbyterm}) we face the sum:
\bea\label{SUM}
\sum_{\ell_1=1}^k\sum_{\ell_2=1}^{k+i(1)}\dots \sum_{\ell_m=1}^{k+i(m-1)}
\eea
in which each term selects a particle $\ell_r$, among the group of $k+i(r-1)$ already created particles, which interacts with a new particle with index $k+i(r-1)+1
$. We fix such a sequence $\ell_1,\dots, \ell_m$ for which, at each step, we have the interaction between the pair of particles $(\ell_r, k+i(r-1)+1)$. Next, we select and fix a sequence $\underline{\sigma}:=\{\sigma_r\}_{r=1}^m$ where $\sigma_r\in \{-1,1\}$ and, more precisely, the choice $\sigma_r=1$ will be associated with the gain part of the $r$-th operator $C_{k+i(r-1),k+i(r)}$ if $i(r)-i(r-1)=j(r)\in \{1,3\}$ while it will be associated to the loss part of $C_{k+i(r-1),k+i(r)}$ if $j(r)=2$. On the other hand, the choice $\sigma_r=-1$ will be associated with the loss part of the $r$-th operator $C_{k+i(r-1),k+i(r)}$ if $j(r)\in \{1,3\}$ while it will be associated to the gain part of $C_{k+i(r-1),k+i(r)}$ if $j(r)=2$.
Now, according to a given choice of $j(1),\dots, j(m)$ (the numbers of particles created at each step), $\ell_1,\dots, \ell_m$ and $\sigma_1,\dots, \sigma_m$, we can construct a sequence of vectors $\{V_{k+i(r)}^r\}_{r=1}^{m}$ where $V_{k+i(r)}^r=(v_{1}^r,\dots, v_{k+i(r)}^r)$ is defined recursively as $V_k^0=V_k$ and:
\bea\label{recursivelyL}
V_{k+i(r)}^r=v_1^{r-1}\dots v_{k+i(r-1)}^{r-1}\, v_{k+i(r-1)+1}\dots v_{k+i(r)}
\eea
if we take into account the loss part of the $r$-th operator $C_{k+i(r-1),k+i(r)}$, namely, $\sigma_r=-1$ and $j(r)\in{1,3}$ or $\sigma_r=1$ and $j(r)=2$. On the other hand,
\bea\label{recursivelyG}
V_{k+i(r)}^r=v_1^{r-1}\dots (v_{\ell_r}^{r-1})'\dots v_{k+i(r-1)}^{r-1}\, v'_{k+i(r-1)+1}\dots v_{k+i(r)}
\eea
if we take into account the gain part of the $r$-th operator $C_{k+i(r-1),k+i(r)}$, namely, $\sigma_r=1$ and $j(r)\in{1,3}$ or $\sigma_r=-1$ and $j(r)=2$.

Then, the contribution to the string (\ref{eq8TERMbyterm}) due to the above choice of 
$\ell_1,\dots, \ell_m$ and $\sigma_1,\dots, \sigma_m$ is:
\bea\label{eq10TERMbyterm}
\!\!\!\!\!\!\!\!\!\!\!\!\left(\prod_{r=1}^m \sigma_r\right) \alpha^{n-m}\!\!\int \!\!dv_{k+1}
\dots\!\int\!\! dv_{k+n}\int \!\!d\omega_1\!\dots \!\int\!\! d\omega_m\! \left(\prod_{r=1}^m B_r^{\omega_r}\right) \left(\prod_{r=1}^m \mathfrak{D}_r\right) \!f_{k+n}^\infty(v_1^m,\dots, v_{k+n}^m),
\eea
where 
\bea\label{eq11TERMbyterm}
B_r^{\omega_r}
=B(v_{\ell_r}^r-v_{k+i(r-1)+1}^r;\omega_r),
\eea
and
\be\label{eq12TERMbyterm}
 \mathfrak{D}_r=\left\{
  \begin{array}{ll}
1\ \, \qquad\qquad\qquad \qquad\qquad\qquad\qquad\qquad\qquad\qquad\qquad\qquad\text{if}\ \ \  j(r)=1, 
\vspace{0.1cm}\\
\delta(v_{\ell_r}^{r-1}-v_{k+i(r-1)+2})+\delta(v_{k+i(r-1)+1}-v_{k+i(r-1)+2}), \ \ \ \ \ \ \, \text{if}\ \ \ j(r)=2,\ \ \sigma_r=-1,\vspace{0.1cm}\\
\delta(v_{\ell_r}^{r-1}-v_{k+i(r-1)+3})\, \delta(v_{k+i(r-1)+1}-v_{k+i(r-1)+2}), \ \ \ \ \ \ \ \ \ \ \text{if}\ \ \ j(r)=3,\ \ \sigma_r=1,
\end{array}
  \right.
\ee
\be\label{eq13TERMbyterm}
 \mathfrak{D}_r=\left\{
  \begin{array}{ll}
1\ \, \, \quad\quad\qquad\qquad \qquad\qquad\qquad\qquad\qquad\qquad\qquad\qquad\qquad \text{if}\ \ \  j(r)=1, 
\vspace{0.1cm}\\
\delta((v_{\ell_r}^{r-1})'-v_{k+i(r-1)+2})+\delta(v'_{k+i(r-1)+1}-v_{k+i(r-1)+2}), \ \  \ \, \text{if}\ \ \ j(r)=2,\ \ \sigma_r=1,\vspace{0.1cm}\\
\delta((v_{\ell_r}^{r-1})'-v_{k+i(r-1)+3})\, \delta(v'_{k+i(r-1)+1}-v_{k+i(r-1)+2}), \ \ \ \ \ \  \ \text{if}\ \ \ j(r)=3,\ \ \sigma_r=-1.
\end{array}
  \right.
\ee
Clearly, to recover the full string (\ref{eq8TERMbyterm}) we have to sum expression (\ref{eq10TERMbyterm}) over all possible choices of $\ell_1,\dots, \ell_m$ and $\sigma_1,\dots, \sigma_m$.\\
The same notations can be used to handle (\ref{eq7TERMbyterm}) and the contribution due to some choice of  
$\ell_1,\dots, \ell_m$ and $\sigma_1,\dots, \sigma_m$ is:
\bea\label{eq15TERMbyterm}
&&\!\!\!\!\!\!\!\!\!\!\!\!\left(\prod_{r=1}^m \sigma_r\right) \alpha^{n-m}\, \phi(N,k)\int \!\!dv_{k+1}
\dots\int\!\! dv_{k+n}\int \!\!d\omega_1\dots \int\!\! d\omega_m\! \left(\prod_{r=1}^m B_r^{\omega_r}\right) \left(\prod_{r=1}^m \mathfrak{D}_r^N\right) \times\nonumber\\
&&\qquad\quad \quad\quad\times\, \left(\prod_{r=1}^m \overline{\chi}_\delta\left(T
_r
^{\omega_r}
V_{k+i(r-1)+1}^r\right)\right) \  f_{k+n}^N(v_1^m,\dots, v_{k+n}^m),
\eea
where
the map $T
_r^{\omega_r}$
is defined as:
\bea\label{MAP}
\left(T
_r
^{\omega_r}
V_{k+i(r-1)+1}^r\right)=(v_{1}^r,\dots, (v_{\ell_r}^{r})', \dots, (v_{k+i(r-1)+1}^r)'),
\eea
$\phi(N,k)\to 1$ as $N\to \infty,\delta\to 0, N\delta^3=\alpha>0$ and $\mathfrak{D}_r^N$ is defined as in (\ref{eq12TERMbyterm}) and (\ref{eq13TERMbyterm}) with the only difference that each Dirac function $\delta(v-w)$ has to be replaced by 
its approximation 
$\frac{\chi_\delta(v,w)}{\delta^3}$ (according to the definitions (\ref{charDECOMPOSITIONgainNL17CASE3BIS}) and (\ref{loss33}) of the operators $C_{k,k+2}^N$ and $C_{k,k+3}^N$).

To conclude the proof of the term by term convergence it is enough to observe that the integrals with respect to the variables involved in the approximation of the Dirac functions are well behaving. In fact, we notice that the characteristic functions $\overline{\chi}_\delta$'s do not depend on such variables, so that we can exploit the uniform convergence of $f_{k+n}^N(v_1^m,\dots, v_{k+n}^m)$ to $f_{k+n}^\infty(v_1^m,\dots, v_{k+n}^m)$ outside the diagonals and the continuity of the cross-section $B$ and the limiting distribution $f_{k+n}^\infty(v_1^m,\dots, v_{k+n}^m)$ to conclude that:
\bea\label{eq16TERMbyterm}
&&\!\!\!\!\!\!\!
\!\!\!\!\! \!\!\!\!\!\!\!\!\!\!\!\!
I_N(V_k,v_{k+1} \dots
, v_{k+i(m-1)+1},\omega_1,\dots,\omega_m):=\int \prod_{r=1}^m\prod_{q=2}^{j(r)} dv_{k+i(r-1)+q}\, \left(\prod_{r=1}^m B_r^{\omega_r}\right)\times\nonumber\\
&&\qquad\qquad\qquad \qquad\qquad\qquad\qquad\qquad \times\, \left(\prod_{r=1}^m \mathfrak{D}_r^N\right)   f_{k+n}^N(v_1^m,\dots, v_{k+n}^m)
\eea
is pointwise converging to:
\bea\label{eq17TERMbyterm}
&&\!\!\!\!\!\!\!
\!\!\!\!\! \!\!\!\!\!\!\!\!\!\!\!\!
I(V_k,v_{k+1} \dots
, v_{k+i(m-1)+1},\omega_1,\dots,\omega_m):=\int \prod_{r=1}^m\prod_{q=2}^{j(r)} dv_{k+i(r-1)+q}\, \left(\prod_{r=1}^m B_r^{\omega_r}\right)\times\nonumber\\
&&\qquad\qquad\qquad \qquad\qquad\qquad\qquad\qquad \times\,
 \left(\prod_{r=1}^m \mathfrak{D}_r\right)   f_{k+n}^\infty(v_1^m,\dots, v_{k+n}^m).
\eea
Therefore, since the term (\ref{eq15TERMbyterm}) can be rewritten as:
\bea\label{eq18TERMbyterm}
&&\!\!\!\!\!\!\!\!\!\!\!\!\!\!\!\!\!\!\!\!\!\!\left(\prod_{r=1}^m \sigma_r\right) \alpha^{n-m}\, \phi(N,k)\int \!\!dv_{k+1}
\dots\int\!\! dv_{k+i(m-1)+1}\int \!\!d\omega_1\dots \int\!\! d\omega_m\!  \left(\prod_{r=1}^m \overline{\chi}_\delta\left(T
_r
^{\omega_r}
V_{k+i(r-1)+1}^r\right)\right)\times\nonumber\\
&&\qquad\qquad\qquad\qquad\times\,   I_N(V_k,v_{k+1} \dots
, v_{k+i(m-1)+1},\omega_1,\dots,\omega_m)=: g_k^N(V_k)
\eea
and the term (\ref{eq10TERMbyterm}), that we denote by $g_k(V_k)$, is equivalent to:
\bea\label{eq19TERMbyterm}
&&\!\!\!\!\!\!\!\!\!\!\!\!\!\!\!\!\!\!\!\!\!\!\!\!\left(\prod_{r=1}^m \sigma_r\right) \alpha^{n-m}\!\!\int \!\!dv_{k+1}
\dots\!\int\!\! dv_{k+i(m-1)+1}\int \!\!d\omega_1\!\dots \!\int\!\! d\omega_m 
 \, I(V_k,v_{k+1} \dots
, v_{k+i(m-1)+1},\omega_1,\dots,\omega_m),
\nonumber\\
&&
\eea
the pointwise convergence of $g_k^N(V_k)$ to $g_k(V_k)$
is just an application of the Lebesgue Dominated Convergence Theorem. In fact, the characteristic functions $\overline{\chi}_\delta$'s are clearly converging to $1$ and, concerning the quantities $I_N$ and $I$, we know that $I_N\to I$ and also that both $I_N$ and $I$  are bounded (uniformly in $N$) by an integrable function. This is a consequence of the assumptions we made on $B$ and the initial data (see Theorem \ref{main}) and also of the fact that, to prove Theorem \ref{main}, we are interested in convergence on compact subsets of $\R^{3k}$. In fact, if $V_k$ belongs to some compact set $\Lambda_k\subset \R^{3k}$, due to the compact support property we required for $B$ (see Assumption $i)$ of Theorem \ref{main}), we are ensured that the velocity 
of any other particle interacting with one the ''first'' $k$ particles will still belong to a compact set. Due to the energy conservation, we also know that $v'_\ell$ and $v'_m$ surely belong to some compact set if (the pre-collisional velocities) $v_\ell$ and $v_m$ are known to belong to some compact set. Then, by such considerations we are guaranteed that $I_N$ and $I$ depend on velocities 
that surely belong to some compact set. 
As a consequence, it is enough to show that $I_N$ and $I$ are uniformly bounded. 
This follows immediately by the 
assumptions on $B$ and the initial data. 

\subsection{Concluding the proof of Theorem \ref{main}}
Let us now evaluate the difference:
$$
d_k^N(t)=f^N_k(t)-f^{\infty}_k(t).
$$
By virtue of expressions (\ref{eq11:REM5.3PILING}),  (\ref{eq13:REM5.3PILING}) and estimates (\ref{eq12:REM5.3PILING}), (\ref{eq14:REM5.3PILING}),
for all compact sets $\Lambda_k \subset \R^{3k}$ we have:
\begin{equation}\label{conclusion}
\limsup_{N \to \infty} 
\sup_{t \in [0,T] } \| d_k^N(t) \|_{L^1 (\Lambda_k)} \leq C_{\Lambda_k}\  \lambda^{M_1}.
\end{equation}
In fact, by (\ref{eq11:REM5.3PILING}) we know that the main contribution to $f^N_k(t)$ is a finite sum of terms like (\ref{eq1TERMbyterm}) for which we proved the pointwise convergence (see the above paragraph 7.2) to the corresponding term (\ref{eq8TERMbyterm}). Indeed, by  (\ref{eq13:REM5.3PILING}) we know that the main contribution to  
 $f^\infty_k(t)$ is a finite sum of terms like (\ref{eq8TERMbyterm}). As a consequence, by the analysis performed in the previous paragraph, we know that the finite sum in (\ref{eq11:REM5.3PILING}) is pointwise converging to the finite sum in (\ref{eq13:REM5.3PILING}).\\
The convergence in  $L^1 (\Lambda_k)$ follows 
by the Dominated Convergence Theorem.
\\
Finally, inequality (\ref{conclusion}) follows by estimates (\ref{eq12:REM5.3PILING}), (\ref{eq14:REM5.3PILING}) and the fact that:
$$
\| \mathscr{E}_k^N(t) \|_{L^1 (\Lambda_k)}\leq C_{\Lambda_k}\  \| \mathscr{E}_k^N(t) \|_{\delta, k}, \qquad
\| \mathscr{E}_k(t) \|_{L^1 (\Lambda_k)}\leq C_{\Lambda_k}\  \| \mathscr{E}_k(t) \|_{\infty }.
$$
Since $M_1$  is arbitrary, the proof of Theorem \ref{main} is concluded.

\section{Initial data}
\label{sec:sec8}
\setcounter{equation}{0}    
\def\theequation{8.\arabic{equation}}

In this section we present and discuss examples of initial data fulfilling hypotheses $\mathbf{1.}-\mathbf{5.}$ of Theorem \ref{main}.

We start by considering a probability distribution $f_{in}=f_{in}(v)$ such that $f_{in}\in C^0(\R^3)$ and:
\begin{eqnarray}\label{indatHYPbis}
 f_{in}(v)\leq G,
 \end{eqnarray}
for some 
$0<G< \frac{1}{\alpha}$. Obviously $f_{in}^{\otimes N}(V_N)$ is not supported on admissible configurations, thus, in order to construct an $N$-particle distribution which tries to conciliate the exclusion constraint with the statistical independence, we introduce the following probability measure:
\begin{eqnarray}\label{inDAT2}
W_0^N(V_N)= 
\frac{
\overline{\chi}_\delta(V_N) f_{in}(v_1)\dots f_{in}(v_N)}{Z_N},
\end{eqnarray}
where $Z_N$ is the normalization factor:
\begin{eqnarray}\label{inDAT3}
Z_N:= \int_{\R^{3}}dv_1\dots \int_{\R^{3}}dv_N\  
\overline{\chi}_\delta(V_N) f_{in}(v_1)\dots f_{in}(v_N).
\end{eqnarray}

The $k$-particle marginal is:
\begin{eqnarray}\label{inDAT2k0}
&&f_k^N(v_1,...,v_k)=\int dV_{k,N}\ 
\frac{
\overline{\chi}_\delta(V_N)\ f_{in}^{\otimes N}(V_{N})
}{Z_N}=\nonumber\\
&& \qquad\qquad =\overline{\chi}_\delta(V_k)f_{in}^{\otimes k}(V_k)
\int dV_{k,N}\, 
\frac{\overline{\chi}_\delta(V_{k,N})\prod_{\substack{1\leq \ell_1\leq k\\ k+1\leq \ell_2\leq N}}\overline{\chi}_\delta(v_{\ell_1},v_{\ell_2}) }{Z_N}\ f_{in}^{\otimes N-k}(V_{k,N}),
\end{eqnarray}
where we recall that $V_{k,N}=v_{k+1}\dots v_{N}$. 
Therefore
\begin{eqnarray}\label{inDAT2k}
f_k^N(v_1,...,v_k):=
 \overline{\chi}_\delta(V_k)f_{in}^{\otimes k}(V_k)\, F^N(V_k),
\end{eqnarray}
where $F^N(V_k)$ is defined by (\ref{inDAT2k0}).

We now pass to analyze the asymptotic behavior of the family $\{f_k^N\}_k$ as $N\to \infty, \delta\to 0, N\delta^3=\alpha \in (0,1)$. However, we need to look first at the asymptotics 
of the partition function $Z_N$. According to (\ref{inDAT3}), we have:
\begin{eqnarray}\label{inDAT3b}
Z_N&=& \int dV_{N-1}\, \  f_{in}^{\otimes N-1}(V_{N-1})\, \overline{\chi}_\delta(V_{N-1})  \int_{\R^{3}}dv_N\  
\prod_{i=1}^{N-1}\overline{\chi}_\delta(v_i,v_N)\,  f_{in}(v_N)=\nonumber\\
&=&\int dV_{N-1} \, \  f_{in}^{\otimes N-1}(V_{N-1})\, \overline{\chi}_\delta(V_{N-1})  \int_{\R^{3}}dv_N\  
\prod_{i=1}^{N-1}\left(1-\chi_\delta(v_i,v_N) \right) f_{in}(v_N)=\nonumber\\
&=&\int dV_{N-1}\, \ f_{in}^{\otimes N-1}(V_{N-1})\, \overline{\chi}_\delta(V_{N-1})  \int_{\R^{3}}dv_N\  
\left(1-\sum_{i=1}^{N-1}	\, \chi_\delta(v_i,v_N) \right) f_{in}(v_N)=\nonumber\\
&=&Z_{N-1}\left(1 - (N-1) \int_{\Delta_i}dv_N\  
 f_{in}(v_N)\right)\geq Z_{N-1}\left(1 - (N-1)\delta^3 ||f_{in}||_\infty\right)\geq\nonumber\\
 &\geq&
 Z_{N-1}\left(1 - \alpha G\right).
\end{eqnarray}
By (\ref{inDAT3b}) we infer:
\begin{eqnarray}\label{inDAT3cBISk}
Z_N\geq Z_{N-k}\left(1 - \alpha G\right)^k\geq \left(1 - \alpha G\right)^N.
\end{eqnarray}
Clearly $Z_N\leq Z_{N-1}$, so that:
\begin{eqnarray}\label{inDAT3f}
\lg Z_N\leq \lg Z_{N-1}\ \ \ \ \text{and} \ \ \ \ N\lg(1-\alpha G)\leq \lg Z_N\leq 0,
\end{eqnarray}
namely:
\begin{eqnarray}\label{inDAT3g}
\frac{1}{N}\lg Z_N\leq \frac{1}{N-1}\lg Z_{N-1}\ \ \ \ \text{and} \ \ \ \ \lg(1-\alpha G)\leq\frac{1}{N}\lg Z_N\leq 0,
\end{eqnarray}
By (\ref{inDAT3g}) it follows that there exists $a\geq 0$ such that:
\begin{eqnarray}\label{inDAT3h}
\lim_{N\to +\infty}\frac{1}{N}\lg Z_N=-a,\ \ \ \ \ \text{with}\ \ \  0\leq  a\leq \lg\left(\frac{1}{1-\alpha G}\right).
\end{eqnarray}
Let us define:
\begin{eqnarray}\label{varphi}
\varphi(N):=\frac{1}{N}\lg Z_N + a.
\end{eqnarray}
We note that, by an explicit calculation, for any $k>0$,
\begin{eqnarray}\label{inDAT3l}
\sum_{M\geq N}^{+\infty} [\varphi(M-k)-\varphi(M)]\to 0,\ \ \text{as}\ \ N\to +\infty.
\end{eqnarray}
As a consequence:
\begin{eqnarray}\label{inDAT3m}
N [\varphi(N-k)-\varphi(N)]\to 0,\ \ \ \text{as}  \ \ N\to +\infty.
\end{eqnarray}
Therefore:
\begin{eqnarray}\label{inDAT3n}
\frac{Z_{N-k}}{Z_N}=e^{ka} e^{-k\varphi(N-k)} e^{N[\varphi(N-k)-\varphi(N)]}\to e^{ka},\ \ \ \ \text{as}  \ \ N\to +\infty.
\end{eqnarray}

Let us come back to the marginals defined in (\ref{inDAT2k}). By (\ref{inDAT2k0}) and the first inequality in (\ref{inDAT3cBISk}) it follows that:
\begin{eqnarray}\label{boundF}
F^N(V_k)\leq \frac{Z_{N-k}}{Z_k}\leq \left(\frac{1}{1-\alpha G}\right)^k
\end{eqnarray}
Note that, definition (\ref{inDAT2k}), assumption (\ref{indatHYPbis}) on $f_{in}$ and estimate (\ref{boundF}) imply that there exists $z_1>0$ such that
assumption $\mathbf{3.}$ of Theorem \ref{main} is satisfied.\\
Moreover, by the first equality in (\ref{inDAT2k0}) we get:
\begin{eqnarray}\label{margNEW}
f_k^N(v_1,...,v_k)=\frac{1}{Z_N}\int dV_{k,N}\ f_{in}(v_1)\, \prod_{i=2}^k \overline{\chi}_\delta(v_1, v_i)\prod_{i=k+1}^N\overline{\chi}_\delta(v_1, v_i)
f_{in}^{\otimes (N-1)}(V_{1,N})
\overline{\chi}_\delta(V_{1,N}).
\end{eqnarray}
Due to the symmetry under permutation of the indeces $k+1,\dots, N$, the contribution of the product over $i\in \{k+1,\dots, N\}$
can be written as follows:
\begin{eqnarray}\label{margNEW1}
\prod_{i=k+1}^N\overline{\chi}_\delta(v_1, v_i)=\prod_{i=k+1}^N\left(1-\chi_\delta(v_1, v_i)\right)
=\sum_{r=0}^{N-k} \frac{(-1)^r}{r!}\, \alpha(N,k,\delta,r)\, \frac{\chi_\delta(v_1, V_{k,k+r})}{\delta^{3r}},
\end{eqnarray}
where
\begin{eqnarray}\label{margNEW2}
\chi_\delta(v_1, V_{k,k+r}):=\prod_{i=k+1}^{k+r}\chi_\delta(v_1, v_i)
\end{eqnarray}
and $\alpha(N,k,\delta,r):=(N-k)(N-k-1)\dots(N-k-r+1)\delta^{3r}$.\\
We observe that, due to the presence of the characteristic function $\overline{\chi}_\delta(V_{1,N})$ in (\ref{margNEW}), all terms 
corresponding to $r\geq 2$ give indeed no contribution. 
Therefore:
\begin{eqnarray}\label{margNEW5}
\overline{\chi}_\delta(V_{1,N})\prod_{i=k+1}^N\overline{\chi}_\delta(v_1, v_i)&=&\overline{\chi}_\delta(V_{1,N})\left(1- \alpha(N,k,\delta,1)\, \frac{\chi_\delta(v_1, v_{k+1})}{\delta^{3}}\right)=\nonumber\\
&=&\overline{\chi}_\delta(V_{1,N})\left(1- (N-k)\delta^3\, \frac{\chi_\delta(v_1, v_{k+1})}{\delta^3}\right).
\end{eqnarray}
Then, by (\ref{margNEW}) we get:
\begin{eqnarray}\label{margNEW8}
&&f_k^N(v_1,...,v_k,0)=\frac{ Z_{N-1}}{Z_N}f_{in}(v_1)\,  \overline{\chi}_\delta(v_1, V_{1,k})\int dV_{k,N}\, 
\frac{f_{in}^{\otimes (N-1)}(V_{1,N})
\overline{\chi}_\delta(V_{1,N})}{Z_{N-1}}+\nonumber\\
&&\ \ \ -  (N-k)\delta^3\frac{Z_{N-1}}{Z_N}f_{in}(v_1)\,  \overline{\chi}_\delta(v_1, V_{1,k})\int dv_{k+1}\, \frac{\chi_\delta(v_1, v_{k+1})}{\delta^3}
\int dV_{k+1,N}\, \frac{f_{in}^{\otimes (N-1)}(V_{1,N})
\overline{\chi}_\delta(V_{1,N})}{Z_{N-1}},\nonumber\\
&&
\end{eqnarray}
where:
\begin{eqnarray}\label{margNEW7}
\overline{\chi}_\delta(v_1, V_{1,k}):=\prod_{i=2}^{k}\overline{\chi}_\delta(v_1, v_i).\nonumber
\end{eqnarray}
We rewrite (\ref{margNEW8}) as an identity for the marginals, i.e.
\begin{eqnarray}\label{margNEW9}
&&f_k^N(v_1,...,v_k,0)=\frac{ Z_{N-1}}{Z_N}f_{in}(v_1)\,  \overline{\chi}_\delta(v_1, V_{1,k})\, f_{k-1}^{N-1}(V_{1,k})+\nonumber\\
&&\ \ \ -  (N-k)\delta^3\frac{Z_{N-1}}{Z_N}f_{in}(v_1)\,  \overline{\chi}_\delta(v_1, V_{1,k})\int dv_{k+1}\, \frac{\chi_\delta(v_1, v_{k+1})}{\delta^3}\, f_k^{N-1}(V_{1,k},v_{k+1}).
\end{eqnarray}
Using the convention $f_0^N\equiv 1$ (for all $N$), for $k=1$ we obtain:
\begin{eqnarray}\label{margNEW10}
f_1^N(v_1,0)=\frac{ Z_{N-1}}{Z_N}f_{in}(v_1)-  (N-1)\delta^3\frac{Z_{N-1}}{Z_N}f_{in}(v_1)\, \int dv_{2}\, \frac{\chi_\delta(v_1, v_{2})}{\delta^3}\, f_1^{N-1}(v_2).
\end{eqnarray}

Now let us define the family $\underline{h}^N$ of sequences of uniformly bounded functions $\{h_k^N\}_{k=0}^\infty=:\underline{h}^N$, such that:
\begin{eqnarray}\label{margNEW17}
&&h_0^N\equiv 1,\ \ \ \ \ \nonumber\\
&&h_1^N=\frac{ Z_{N-1}}{Z_N}f_{in}(v_{1}),\nonumber\\
&& h_k^N\equiv 0,\ \ \ \text{for}\ \ k\geq 2
\end{eqnarray}
Moreover we define the family $\underline{f}^N$ of sequences $\{f_k^N\}_{k=0}^\infty=:\underline{f}^N$, where $f_0^N=1$ and $f_k^N\equiv 0$  for $k\geq N+1$.
Using these definitions, by (\ref{margNEW9}) we get the following identity:
\begin{eqnarray}\label{margNEW18}
\underline{f}^N=\underline{h}^N + \mathcal{K}_N \underline{f}^{N-1}
\end{eqnarray}
where the operator $\mathcal{K}_N$ is defined as follows:
\begin{eqnarray}\label{margNEW19}
&&\left(\mathcal{K}_N \underline{f}^{N-1}\right)_0\equiv 0,\nonumber\\
&&\left(\mathcal{K}_N \underline{f}^{N-1}\right)_1:=-  (N-1)\delta^3\frac{Z_{N-1}}{Z_N}f_{in}(v_1)\, \int dv_{2}\, \frac{\chi_\delta(v_1, v_{2})}{\delta^3}\, f_1^{N-1}(v_2)\nonumber\\
&&\left(\mathcal{K}_N \underline{f}^{N-1}\right)_k:=\frac{ Z_{N-1}}{Z_N}f_{in}(v_1)\,  \overline{\chi}_\delta(v_1, V_{1,k})\, f_{k-1}^{N-1}(V_{1,k})+\nonumber\\
&&\ \ 	\ \ \ -  (N-k)\delta^3\frac{Z_{N-1}}{Z_N}f_{in}(v_1)\,  \overline{\chi}_\delta(v_1, V_{1,k})\int dv_{k+1}\, \frac{\chi_\delta(v_1, v_{k+1})}{\delta^3}\, f_k^{N-1}(V_{1,k},v_{k+1})
,\ \ \ \text{for}\ \ 2\leq k\leq N
\nonumber\\
&&\left(\mathcal{K}_N \underline{f}^N\right)_k\equiv 0,\ \ \ \text{for}\ \ k\geq N+1.
\end{eqnarray}
Now, we can iterate in formula (\ref{margNEW18}) obtaining:
\begin{eqnarray}\label{margNEW24}
\underline{f}^N=\sum_{n=0}^{N-1}\mathcal{K}_N\mathcal{K}_{ N-1}\dots \mathcal{K}_{ N-n+1}\, \underline{h}^{ N-n}.
\end{eqnarray}
To control the sum (\ref{margNEW24}) uniformly in $N$, we introduce the space:
$$
\mathcal{H}_\xi:=\{\underline{g}:=\{g_k\}_{k=0}^\infty\ \ \text{such\ that}\ \ g_0\equiv 1,\ \ \text{and}\ \ ||\underline{g}||_{\xi}<+\infty\}
$$
where
\begin{eqnarray}\label{margNEW25}
||\underline{g}||_\xi=:\sup_k \xi^{-k}||g_k||_\infty,
\end{eqnarray}
for some $\xi>1$. Then we estimate the operator norm $\left\|\mathcal{K}_N\right\|$ of $\mathcal{K}_N:\mathcal{H}_\xi\to \mathcal{H}_\xi$. We have:
\begin{eqnarray}\label{margNEW27}
&&\xi^{-1}\left|\left(\mathcal{K}_N \underline{g}\right)_1(v_1)\right|\leq \xi^{-1} (N-1)\delta^3\frac{Z_{N-1}}{Z_N}\left|f_{in}(v_1)\right|\, \frac{1}{\delta^3}\int_{\Delta_1} dv_{2}\, \, \left|g_1(v_2)\right|\leq\nonumber\\
&&\ \ \ \leq \xi^{-1} \alpha\, \frac{Z_{N-1}}{Z_N}\,  G\,  \left\|g_1\right\|_\infty\leq \alpha\, \frac{Z_{N-1}}{Z_N}\,  G\, \left\|\underline{g}\right\|_\xi\leq \frac{\alpha G}{1-\alpha G}\,  \left\|\underline{g}\right\|_\xi,
\end{eqnarray}
where in the last inequality we used (\ref{inDAT3b}).\\
Moreover, for $2\leq k\leq N$ we have:
\begin{eqnarray}\label{margNEW30}
&&\xi^{-k}\left|\left(\mathcal{K}_N \underline{g}\right)_k(V_k)\right|\leq \xi^{-k} \frac{G}{1-\alpha G}\, \xi^{-(k-1)}||g_{k-1}||_\infty\, \xi^{k-1}+\nonumber\\
&&\ \ 	\ \ \ + \xi^{-k}\frac{\alpha G}{1-\alpha G} \, \frac{1}{\delta^3}\, \int_{\Delta_1} dv_{k+1}\, \left|g_k(V_{1,k},v_{k+1})\right|
\leq  \frac{G}{1-\alpha G}\, \left(\frac{1}{\xi}+ \alpha \right)\, ||\underline{g}||_\xi.
\end{eqnarray}
As a consequence, $\left\|\mathcal{K}_N\right\|<1$ if the condition:
\begin{eqnarray}\label{margNEW34}
\frac{G}{1-\alpha G}\, \left(\frac{1+ \alpha\xi}{\xi} \right)<1
\end{eqnarray}
is fulfilled. For example, (\ref{margNEW34}) holds for:
\begin{equation}\label{alphaXI}
G<\frac{1}{3\alpha}\ \ \ \text{and}\ \ \ \frac{2G}{1-\alpha G}<\xi<\frac{1}{\alpha}.
\end{equation}
Under such conditions we introduce the formal limit of (\ref{margNEW24}), namely:
\begin{eqnarray}\label{margNEW42}
\underline{f}^\infty=\sum_{n=0}^{\infty}\mathcal{K}_\infty^n\, \underline{h}^{ \infty},
\end{eqnarray}
where the sequence $\underline{h}^\infty$ of uniformly bounded functions $\{h_k^\infty\}_{k=0}^\infty=:\underline{h}^\infty$ is defined as:
\begin{eqnarray}\label{margNEW40}
&&h_0^\infty\equiv 1,\ \ \ \ \ \nonumber\\
&&h_1^\infty=e^a\, f_{in}(v_{1}),\nonumber\\
&& h_k^\infty\equiv 0,\ \ \ \text{for}\ \ k\geq 2
\end{eqnarray}
while the operator $\mathcal{K}_\infty$ is defined by:
\begin{eqnarray}\label{margNEW41}
&&\left(\mathcal{K}_\infty \underline{f}^{\infty}\right)_0\equiv 0,\nonumber\\
&&\left(\mathcal{K}_\infty \underline{f}^{\infty}\right)_1:=-  \alpha\, e^a\, f_{in}(v_1)\,  f_1^{\infty}(v_1)\nonumber\\
&&\left(\mathcal{K}_\infty \underline{f}^{\infty}\right)_k:=e^a\, f_{in}(v_1)\, f_{k-1}^{\infty}(V_{1,k})-  \alpha\, e^a\, f_{in}(v_1)\,  f_k^{\infty}(V_{1,k},v_{1})
,\ \ \ \text{for}\ \ k\geq 2.
\end{eqnarray}
Proceeding as above we readily deduce that $\left\|\mathcal{K}_\infty\right\|<1$ and the series (\ref{margNEW42}), which is absolutely convergent, 
defines the sequence $\underline{f}^\infty$. Moreover, since:
 \begin{eqnarray}\label{margNEW43}
||\underline{f}^\infty||_\xi<+\infty,
\end{eqnarray}
and $\xi<\frac{1}{\alpha}$,
we are ensured that assumption $\mathbf{4.}$ of our Theorem \ref{main} is satisfied.

It remains to show that assumptions $\mathbf{2.}$ and $\mathbf{5.}$ are verified. Concerning the first one, by the uniform boundedness of series (\ref{margNEW24}) and (\ref{margNEW42}), it is enough to show that term by term convergence holds, namely:
\begin{eqnarray}\label{margNEW56}
\left|\left(\mathcal{K}_N\mathcal{K}_{N-1}\dots \mathcal{K}_{N-n+1}\, \underline{h}^{N-n}\right)_k(V_k) - \left(\mathcal{K}_\infty^n\, \underline{h}^\infty\right)_k(V_k)\right|\to 0,
\end{eqnarray}
uniformly outside the diagonals. The proof of term by term convergence proceeds by direct inspection. For the sake of brevity we omit the calculations and we just notice that the only tools that are needed are property (\ref{inDAT3n}) for the quotient of partition functions and the Mean Value Theorem.

In principle we can compute the limiting sequence $\{f_k^\infty\}_k$. For instance, for $k=1$ we have:
\begin{eqnarray}\label{margNEW77ante}
 f_1^\infty(v)=e^a\, f_{in}(v)\sum_{n\geq 0} (-1)^n\, \alpha^n\, \left[e^a\, f_{in}(v)\right]^{n}=\frac{e^a\, f_{in}(v)}{1+\alpha\, e^a\, f_{in}(v)}=\frac{ f_{in}(v)}{e^{-a}+\alpha\,f_{in}(v)},\nonumber
\end{eqnarray}
where we remind that $0\leq a\leq \lg \frac{1}{1-\alpha\, G}$.
The coefficient $a$ can be explicitly determined by the normalization conditions $\int dv\,  f_{in}(v)=1$ and
$$
\int dv\, \left(\frac{ f_{in}(v)}{e^{-a}+\alpha\,f_{in}(v)}\right)=1.
$$

For $k\geq 2$ we have a similar but more complicated expression. 
It is easy to check that assumption $\mathbf{5.}$ of Theorem \ref{main} is satisfied because the distributions $f_k^\infty$'s turn out to be linear combinations of products of $f_{in}$ (see (\ref{margNEW42}) and (\ref{margNEW41})). Since we required this to be continuous, such a property is authomatically inherited by the limiting distributions. 
\\

The sequence $\{f_k^N\}_k$ we have introduced and analyzed so far do not factorize as $N\to \infty,\delta\to 0, N\delta^3=\alpha>0$. The reason for such a permanence of correlations in the limit is easy to explain: considering an $N$-particle system where the velocities $v_1,\dots, v_N$ are distributed according to the measure $W_0^N$ defined in (\ref{inDAT2}), one finds that
a given particle, say particle $1$, has a forbidden volume in the phase space whose
measure is $(N-1)\delta^3 \approx  \alpha>0$. As a consequence, it cannot be distributed independently of the other particles implying that
there are intrinsic correlations due to the exclusion constraint. Since the ''size'' of such correlations is proportional to $N\delta^3=\alpha$ and $\alpha$ is not vanishing as $N\to \infty$, such correlations are present even in the limit,
preventing the possibility to have statistical independence, i.e., factorization. This is in contrast with the case of the low density regime (see e.g \cite{GSRT} and \cite{PSS}).

However, we can construct a family of initial states which converge to a product state as $N\to\infty$. For instance, choose a continuous one particle probability distribution $f_{in}$ satisfying the bound $||f_{in}||_\infty\leq \frac{1}{\alpha}$. Then, consider an $N$-particle system such that the velocities $v_1,\dots, v_N$ are distributed according to 
the symmetric probability measure:
\begin{eqnarray}\label{initialFACT}
W_N(V_N)=\frac{1}{N!}\sum_{\pi \in \mathscr{P}_N}\frac{1}{\delta^{3N}}\prod_{i=1}^N \chi_{\overline{\Delta}_{\pi(i)}}(v_i),
\end{eqnarray}
where $ \mathscr{P}_N$ is the group of permutations on $\{1,\dots, N\}$ and $\overline{\Delta}_{1}\dots\overline{\Delta}_{N}$ is a fixed sequence of different cells of volume $\delta^3$ (to be specified later on, according to $f_{in}$). Again, $W_N$ is supported on $\mathcal{A}^N_\delta$. The corresponding marginals can be easily computed and, for any $k=1,\dots, N$, we get:
\begin{eqnarray}\label{initialFACTmarg}
f_k^N(V_k)=\frac{1}{N(N-1)\dots (N-k+1)}\,  \sum_{\substack{i_1\dots i_k:\\i_r\neq i_s,\, r\neq s}}\frac{\chi_{\overline{\Delta}_{i_1}}(v_1)\dots \chi_{\overline{\Delta}_{i_k}}(v_k)}{\delta^{3k}}.
\end{eqnarray}
In particular, 
\begin{eqnarray}\label{initialFACTmargk=1}
f_1^N(v)=\frac{1}{N}\,  \sum_{\substack{i=1}}^N\frac{\chi_{\overline{\Delta}_{i}}(v)}{\delta^3}
\end{eqnarray}
is ''almost'' an empirical measure. Therefore, the sequence of cells $\overline{\Delta}_{1}\dots\overline{\Delta}_{N}$ is chosen in such a way that, for any $\varphi \in C_b^0(\R^3)$,
\begin{eqnarray}\label{llnAPPROX}
\int dv\, f_1^N(v)\, \varphi(v)\to \int dv\, f_{in}(v)\,  \varphi(v),\ \ \ \ \ \text{as}\ \ \ N\to \infty,\delta\to 0, N\delta^3=\alpha>0.
\end{eqnarray}
By (\ref{llnAPPROX}) it follows that for $k\geq 2$
we have:
\begin{eqnarray}\label{llnAPPROXkLARGER2}
\int dV_k\, f_k^N(V_k)\, \varphi_k(V_k)\to \int dV_k\, f_{in}^{\otimes  k}(V_k)\,  \varphi_k(V_k),\ \ \ \ \ \text{for any $\varphi_k\in C_b^0(\R^{3k})$},
\end{eqnarray}
namely, $f_k^N$ is weakly converging to $f_{in}^{\otimes  k}$. Unfortunately, since convergence is only weak, such an example eludes the hypotheses of Theorem \ref{main} (for which we need uniform convergence outside the diagonals).

Nevertheless, we can adapt the previous example to construct another probability distribution exhibiting factorization in the limit and fulfilling the hypotheses of Theorem \ref{main}.\\
We first introduce a sequence $\alpha_N\to \alpha$ such that:
\begin{eqnarray}\label{alphaN}
\delta=\left(\frac{\alpha_N}{N}\right)^{\frac{1}{3}}
\end{eqnarray}
and $(\sqrt\delta)^{-1}$ is an integer. Therefore, we consider a partition of $\R^3$ in cells $\widetilde{\Delta}$ of side $\widetilde{\delta}=\sqrt \delta$. By construction, each cell $\widetilde{\Delta}$ can be partitioned in cells $\Delta$ of the finer grid (i.e. $|\Delta|=\delta^3$). \\
Now, given $f_{in}$ as before, we fix an integer $M<N$, a sequence $\widetilde{\Delta}_1,\dots, \widetilde{\Delta}_M$ of cells in the larger grid and a sequence $\{N(\widetilde{\Delta}_j)\}_{j=1}^M$ of numbers such that $N(\widetilde{\Delta}_j)\geq 1$ is the number of particles whose velocity belong to $\widetilde{\Delta}_j$ (clearly, $\sum_{i=1}^M N(\widetilde{\Delta}_j)=N$). Moreover, we require that:
\begin{eqnarray}\label{linkconFIN}
\frac{N(\widetilde{\Delta}_j)}{N\delta^{3/2}} - \frac{1}{\delta^{3/2}}\int_{\widetilde{\Delta}_j} dv\, f_{in}(v)\to 0,\ \ \ \text{as}\ \ N\to 0, \delta\to 0, N\delta^3=\alpha.
\end{eqnarray}
Next, we partition the set $I_N=\{1,\dots, N\}$ into subsets $I_1,\dots, I_M$ such that $|I_j|=N(\widetilde{\Delta}_j)$, where $j=1,\dots, M$.  We introduce the symmetric probability measure:
\begin{eqnarray}\label{svolta}
\widetilde{W}_N(V_N)=\sum_{\substack{I_1\dots I_M:\\ | I_j|=N(\widetilde{\Delta}_j) }}\frac{N(\widetilde{\Delta}_1)!\dots N(\widetilde{\Delta}_M)!}{N!}\prod_{j=1}^M W_{N(\widetilde{\Delta}_j)}(V_{I_j})
\end{eqnarray}
where, for any $j=1,\dots, M$,  $V_{I_j}=\{v_s\}_{s\in I_j}$ and
\begin{eqnarray}\label{adaptation}
W_{N(\widetilde{\Delta}_j)}(V_{I_j})= 
\frac{\overline{\chi}_\delta(V_{I_j})\, \left(\frac{1}{\delta^{3/2}}\right)^{N(\widetilde{\Delta}_j)}\prod_{s\in I_j} \chi_{\widetilde{\Delta}_j}(v_s)}
{Z_{N(\widetilde{\Delta}_j)}},
\end{eqnarray}
being $Z_{N(\widetilde{\Delta}_j)}$ the normalization factor:
\begin{eqnarray}\label{normADAPT}
Z_{N(\widetilde{\Delta}_j)}:= \int dV_{I_j}\, \overline{\chi}_\delta(V_{I_j})\, \left(\frac{1}{\delta^{3/2}}\right)^{N(\widetilde{\Delta}_j)}\prod_{s\in I_j} \chi_{\widetilde{\Delta}_j}(v_s).
\end{eqnarray}
Notice that $W_{N(\widetilde{\Delta}_j)}$ is just the ''restriction'' of (\ref{inDAT2}) to $\widetilde{\Delta}_j$ with $f_{in}(v)$ replaced by $\frac{\chi_{\widetilde{\Delta}_j}(v)}{\delta^{3/2}}$.\\
In other words, once the sequence $\{N(\widetilde{\Delta}_j)\}_{j=1}^M$ is fixed, in any (big) cell $\widetilde{\Delta}_j$ the particles are uniformly distributed and have to satisfy the exclusion constraint with respect to the $ \delta^{-\frac{3}{2}}$ smaller cells $\Delta$ in $\widetilde{\Delta}_j$. On the other hand, particles whose velocities belong to different (big) cells $\widetilde{\Delta}_j$ are independently distributed.

Let us compute the one particle marginal of (\ref{svolta}). If $\widetilde{\Delta}_i$ is such that $v_1\in \widetilde{\Delta}_i$, then:
\begin{eqnarray}\label{svoltaMARG}
\widetilde{f}_1^N(v_1)=
\frac{N(\widetilde{\Delta}_i)}{N}\, \frac{\chi_{\widetilde{\Delta}_i}(v_1)}{\delta^{3/2}} \ F_i^N(v_1),
\end{eqnarray}
where
\begin{eqnarray}\label{svoltaMARGI}
&&F_i^N(v_1)=\sum_{\substack{I_1\dots I_M:\\ | I_i|=N(\widetilde{\Delta}_i)-1\\| I_j|=N(\widetilde{\Delta}_j),\ j\neq i }}\frac{N(\widetilde{\Delta}_1)!\dots( N(\widetilde{\Delta}_i)-1)!\dots N(\widetilde{\Delta}_M)!}{(N-1)!}\nonumber\\
&&\int dV_{2,N}\, W_{N(\widetilde{\Delta}_1)}(V_{I_1})\dots\frac{\overline{\chi}_\delta(V_{I_i})\, \left(\frac{1}{\delta^{3/2}}\right)^{N(\widetilde{\Delta}_i)-1}\prod_{\substack{s\in I_i\\ s\neq 1}} \chi_{\widetilde{\Delta}_i}(v_s)}{Z_{N(\widetilde{\Delta}_i)}}\dots W_{N(\widetilde{\Delta}_M)}(V_{I_M})=\nonumber\\
&&= \int dV_{I_i -1}\, \frac{\overline{\chi}_\delta(V_{I_i})\, \left(\frac{1}{\delta^{3/2}}\right)^{N(\widetilde{\Delta}_i)-1}\prod_{\substack{s\in I_i\\ s\neq 1}} \chi_{\widetilde{\Delta}_i}(v_s)}{Z_{N(\widetilde{\Delta}_i)}}=C_i.
\end{eqnarray}
Hence, $F_i^N(v_1)$ is just constant on $\widetilde{\Delta}_i$.
Now, by definition of marginals,
$$
\int_{\widetilde{\Delta}_i} \widetilde{f}_1^N(v_1)dv_1=\frac{N(\widetilde{\Delta}_i)}{N},\ \ \ \ \forall\ \, i=1,\dots, M
$$
that is the probability that the velocity $v_1$ of particle $1$ belongs to $\widetilde{\Delta}_i$. Thus, by (\ref{svoltaMARG}) and (\ref{svoltaMARGI}) we conclude that $C_i=1$ and
\begin{eqnarray}\label{svoltaMARGIII}
\widetilde{f}_1^N(v_1)=\sum_{i=1}^M \frac{N(\widetilde{\Delta}_i)}{N}\, \frac{\chi_{\widetilde{\Delta}_i}(v_1)}{\delta^{3/2}},
\end{eqnarray}
that, by condition (\ref{linkconFIN}),  implies:
$$
 ||\widetilde{f}_1^N\to f_{in}||_\infty\to 0,\ \ \ \ \text{as $N\to \infty, \delta\to 0, N\delta^3=\alpha>0$}. 
$$
In a similar way we can deal with the $k$-particle marginals, with $k\geq 2$. In fact, 
for $v_1,\dots, v_k$ such that $v_\ell\neq v_m$ if $l\neq m$, we are guaranteed that, if $N$ is sufficiently large, $v_1,\dots, v_k$
 belongs to different cells $\widetilde{\Delta}_{i_1}, \dots, \widetilde{\Delta}_{i_k}$ (in the larger grid). Then, by computations analogous to those we performed previously, we get:
\begin{eqnarray}\label{svoltaMARGV}
\widetilde{f}_k^N(V_k)=\sum_{i_1=1}^M \sum_{\substack{i_2=1\\ i_2\neq i_1}}^M\dots \sum_{\substack{i_k=1:\\i_k\neq i_s,\\ 1\leq s\leq k+1}}^M
\frac{N(\widetilde{\Delta}_{i_1})\dots N(\widetilde{\Delta}_{i_k})}{N(N-1)\dots(N-k+1)}\  \frac{\chi_{\widetilde{\Delta}_{i_1}}(v_1)}{\delta^{3/2}}\dots \frac{\chi_{\widetilde{\Delta}_{i_k}}(v_k)}{\delta^{3/2}},
\end{eqnarray}
so that, by condition (\ref{linkconFIN}), we find:
$$
\sup_{V_k\in K}\left| f_k^N(V_k)\to f_{in}^{\otimes k}(V_k)\right|\to 0,\ \ \ \ \text{as $N\to \infty, \delta\to 0, N\delta^3=\alpha>0$}. 
$$
for any compact set $K\subset \mathcal{A}^k$ (i.e., $f_k^N\to f_{in}^{\otimes k}$ uniformly outside the diagonals $v_\ell=v_m$). 

In conclusion, we have presented an example of initial data that are asymptotically factorized and, moreover, they fulfill the hypotheses of Theorem \ref{main}.

\label{appendixBBGKY}
\section*{Appendix}
\setcounter{equation}{0}    
\def\theequation{A.\arabic{equation}}

This section is devoted to the derivation of the BBGKY hierarchy of equations solved by the marginals $f_k^N(t)$ of the $N$-particle distribution $W^N(t)$.
We have:
\be
\partial_t f_k^N=\frac{1}{N}\sum_{1\leq i<j\leq N}\int dV_{k,N} \int d\omega B(v_i-v_j; \omega)[\underbrace{\overline{\chi}_\delta(V_N) W^N(V_N^{i.j})}_{\rm G}-\underbrace{\overline{\chi}_\delta(V_N^{i.j}) W^N(V_N)}_{\rm L}] \label{LNridAPP}
\ee
where we recall that $V_{k,N}=v_{k+1}...\, v_{N} $.

Let us focus on the $G$ term. Then, the computations for the $L$ term will work in the same way. \\
We have:
\begin{eqnarray}
G:=\frac{1}{N}\sum_{1\leq i<j\leq N}\int dV_{k,N} \int d\omega\,  B_{i.j}^{\omega}\ \overline{\chi}_\delta(V_N) W^N(V_N^{i.j}) \label{gain}
\end{eqnarray}
where $B_{i,j}^{\omega}=B(v_i-v_j; \omega)$.
Three cases must be considered:

\begin{itemize}

\item $k<i<j$ 

\item $i<j\leq k$ 

\item $i\leq k<j$  

\end{itemize} 
which yield, respectively, the contributions $G_1$, $G_2$ and $G_3$ and
the corresponding contributions $L_1$, $L_2$ and $L_3$ for the \textit{Loss} term in (\ref{LNrid}).  
The first case can be disregarded as 
$G_1$ and $L_1$ compensate exactly. \\
Then, let us consider, separately, the $G_2$ and the $G_3$ terms.

\subsubsection*{A.1 \ \  The term $G_2$}

In this paragraph, we consider the second case $i<j\leq k$.
According to (\ref{CHAR}), the characteristic function $\overline{\chi}_\delta(V_N)$ can be written as follows:
\begin{eqnarray}
\overline{\chi}_\delta(V_N)=\prod_{1\leq \ell_1<\ell_2\leq k}\overline{\chi}_\delta(v_{\ell_1},v_{\ell_2})\ \prod_{k+1\leq\ell_1<\ell_2\leq N}\overline{\chi}_\delta(v_{\ell_1},v_{\ell_2})\ \prod_{\substack{1\leq \ell_1\leq k\\k+1\leq \ell_2\leq N}}\overline{\chi}_\delta(v_{\ell_1},v_{\ell_2}),
\label{charDECOMPOSITION}
\end{eqnarray}
that
gives rise to:
\begin{eqnarray}
\overline{\chi}_\delta(V_N)=\overline{\chi}_\delta(V_k)\ \overline{\chi}_\delta(V_{k,N})\ \prod_{\substack{1\leq \ell_1\leq k\\k+1\leq \ell_2\leq N}}\overline{\chi}_\delta(v_{\ell_1},v_{\ell_2}).
\label{charDECOMPOSITION1}
\end{eqnarray}
By expanding the last product of $\overline{\chi}_\delta$'s we get:
\begin{eqnarray}
&&\overline{\chi}_\delta(V_k)\ \overline{\chi}_\delta(V_{k,N})\ \prod_{1\leq \ell_1\leq k}\ \prod_{k+1\leq \ell_2\leq N}\left(1-\chi_\delta(v_{\ell_1},v_{\ell_2})\right)=\nonumber\\
&&\qquad\qquad =\overline{\chi}_\delta(V_k)\ \overline{\chi}_\delta(V_{k,N})\ \prod_{1\leq \ell_1\leq k}\left(1-\sum_{k+1\leq \ell_2\leq N}\chi_\delta(v_{\ell_1},v_{\ell_2})\right),
\label{charDECOMPOSITION2}
\end{eqnarray}
where the equality follows by the admissibility of the $N-k$-particle configuration $V_{k,N}$. 
Now, by analogous considerations (\ref{charDECOMPOSITION2}) yields:
\begin{eqnarray}
&&\overline{\chi}_\delta(V_k)\ \overline{\chi}_\delta(V_{k,N})\ \prod_{1\leq \ell_1\leq k}\left(1-\sum_{k+1\leq \ell_2\leq N}\chi_\delta(v_{\ell_1},v_{\ell_2})\right)=\nonumber\\
&&=\overline{\chi}_\delta(V_k)\ \overline{\chi}_\delta(V_{k,N})\ \left(1-\sum_{\substack{1\leq \ell_1\leq k\\k+1\leq \ell_2\leq N}}\chi_\delta(v_{\ell_1},v_{\ell_2})+\right.\nonumber\\
&&\ \ \left.+\sum_{2\leq m\leq \min\{k,N-k\}}(-1)^m \sum_{\substack{1\leq \ell_1,\dots, \ell_m\leq k,\\ \ell_p\neq \ell_q}}\ \sum_{\substack{k+1\leq \overline{\ell}_1,\dots, \overline{\ell}_m\leq N,\\ \overline{\ell}_p\neq \overline{\ell}_q}}\chi_\delta(v_{\ell_1},v_{\overline{\ell}_1})\dots\chi_\delta(v_{\ell_m},v_{\overline{\ell}_m})\right).
\label{charDECOMPOSITION3}
\end{eqnarray}
In view of the fact that (\ref{charDECOMPOSITION3}) will be used in situations in which $V_{N}^{i,j}$ is known to be admissible, we readily realize that, in this case, 
\begin{eqnarray}
&&\chi_\delta(v_{\ell_1},v_{\overline{\ell}_1})\dots\chi_\delta(v_{\ell_m},v_{\overline{\ell}_m})=0,\ \ \ \ \text{if}\ \ m>2\nonumber\\
&& \chi_\delta(v_{\ell_1},v_{\overline{\ell}_1})\chi_\delta(v_{\ell_2},v_{\overline{\ell}_2})=0,\ \ \ \ \text{unless}\ \ \ell_1=i,j,\, \ell_2=j,i
\label{charNEWfinale}
\end{eqnarray}
Thus, by (\ref{charDECOMPOSITION3}) and (\ref{gain}) we have:
\begin{eqnarray}
&&\!\!\!\!\!\!\!\!\!\!\!\!\!\!G_2:=\frac{1}{N}\sum_{1\leq i<j\leq k}\int dV_{k,N} \int d\omega \, B_{i.j}^{\omega}\ W^N(V_N^{i.j})\overline{\chi}_\delta(V_k)\ \overline{\chi}_\delta(V_{k,N})\times\nonumber\\
&&\qquad\qquad\times\, \left(1-\sum_{\substack{1\leq \ell_1\leq k\\k+1\leq \ell_2\leq N}}\chi_\delta(v_{\ell_1},v_{\ell_2})+2\sum_{\substack{k+1\leq \overline{\ell}_1,\overline{\ell}_2\leq N
,\\ \overline{\ell}_1\neq \overline{\ell}_2}}
\ \chi_\delta(v_{i},v_{\overline{\ell}_1})\chi_\delta(v_{j},v_{\overline{\ell}_2})\right).
\label{charDECOMPOSITIONgain}
\end{eqnarray}
The term involving the $1$ in the above parentesis gives rise to:
\begin{eqnarray}
&&G_2^{(1)}:=\frac{1}{N}\sum_{1\leq i<j\leq k}\int dv_{k+1}...\int dv_{N} \int d\omega\,  B_{i.j}^{\omega}\ \overline{\chi}_\delta(V_k)\ \overline{\chi}_\delta(V_{k,N})W^N(V_N^{i.j})=\nonumber\\
&&\ \ \ \ \ \ \ =\frac{1}{N}\sum_{1\leq i<j\leq k} \int d\omega\,  B_{i.j}^{\omega}
\ \overline{\chi}_\delta(V_k)\  \int dv_{k+1}...\int dv_{N} W^N(V_N^{i.j}).
\label{charDECOMPOSITIONgain1}
\end{eqnarray}
Therefore, defining $G_2^{(1)}:=\frac{1}{N}\left(L_k^{N,+} f_k^N\right)(V_k)$, we have
\begin{eqnarray}
\left(L_k^{N,+} f_k^N\right)(V_k)=\sum_{1\leq i<j\leq k} \int d\omega\,  B(v_i-v_j; \omega)\ \overline{\chi}_\delta(V_k)
f_k^N(V_k^{i.j})
\label{charDECOMPOSITIONgain16app}
\end{eqnarray}
and it is easy to check that (\ref{charDECOMPOSITIONgain16app}) is exactly the $k$-particle version of the gain part associated with the operator $L_N$ appearing in (\ref{ME}).

Let us consider the term in (\ref{charDECOMPOSITIONgain}) which is linear in the two-particle functions $\chi_\delta$'s. We have:
\begin{eqnarray}
G_2^{(2)}:=-\frac{1}{N}\sum_{1\leq i<j\leq k}\int dV_{k,N} \int d\omega\,  B_{i.j}^{\omega}\ W^N(V_N^{i.j})\ \overline{\chi}_\delta(V_k)\ \overline{\chi}_\delta(V_{k,N})\ \sum_{\substack{1\leq \ell_1\leq k\\k+1\leq \ell_2\leq N}}\chi_\delta(v_{\ell_1},v_{\ell_2})
\label{charDECOMPOSITIONgainLIN}
\end{eqnarray}
By the symmetry of $W^N$ under permutation of particles it follows that each term in the sum over $k+1\leq \ell_2\leq N$ gives the same contribution, thus:
\begin{eqnarray}
&&\!\!\!\!\!\!\!\!\!\!G_2^{(2)}=-\frac{N-k}{N}\sum_{1\leq i<j\leq k}\int dV_{k,N}\int d\omega\,  B_{i.j}^{\omega}\ W^N(V_N^{i.j}) \overline{\chi}_\delta(V_k)\ \sum_{1\leq \ell_1\leq k}\chi_\delta(v_{\ell_1},v_{k+1})=\nonumber\\
&&\!\!\!\!\!\!\!\!\!\!\ \ \ \ \ \ \ =-\frac{N-k}{N}\sum_{1\leq i<j\leq k}\int dV_{k,N} \int d\omega\,  B_{i.j}^{\omega}\ W^N(V_N^{i.j})\overline{\chi}_\delta(V_k)\ \left(\chi_\delta(v_{i},v_{k+1})+\chi_\delta(v_{j},v_{k+1})\right),\nonumber
\label{charDECOMPOSITIONgainLIN2}
\end{eqnarray}
by the same argument leading to (\ref{charNEWfinale}).
Therefore, defining $G_2^{(2)}:=\frac{1}{N}\left(L_{k,k+1}^{N,+} f_{k+1}^N\right)(V_k)$, we have
\begin{eqnarray}
&&\!\!\!\!\!\!\!\!\!\!\!\!\!\!\left(L_{k,k+1}^{N,+} f_{k+1}^N\right)(V_k)=-(N-k)\sum_{1\leq i<j\leq k}\int dv_{k+1}\int d\omega\,  B_{i.j}^{\omega,\delta}\ f_{k+1}^N(V_{k+1}^{i.j})\times\nonumber\\
&&\qquad \qquad\qquad \quad \qquad\times \ \overline{\chi}_\delta(V_k)\ \left(\chi_\delta(v_{i},v_{k+1})+\chi_\delta(v_{j},v_{k+1})\right).
\label{charDECOMPOSITIONgainLIN5}
\end{eqnarray}

Let us now consider the term in (\ref{charDECOMPOSITIONgain}) which is nonlinear in the two-particle functions $\chi_\delta$'s. We have:
\begin{eqnarray}
G_2^{(3)}:=\frac{2}{N}\sum_{1\leq i<j\leq k}\int dV_{k,N} \int d\omega\,  B_{i.j}^{\omega}\ W^N(V_N^{i.j})\overline{\chi}_\delta(V_k)\ \sum_{\substack{k+1\leq \overline{\ell}_1,\overline{\ell}_2\leq N
,\\ \overline{\ell}_1\neq \overline{\ell}_2}}
\ \chi_\delta(v_{i},v_{\overline{\ell}_1})\chi_\delta(v_{j},v_{\overline{\ell}_2}),
\label{charDECOMPOSITIONgainNL2}
\end{eqnarray}
that, using the symmetry of $W^N$ under permutation of particles, yields: 
\begin{eqnarray}
G_2^{(3)}=\frac{2(N-k)(N-k-1)}{N}\sum_{1\leq i<j\leq k}\int dV_{k,N} \int d\omega\,  B_{i.j}^{\omega}\ W^N(V_N^{i.j})\overline{\chi}_\delta(V_k)
\, 
\chi_\delta(v_{i},v_{k+1})\chi_\delta(v_{j},v_{k+2}).\nonumber
\label{charDECOMPOSITIONgainNL3}
\end{eqnarray}
Therefore, defining $G_2^{(3)}:=\frac{1}{N}\left(L_{k,k+2}^{N,+}f_{k+2}^N\right)(V_k)$, we get:
\begin{eqnarray}
&&\left(L_{k,k+2}^{N,+} f_{k+2}^N\right)(V_k)=2(N-k)(N-k-1)\sum_{1\leq i<j\leq k}\int dv_{k+1}\int dv_{k+2} \int d\omega\,  B_{i.j}^{\omega}\ f_{k+2}^N(V_{k+2}^{i.j})\, \times\nonumber\\
&&\qquad\qquad\qquad\qquad\qquad \times \, \overline{\chi}_\delta(V_k)
\ \chi_\delta(v_{i},v_{k+1})\chi_\delta(v_{j},v_{k+2}).
\label{charDECOMPOSITIONgainNL4}
\end{eqnarray}

\subsubsection*{A.2 \ \  The term $G_3$}

Let us now address the case $i\leq k<j$. By (\ref{charDECOMPOSITION1}) and (\ref{gain}) we have:
\begin{eqnarray}
&&G_3:=\frac{N-k}{N}\sum_{1\leq i\leq k}\int dV_{k,N} \int d\omega\,  B_{i.k+1}^{\omega}\ W^N(V_N^{i.k+1})\overline{\chi}_\delta(V_k)\ \overline{\chi}_\delta(V_{k,N})\prod_{\substack{1\leq \ell_1\leq k\\k+1\leq \ell_2\leq N}}\overline{\chi}_\delta(v_{\ell_1},v_{\ell_2})=\nonumber\\
&&\ \ \ \ \ \  =\frac{N-k}{N}\sum_{1\leq i\leq k}\int dv_{k+1}...\int dv_{N} \int d\omega\,  B_{i.k+1}^{\omega}\ W^N(V_N^{i.k+1})\overline{\chi}_\delta(V_k)\ \overline{\chi}_\delta(V_{k,N})\times\nonumber\\
&&\qquad \qquad\qquad\qquad\qquad \times 
\prod_{\substack{1\leq \ell_1\leq k}}
\overline{\chi}_\delta(v_{\ell_1},v_{k+1})\, \prod_{\substack{1\leq \ell_1\leq k\\k+2\leq \ell_2\leq N}}\overline{\chi}_\delta(v_{\ell_1},v_{\ell_2}).
\label{charDECOMPOSITIONgainsostCASE3}
\end{eqnarray}
Now, we expand the product $\prod_{\substack{1\leq \ell_1\leq k\\k+2\leq \ell_2\leq N}}\overline{\chi}_\delta(v_{\ell_1},v_{\ell_2})$ as we did previously (see (\ref{charDECOMPOSITION2}) and (\ref{charDECOMPOSITION3})), i.e.
\begin{eqnarray}
&&\overline{\chi}_\delta(V_k)\ \overline{\chi}_\delta(V_{k,N})\ 
\prod_{\substack{1\leq \ell_1\leq k}}
\overline{\chi}_\delta(v_{\ell_1},v_{k+1})\, \prod_{1\leq \ell_1\leq k}\left(1-\sum_{k+2\leq \ell_2\leq N}\chi_\delta(v_{\ell_1},v_{\ell_2})\right)=\nonumber\\
&&=\overline{\chi}_\delta(V_k)\ \overline{\chi}_\delta(V_{k,N})\ 
\prod_{\substack{1\leq \ell_1\leq k}}
\overline{\chi}_\delta(v_{\ell_1},v_{k+1})\ \left(1-\sum_{\substack{1\leq \ell_1\leq k\\k+2\leq \ell_2\leq N}}\chi_\delta(v_{\ell_1},v_{\ell_2})\right).
\label{charDECOMPOSITION3bis}
\end{eqnarray}
Note that, in contrast with expansion (\ref{charDECOMPOSITIONgain}), the term $\chi_\delta(v_{\ell_1}, v_{\overline{\ell}_1})\chi_\delta(v_{\ell_2}, v_{\overline{\ell}_2})$ with $1\leq \ell_1\neq \ell_2\leq k$ and $k+2\leq \overline{\ell}_1\neq \overline{\ell}_2\leq N$, is now vanishing as $V_{N}^{i,k+1}$ is admissible.
Thus
\begin{eqnarray}
&&\!\!\!\!\!G_3=\frac{N-k}{N}\sum_{1\leq i\leq k}\int dV_{k,N} \int d\omega\,  B_{i.k+1}^{\omega}\ W^N(V_N^{i.k+1})\overline{\chi}_\delta(V_k)\ \overline{\chi}_\delta(V_{k,N})
\times\nonumber\\
&&\qquad\qquad\qquad\qquad\times\, \prod_{\substack{1\leq \ell_1\leq k}}
\overline{\chi}_\delta(v_{\ell_1},v_{k+1})\left(1-\sum_{\substack{1\leq \ell_1\leq k\\k+2\leq \ell_2\leq N}}\chi_\delta(v_{\ell_1},v_{\ell_2})\right).
\label{charDECOMPOSITIONgainsostIIICASE3}
\end{eqnarray}

In order to recover some marginal $f_p^N$ (for a suitable index $p$), we need to understand the effect of the presence of the characteristic function $\overline{\chi}_\delta(V_{k,N})$ (that involves the integrated variables $v_{k+1}\dots v_N$). In fact, here it cannot be ignored - as it was in the situation $1\leq i\neq j\leq k$ - since it involves one of the colliding particles (say, that with velocity $v_{k+1}$). Thus, we expand:
\begin{eqnarray}
&&\overline{\chi}_\delta(V_{k,N})=\overline{\chi}_\delta(V_{k+1,N})\prod_{k+2\leq n\leq N}\overline{\chi}_\delta(v_n, v_{k+1})=\overline{\chi}_\delta(V_{k+1,N})\prod_{k+2\leq n\leq N}\left(1-\chi_\delta(v_n, v_{k+1})\right)=\nonumber\\
&&\ \ \ \ \ \ \ \ \ \ \ \ \, =\overline{\chi}_\delta(V_{k+1,N})\left(1-\sum_{k+2\leq n\leq N}\chi_\delta(v_n, v_{k+1})\right).
\label{charDECOMPOSITIONgainNL4case3}
\end{eqnarray}
So, by (\ref{charDECOMPOSITIONgainsostCASE3}) and (\ref{charDECOMPOSITIONgainNL4case3}) we get:
\begin{eqnarray}
&&G_3=\frac{N-k}{N}\sum_{1\leq i\leq k}\int dv_{k+1}...\int dv_{N} \int d\omega \, B_{i.k+1}^{\omega}\ W^N(V_N^{i.k+1})\overline{\chi}_\delta(V_k)\ \overline{\chi}_\delta(V_{k+1,N}) \times\nonumber\\
&&\times\prod_{\substack{1\leq \ell_1\leq k}}
\overline{\chi}_\delta(v_{\ell_1},v_{k+1})\left(1-\sum_{k+2\leq n\leq N}\chi_\delta(v_n, v_{k+1})\right)\left(1-\sum_{\substack{1\leq \ell_1\leq k\\k+2\leq \ell_2\leq N}}\chi_\delta(v_{\ell_1},v_{\ell_2})\right),
\label{charDECOMPOSITIONgainNL5CASE3}
\end{eqnarray}
that, without loss of generality, can be written as:
\begin{eqnarray}
&&G_3=\frac{N-k}{N}\sum_{1\leq i\leq k}\int dv_{k+1}...\int dv_{N} \int d\omega\,  B_{i.k+1}^{\omega}\ W^N(V_N^{i.k+1})\overline{\chi}_\delta(V_k)\ \overline{\chi}_\delta(V_{k+1,N})\times\nonumber\\
&&\times\prod_{\substack{1\leq \ell_1\leq k}}
\overline{\chi}_\delta(v_{\ell_1},v_{k+1})\left(1-(N-k-1)\chi_\delta(v_{k+2}, v_{k+1})\right)\left(1-\sum_{\substack{1\leq \ell_1\leq k\\ k+2\leq \ell_2\leq N}}\chi_\delta(v_{\ell_1},v_{\ell_2})\right).
\label{charDECOMPOSITIONgainNL6CASE3}
\end{eqnarray}
The first term we have to deal with is:
\begin{eqnarray}
G_3^{(1)}:=\frac{N-k}{N}\sum_{1\leq i\leq k}\int dV_{k,N} \int d\omega \, B_{i.k+1}^{\omega}\ \overline{\chi}_\delta(V_k)\ \overline{\chi}_\delta(V_{k+1,N})\prod_{\substack{1\leq \ell_1\leq k}}
\overline{\chi}_\delta(v_{\ell_1},v_{k+1})W^N(V_N^{i.k+1}).
\nonumber
\label{charDECOMPOSITIONgainNL2case3}
\end{eqnarray}
The admissibility of the $N-k-1$-particle configuration $V_{k+1,N}$ is ensured by the fact that the probability distribution $W^N$ is evaluated on $V_N^{i.k+1}$, so that the characteristic function $\overline{\chi}_\delta(V_{k+1,N})$ can be ignored and we get:
\begin{eqnarray}
\left(C_{k,k+1}^{N,+} f_{k+1}^N\right)(V_k)=\frac{N-k}{N}\sum_{1\leq i\leq k}\int dv_{k+1} \int d\omega\,  B_{i.k+1}^{\omega}\  \overline{\chi}_\delta(V_{k+1})\ 
f_{k+1}^N(V_{k+1}^{i.k+1}),
\label{charDECOMPOSITIONgainNL8case3}
\end{eqnarray}
where $\left(C_{k,k+1}^{N,+} f_{k+1}^N\right)(V_k):=G_3^{(1)}$.

The second term we have to consider is:
\begin{eqnarray}
&&\!\!\!\!\!\!\!G_3^{(2)}:=-\frac{(N-k)}{N}\sum_{1\leq i\leq k}\int dv_{k+1}...\int dv_{N} \int d\omega\,  B_{i.k+1}^{\omega}\ W^N(V_N^{i.k+1})\overline{\chi}_\delta(V_k)\ \times\nonumber\\
&&\qquad\qquad\qquad\times\overline{\chi}_\delta(V_{k+1,N})\prod_{\substack{1\leq \ell_1\leq k}}
\overline{\chi}_\delta(v_{\ell_1},v_{k+1})\sum_{\substack{1\leq \ell_1\leq k\\ k+2\leq \ell_2\leq N}}\chi_\delta(v_{\ell_1},v_{\ell_2}),
\label{charDECOMPOSITIONgainNL9CASE3}
\end{eqnarray}
that, by the admissibility of configuration $V_N^{i.k+1}$ and the symmetry of $W^N$, yields:
\begin{eqnarray}
&&G_3^{(2)}=-\frac{(N-k)(N-k-1)}{N}\sum_{1\leq i\leq k}\int dv_{k+1}...\int dv_{N} \int d\omega\,  B_{i.k+1}^{\omega}\ W^N(V_N^{i.k+1})\times\nonumber\\
&&\qquad\qquad\qquad\qquad \times\overline{\chi}_\delta(V_{k})\, \prod_{\substack{1\leq \ell_1\leq k}}
\overline{\chi}_\delta(v_{\ell_1},v_{k+1})\, \chi_\delta(v_{i},v_{k+2}).
\label{charDECOMPOSITIONgainNL10CASE3}
\end{eqnarray}

According to (\ref{charDECOMPOSITIONgainNL6CASE3}), the next term we take into account is:
\begin{eqnarray}
&&G_3^{(3)}:=-\frac{(N-k)(N-k-1)}{N}\sum_{1\leq i\leq k}\int dV_{k,N} \int d\omega\,  B_{i.k+1}^{\omega}\ W^N(V_N^{i.k+1})\overline{\chi}_\delta(V_k)\ \overline{\chi}_\delta(V_{k+1,N})\times\nonumber\\
&&\qquad \qquad\qquad\times\prod_{\substack{1\leq \ell_1\leq k}}
\overline{\chi}_\delta(v_{\ell_1},v_{k+1})\chi_\delta(v_{k+2}, v_{k+1})=\nonumber\\
&&\ \ \ \ \ \ \ =-\frac{(N-k)(N-k-1)}{N}\sum_{1\leq i\leq k}\int dv_{k+1}\int dv_{k+2} \int d\omega \, B_{i.k+1}^{\omega}\ f_{k+2}^N(V_{k+2}^{i.k+1})\overline{\chi}_\delta(V_k)\times\nonumber\\
&&\qquad \qquad\qquad\times\prod_{\substack{1\leq \ell_1\leq k}}
\overline{\chi}_\delta(v_{\ell_1},v_{k+1})\chi_\delta(v_{k+2}, v_{k+1}).
\label{charDECOMPOSITIONgainNL16CASE3}
\end{eqnarray}
Defining $G_3^{(2)}+G_3^{(3)}:=\left(C_{k,k+2}^{N,+} f_{k+2}^N\right)(V_k)$, we finally obtain:
\begin{eqnarray}
&&\left(C_{k,k+2}^{N,+} f_{k+2}^N\right)(V_k)=-\frac{(N-k)(N-k-1)}{N}\sum_{1\leq i\leq k}\int dv_{k+1}\int dv_{k+2} \int d\omega\,  B_{i.k+1}^{\omega}\ f_{k+2}^N(V_{k+2}^{i.k+1})\times\nonumber\\
&&\qquad \qquad\qquad\qquad\qquad\times\, \overline{\chi}_\delta(V_{k+1})\, 
\left(\chi_\delta(v_{i}, v_{k+2})+\chi_\delta(v_{k+2}, v_{k+1})\right).
\label{charDECOMPOSITIONgainNL17CASE3}
\end{eqnarray}

Let us pass to the terms that are nonlinear with respect to the functions $\chi_\delta$'s. The first one is:
\begin{eqnarray}
&&G_3^{(4)}:=\frac{(N-k)(N-k-1)}{N}\sum_{1\leq i\leq k}\int dV_{k,N} \int d\omega\,  B_{i.k+1}^{\omega}\ W^N(V_N^{i.k+1})\overline{\chi}_\delta(V_k)\ \overline{\chi}_\delta(V_{k+1,N})\times\nonumber\\
&&\qquad\qquad\qquad\times\prod_{\substack{1\leq \ell_1\leq k}}
\overline{\chi}_\delta(v_{\ell_1},v_{k+1})\chi_\delta(v_{k+2}, v_{k+1})\sum_{\substack{1\leq \ell_1\leq k\\ k+2\leq \ell_2\leq N}}\chi_\delta(v_{\ell_1},v_{\ell_2}).
\label{g4CASE3}
\end{eqnarray}
By the admissibility of configuration $V_N^{i.k+1}$, we get:
\begin{eqnarray}
&&G_3^{(4)}=\frac{(N-k)(N-k-1)}{N}\sum_{1\leq i\leq k}\int dV_{k,N} \int d\omega\,  B_{i.k+1}^{\omega}\ W^N(V_N^{i.k+1})\overline{\chi}_\delta(V_k)\ \overline{\chi}_\delta(V_{k+1,N})\times\nonumber\\
&&\qquad\qquad\qquad\times\prod_{\substack{1\leq \ell_1\leq k}}
\overline{\chi}_\delta(v_{\ell_1},v_{k+1})\chi_\delta(v_{k+2}, v_{k+1})\sum_{ k+2\leq \ell_2\leq N}\chi_\delta(v_{i},v_{\ell_2}),
\label{g41CASE3}
\end{eqnarray}
namely
\begin{eqnarray}
&&\!\!\!\!\!\!\!\!\!\!\!\!\!\!\!\!\!\!\!\!G_3^{(4)}=\frac{(N-k)(N-k-1)}{N}\sum_{1\leq i\leq k}\int dV_{k,N} \int d\omega \, B_{i.k+1}^{\omega}\ W^N(V_N^{i.k+1})\overline{\chi}_\delta(V_k)\ \overline{\chi}_\delta(V_{k+1,N})\times\nonumber\\
&&\!\!\!\!\!\!\!\!\!\!\!\!\!\!\!\!\!\!\!\!\qquad\qquad\qquad\times\prod_{\substack{1\leq \ell_1\leq k}}
\overline{\chi}_\delta(v_{\ell_1},v_{k+1})\chi_\delta(v_{k+2}, v_{k+1})\left(\chi_\delta(v_{i},v_{k+2})+\sum_{ k+3\leq \ell_2\leq N}\chi_\delta(v_{i},v_{\ell_2})\right),
\label{g42CASE3}\nonumber
\end{eqnarray}
and, thanks to the simmetry of $W^N$,
\begin{eqnarray}
&&\!\!\!\!\!\!\!\!\!\!\!\!\!\!\!\!\!\!\!\!G_3^{(4)}:=\frac{(N-k)(N-k-1)}{N}\sum_{1\leq i\leq k}\int dV_{k,N}\int d\omega\,  B_{i.k+1}^{\omega}\ W^N(V_N^{i.k+1})\overline{\chi}_\delta(V_k)\ \overline{\chi}_\delta(V_{k+1,N})\times\nonumber\\
&&\!\!\!\!\!\!\!\!\!\!\!\!\!\!\!\!\!\!\!\!\qquad\qquad\qquad\times\prod_{\substack{1\leq \ell_1\leq k}}
\overline{\chi}_\delta(v_{\ell_1},v_{k+1})\chi_\delta(v_{k+2}, v_{k+1})\left[\chi_\delta(v_{i},v_{k+2})+(N-k-2)\chi_\delta(v_{i},v_{k+3})\right].
\label{g43CASE3}
\end{eqnarray}
Let us focus on the term:
\begin{eqnarray}
&&\frac{(N-k)(N-k-1)}{N}\sum_{1\leq i\leq k}\int dv_{k+1}...\int dv_{N} \int d\omega \, B_{i.k+1}^{\omega}\ W^N(V_N^{i.k+1})\overline{\chi}_\delta(V_k)\ \overline{\chi}_\delta(V_{k+1,N})\times\nonumber\\
&&\qquad\qquad\qquad\times\prod_{\substack{1\leq \ell_1\leq k}}
\overline{\chi}_\delta(v_{\ell_1},v_{k+1})\chi_\delta(v_{k+2}, v_{k+1})\, \chi_\delta(v_{i},v_{k+2}).
\label{g44CASE3}
\end{eqnarray}
The product
$$
\chi_\delta(v_{k+2}, v_{k+1})\, \chi_\delta(v_{i},v_{k+2})
$$
is different from zero if and only if $v_{k+1}$ and $v_i$ are in the same cell but this is forbidden by the presence of factor 
$$
\prod_{\substack{1\leq \ell_1\leq k}}
\overline{\chi}_\delta(v_{\ell_1},v_{k+1}).
$$
Therefore (\ref{g44CASE3}) does not give any contribution and (\ref{g43CASE3}) yields:
\begin{eqnarray}
&&G_3^{(4)}=\frac{(N-k)(N-k-1)(N-k-2)}{N}\sum_{1\leq i\leq k}\int dv_{k+1}...\int dv_{N} \int d\omega\,  B_{i.k+1}^{\omega}\ W^N(V_N^{i.k+1})\times\nonumber\\
&&\qquad\qquad\qquad\times\overline{\chi}_\delta(V_k)\overline{\chi}_\delta(V_{k+1,N})\prod_{\substack{1\leq \ell_1\leq k}}
\overline{\chi}_\delta(v_{\ell_1},v_{k+1})\ \chi_\delta(v_{k+2}, v_{k+1})\chi_\delta(v_{i},v_{k+3}).
\label{g45CASE3}
\end{eqnarray}
Thus, ignoring the characteristic function $\overline{\chi}_\delta(V_{k+1,N})$ and defining $G_3^{(4)}:=\left(C_{k,k+3}^{N,+} f_{k+3}^N\right)(V_k)$, we get:
 \begin{eqnarray}
&&\left(C_{k,k+3}^{N,+} f_{k+3}^N\right)(V_k)=\frac{(N-k)(N-k-1)(N-k-2)}{N}\sum_{1\leq i\leq k}\int dv_{k+1}\int dv_{k+2}\int dv_{k+3} \, \times\nonumber\\
&&\qquad\qquad\qquad\times\, \int d\omega\,  B_{i.k+1}^{\omega}\, f_{k+3}^N(V_{k+3}^{i.k+1})\overline{\chi}_\delta(V_{k+1})
\ \chi_\delta(v_{k+2}, v_{k+1})\chi_\delta(v_{i},v_{k+3}).
\label{g46CASE3}
\end{eqnarray}

Therefore, the gain term $G$ in (\ref{gain}) 
can be finally cast in the form:
\be
G=\frac{1}{N}L_{k}^{N,+} f_{k}^N+\frac{1}{N}\sum_{s=1}^{2}L_{k,k+s}^{N,+} f_{k+s}^N+\sum_{s=1}^{3}C_{k,k+s}^{N,+} f_{k+s}^N, \label{gainBBGKY}
\ee
where the operators $L_k^{N,+}, \, L_{k,k+1}^{N,+},\, L_{k,k+2}^{N,+},\,  C_{k,k+1}^{N,+}, \, C_{k,k+2}^{N,+},\, C_{k,k+3}^{N,+}$ are given by (\ref{charDECOMPOSITIONgain16app}), (\ref{charDECOMPOSITIONgainLIN5}), (\ref{charDECOMPOSITIONgainNL4}), (\ref{charDECOMPOSITIONgainNL8case3}), (\ref{charDECOMPOSITIONgainNL17CASE3}), (\ref{g46CASE3}) respectively.

Analogously we can handle the loss terms to get the desired result.


\begin{thebibliography}{2}

\bibitem{BCEP1}  D. Benedetto, F. Castella, R. Esposito, M. Pulvirenti, \emph{Some considerations on the derivation of the nonlinear quantum Boltzmann equation}.\  J. Statist. Phys. {\bf 116},  no. 1-4 (2004),  pp 381-410.


\bibitem{BCEP} D. Benedetto, F. Castella, R. Esposito, M. Pulvirenti, \emph{On the Weak-Coupling Limit for Bosons and Fermions}.\ Mathematical Models and Methods in Applied Sciences
\  {\bf 15}, no. 12 (2005), pp 1811-1843.

\bibitem{BCEP3}  D. Benedetto, F. Castella, R. Esposito, M. Pulvirenti, \emph{Some considerations on the derivation of the nonlinear quantum Boltzmann equation. II. The low density regime}. \ J. Stat. Phys. {\bf 124}, no. 2-4 (2006),  pp 951-996.


\bibitem{BCEP4}  D. Benedetto, F. Castella, R. Esposito, M. Pulvirenti, \emph{From the N-body Schr\"odinger equation to the quantum Boltzmann equation: a term-by-term convergence result in the weak coupling regime}. \ Comm. Math. Phys. {\bf 277},  no. 1 (2008),  pp 1-44.

\bibitem{DarioMario}  D. Benedetto, M. Pulvirenti, \emph{The Classical Limit for the Uehling-Uhlenbeck Operator}.\ 
Bulletin of the Institute of Mathematics Academia Sinica (New Series)
{\bf 2}, no. 4 (2007)  pp 907-920

\bibitem{BPS} A. Bobylev, M. Pulvirenti, C. Saffirio, \emph{From Particle Systems to the Landau Equation: A Consistency Result}.  Comm. Math. Phys. {\bf 319}\ (2013), pp 683-702.


\bibitem{Pulv} C. Cercignani, R. Illner, M. Pulvirenti, \emph{The Mathematical Theory of Dilute Gases}. Springer-Verlag, Berlin (1994).

\bibitem{DEP}  A. De Masi, R. Esposito, E.Presutti, \emph{Kinetic limits of the HPP cellular automaton}. J. Stat. Phys. {\bf 66} (1992), pp 403-464.

\bibitem{DOLB} J. Dolbeault, \emph{Kinetic models and quantum effects: A modified Boltzmann equation
for Fermi-Dirac particles}. Arch. Rat. Mech. Anal. {\bf 127} (1994), pp 101-131.

\bibitem {ESY} L. Erd\"os, M. Salmhofer, H.-T. Yau,\ \emph{On the quantum Boltzmann equation}.\  J. Stat. Phys.  \textbf {116} (2004), pp 367-380.

\bibitem{GSRT} I. Gallagher, L. Saint Raymond and B. Texier, \emph{From Newton to Boltzmann: the
case of short-range potentials}.\  preprint arXiv:1208.5753v2 (2013).

\bibitem{HS} E. Hewitt, L. J.  Savage, \emph{Symmetric measures on Cartesian products}.\ Trans. Amer. Math. Soc. {\bf 80} (1955), pp 470-501.

\bibitem{IP} R. Illner and M. Pulvirenti, \emph{Global Validity of the Boltzmann equation for a Two-
Dimensional Rare Gas in the Vacuum}.\  Comm. Math. Phys. {\bf 105} (1986), pp 189-203.

\bibitem{KacA} M.\ Kac, \emph{Foundations of kinetic theory}.\ in ''Proceedings of the Third Berkeley Symposium on Mathematical Statistics and Probability'', University of California Press, Berkeley and Los Angeles (1956).



\bibitem{King} F. King, \emph{BBGKY Hierarchy for Positive Potentials}.\  Ph.D. Thesis, Department of
Mathematics, Univ. California, Berkeley (1975).

\bibitem{LANFORD} O. E.  Lanford,\ \emph{Time evolution of large classical system}.\ E. J. Moser (ed.), Lecture
Notes in Phys. {\bf 38}, Springer, New York (1975), pp 70-111. 

\bibitem{MM} S. Mishler, C. Mouhot, \emph{Kac's Program in Kinetic Theory}.\  Inventiones mathematicae
{\bf193}, Issue 1  (2013), pp 1-147.

\bibitem{nord} L. W. Nordheim, \emph{On the Kinetic Method in the New Statistics and Its Application in the Electron Theory of Conductivity}.\  Proc. Roy. Soc. London. Ser. A, {\bf 119}, no. 783 (1928), pp 689-698.

\bibitem{Pulvi} M. Pulvirenti,\ \emph{The weak-coupling limit of large classical and quantum systems}.\  International Congress of Mathematicians. Vol. {\bf III},  Eur. Math. Soc., Z\"urich  (2006), pp 229-256.

\bibitem{PSS} M. Pulvirenti, C. Saffirio, S. Simonella,\ \emph{On the validity of the Boltzmann equation
for short range potentials}.\ preprint arXiv:1301.2514v1 (2013).


\bibitem{PWZR} M. Pulvirenti, W. Wagner, M.B. Zavelani Rossi, \ \emph{Convergence of particle schemes for the Boltzmann equation}.\  Eur. J. Mech. B/Fluids\  {\bf 13} (1994), pp 339-351.

\bibitem{RUELLE} D. Ruelle, \ \emph{Statistical mechanics: Rigorous results}.\  W. A. Benjamin, Inc., New York-Amsterdam (1969).


\bibitem{S} H. Spohn, \emph{Boltzmann equation and Boltzmann hierarchy}.\  In ''Kinetic Theories and
the Boltzmann equation'', Lecture Notes in Mathematics {\bf 1048} ed. C.
Cercignani, Springer-Verlag, Berlin (1984), pp 207-220. 

\bibitem{S1} H. Spohn,  \emph{Large Scale Dynamics of Interacting Particles}.\  Texts and Monographs in Physics, Springer-Verlag, Heidelberg (1991).

\bibitem{S2} H. Spohn,  \emph{Boltzmann hierarchy and Boltzmann equation}.\   Kinetic
theories and the Boltzmann
equation (Montecatini, 1981), pp 207-220.


\bibitem{U} K. Uchiyama, \emph{Derivation of the Boltzmann equation from particle dynamics}.\  Hiroshima Math. J. {\bf 18} (1988), pp 245-297.

\bibitem{UU} E. A. Uehling and G. E. Uhlenbeck, {\em Transport phenomena in Einstein-Bose and
Fermi-Dirac gases}. Phys. Rev. {\bf 43} (1933), pp 552-561.



\end{thebibliography}
\end{document}